%% file: main.tex
\documentclass[preprint,12pt]{elsarticle} 

\usepackage{svg}
\usepackage{graphicx}
\usepackage{amsmath, amssymb}
\usepackage{newtxtext,newtxmath}
\usepackage{hyperref}
\usepackage{color}
\usepackage{lineno}
\usepackage{booktabs}
\usepackage{multirow}
\usepackage{float}
\usepackage{listings}
\usepackage{xcolor}
\usepackage{caption}
\usepackage{subcaption}
\usepackage{soul}
\usepackage{rotating}
\usepackage{pdflscape}
\usepackage{array}
\usepackage{longtable}
\usepackage{makecell}
\usepackage{pdflscape}
\usepackage{ragged2e}

\newcolumntype{L}[1]{%
  >{\RaggedRight\arraybackslash}p{#1}%
}

\newcolumntype{C}[1]{%
  >{\Centering\arraybackslash}p{#1}%
}

\newcommand{\asbaseline}{As baseline}

\newcommand{\tablesection}[1]{%
  \addlinespace[0.55em]
  \multicolumn{6}{@{}l}{\textbf{#1}}\\[-0.15em]
  \midrule
}

\usepackage{numprint}
\npdecimalsign{\ensuremath{.}}

\journal{Journal Name (e.g., Simulation Modelling Practice and Theory)}

\definecolor{mygreen}{RGB}{0,200,0} 

\newcommand{\mv}[1]{MultiVeStA{}}

\newcommand{\come}[1]{{#1}}
\newcommand{\comee}[1]{{#1}}

\begin{document}

\begin{frontmatter}

\title{Reliable and explainable analysis of AVS/RS 
by Statistical Model Checking and  Process Mining}

\author[a]{Emanuele Guerrazzi\corref{cor1}}
\ead{emanuele.guerrazzi@santannapisa.it}

\author[a]{Andrea Vandin}
\ead{andrea.vandin@santannapisa.it}

\cortext[cor1]{Corresponding author}

\address[a]{L'EMbeDS, Sant'Anna School of Advanced Studies, Pisa, Italy}

\begin{abstract}
Automated Vehicle Storage and Retrieval Systems (AVS/RS) are increasingly adopted to boost warehouse throughput while containing operational costs. \come{We present a tool-supported analysis workflow that combines discrete-event simulation, Statistical Model Checking (SMC), and Process Mining (PM) techniques to quantify and explain AVS/RS performance. }
SMC, implemented through the \mv{} tool, delivers statistically reliable estimates of key indicators, such as cycle times \come{and inventory consolidation}, across alternative stocking and picking policies with user-specified confidence levels. \mv{} orchestrates multiple independent simulations in a black-box manner, enabling automated verification. PM ingests the resulting event logs to uncover the behavioural patterns and bottlenecks that drive those indicators, closing the interpretability gap typical of simulation studies. We instantiate the framework on a new open-source AVS/RS simulator configured according to recent industrial guidelines and evaluate multiple control strategies \come{and different scenarios}. 
\come{Results show that the coordinated use of SMC and PM supports both predictive what-if analyses and post-hoc diagnostics. To the best of our knowledge, this is the first application of a combined SMC-PM workflow to AVS/RS operations, complemented by a reusable open-source simulation and analysis toolchain.}

\end{abstract}


\end{frontmatter}



\section{Introduction}
Automated Vehicle Storage and Retrieval System (AVS/RS) is a type of Autonomous Storage and Retrieval Systems (AS/RS) that gained remarkable popularity over the last two decades \citep{Marchet2013} and remains widely adopted today thanks to its high throughput, scalability, and energy efficiency \citep{Yang2023}. 
The throughput of such systems is higher compared to previous AS/RS, such as Crane-Based Storage and Retrieval Systems, 
thanks to the parallelization of operations across multiple vehicles that can work simultaneously on different levels \citep{Lerher2021}. The same holds for scalability, as additional shuttles can be easily added to a working system \citep{Ekren2024},  
and for energy efficiency, as 
shuttles are less energy-intensive than traditional stacker cranes \citep{Guerrazzi2019, YetkinEkren2021}. 
These characteristics make AVS/RS particularly suitable for a wide range of warehouse applications, from handling small e-commerce items to managing heavier pallets \citep{Boysen2019, Lupi2024}. 
Indeed, another key advantage of AVS/RS is the breadth of feasible layouts and design variants. 

%

%

\begin{figure}[t]
    \centering
    \includegraphics[width=1\linewidth]{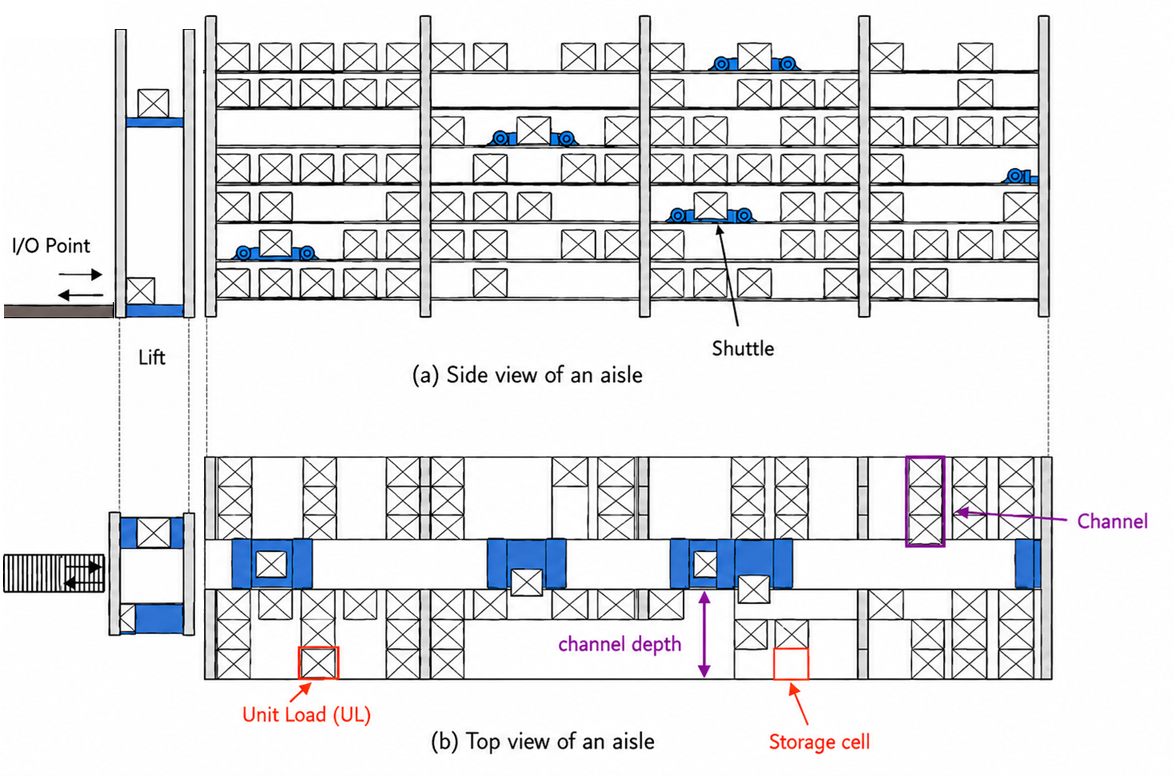}
    \caption{Basic layout of a multi-deep AVS/RS. The figure shows a storage rack arranged in aisles and levels, shuttles traveling horizontally along each level, and lifts providing vertical transfers between levels and I/O stations. Adapted from \citep{Kucukyasar2021}}
    \label{fig:avsrs_layout}
\end{figure}

\come{Figure ~\ref{fig:avsrs_layout} illustrates the physical layout of a representative multi-deep AVS/RS. The warehouse is organized into aisles, each divided into multiple levels connected by a lift serving the I/O station. At each level, storage is arranged into multiple channels, \come{hereafter also referred to as tunnels, with the two terms used interchangeably throughout the paper}.
Each channel consists of successive storage cells along its depth, where Unit Loads (ULs) are stored. Since each unit load contains a single SKU in the considered case study, the remainder of the paper refers to unit loads simply as SKUs. Shuttles travel horizontally along each level, while the lift enables vertical movements between levels and the I/O station. Figure ~\ref{fig:avsrs_layout}(b) highlights the spatial relationship between aisles, levels, channels, and storage cells.}

Due to the inherent complexity of these systems, simulation-based approaches are commonly employed to evaluate their performance under different configurations and operational scenarios \cite{10.5555/554952}. 
%
For simulation-based studies to be meaningful,  it is essential for them to be 
statistically valid \citep{Ferrari2024}.
In this regard, Statistical Model Checking (SMC) \come{comprises a family of simulation-based verification techniques that combine sampled system executions with statistical inference to} estimate quantitative \come{properties or assess probabilistic specifications with explicit statistical confidence} \citep{Agha2018,legay2010, Sen2005}. 
A particularly suitable variant for engineering applications is black-box SMC, which does not require explicit access to the internal structure of the model and can therefore be applied directly to existing simulators without reformulating their logic into a formal language \citep{Sen2005}.
This is supported by a number of tools, including \mv{} \citep{Vandin2022,DBLP:conf/ifm/GilmoreRV17,sebastio2013multivesta}
\footnote{\url{https://github.com/andrea-vandin/MultiVeStA/wiki}} 
a general-purpose SMC software capable of tool-chaining with external simulators written multiple languages (e.g., MATLAB, Python, C++, or Java) 
\citep{Vandin2022}. 
\mv{} enables the automated statistical analysis of complex simulation models while enhancing their reproducibility and accuracy, and it reduces potential biases of simulation-based approaches \citep{Vandin2022, Casaluce2024}.
Considering specifically AVS/RS models, the integration with \mv{} enables rigorous simulation-driven performance evaluation of AVS/RS under different configurations, control strategies, and environmental conditions. 

However, a well-known limitation of traditional SMC is that results are provided only in the form of numerical estimates, without additional insights into the behaviours that generated them \citep{Agha2018,Casaluce2024}. If a performance metric unexpectedly degrades, the modeller cannot easily determine whether this is due to a true system bottleneck or a modelling flaw.
To overcome this black-box limitation, recent works have integrated SMC with Process Mining (PM) \citep{Casaluce2024,DBLP:conf/bpm/CasaluceBCV22}. 
PM is a family of techniques from process-oriented data science that analyse the executions (i.e., traces, or logs) of activities generated by information systems with the goal of identifying process patterns, bottlenecks, and other issues in a model~\cite{VanderAalst2016}. PM is gaining more and more popularity in industry because visualizing the flow of activities can help to identify opportunities for improvement (see, e.g., the recent study by Deloitte on the industrial adoption of PM~\cite{deloitte}). 
The idea behind the integration of SMC and PM is that, by analysing logs generated by simulations, PM can reconstruct a compact behavioural model (a graphical representation of the behaviour) that exposes how the system actually operates. This enables automatic detection of unexpected dynamics, unreachable transitions, and performance-critical paths, improving both model interpretability and trust in the results.
So far, the integration of PM and SMC has been investigated in specific modelling domains such as software product lines, collective adaptive systems, or threat modelling \citep{Casaluce2024,DBLP:conf/isola/CasaluceTV24,DBLP:conf/damocles/CasaluceBCLV24}. 

Related approaches have connected process-oriented modelling and PM artifacts with formal verification without relying on SMC. For example, Cosma et al. ~\cite{Cosma2023DCRPetriNets} propose a semantics-preserving transformation from Dynamic Condition Response graphs to
safe Petri nets, enabling process specifications derived from texts and event logs
to be analysed through mature Petri-net and model-checking tools. 

To the best of the authors' knowledge, only a few SMC approaches to analyse warehousing have been proposed \cite{Ferrari2024}, while PM-based approaches are more widespread (see, e.g.,~\cite{Liiv2014, Kedem2018, Montianrat2020}). However, no approaches combining SMC and PM have been proposed for warehousing. Furthermore, no SMC or PM alone approaches have ever been applied to AVS/RS systems in warehousing.
%
The goal of this work is to fill these gaps by investigating the suitability of combining SMC and PM for the quantitative and behavioural analysis of AVS/RS simulators. 
In this way, we aim to obtain \comee{a} statistically sound and interpretable analysis. 

\come{To investigate the robustness} and practical relevance of our approach, we created a new open source \come{tier-to-tier multi-deep} AVS/RS simulator consistent with the operational parameters and design configurations reported in the literature \citep{Marolt2023}. 
\come{The experimental campaign contains a controlled synthetic baseline and three one-factor robustness variants, each designed to isolate a different source of operational variability while keeping all remaining model parameters unchanged. The first scenario adopts a literature-inspired workload calibrated to the number of stocking and picking orders reported in the industrial AVS/RS case study by \cite{Lupi2024}. The second introduces slower vehicle kinematics inspired by the operating parameters considered by \cite{Battarra2022}, thereby assessing the sensitivity of policy performance to reduced shuttle, satellite, and lift dynamics. The third is a stress-test scenario derived from the baseline configuration but initialized \comee{with a denser inventory configuration, allowing us to examine policy behaviour under a more constrained initial storage state.}. Since complete industrial event logs are not publicly available, these scenarios should be interpreted as external-data-informed robustness and sensitivity assessments rather than as full empirical validation.}

The simulator is publicly available at 
\citep{zenodo_simulator}.
%
This is actually an additional notable contribution of our work, as we are not aware of any other available open source AVS/RS simulator.

\come{The contribution of this study is integrative and application-oriented rather than algorithmic. We do not introduce a new DES formalism, SMC procedure, or PM algorithm. Specifically, the paper contributes: (i) a coordinated analysis workflow in which black-box SMC provides confidence-controlled KPI estimates from independent simulation replications, while PM analyses the generated event logs to characterise control-flow, performance, and resource-use patterns; (ii) the application of this workflow, to the best of our knowledge for the first time, to a tier-to-tier, multi-deep AVS/RS across multiple operating policies and robustness scenarios; and (iii) an open-source, \comee{parameterisable} AVS/RS simulator together with the adapter and supporting analysis artifacts required to reproduce the study. The methodological value therefore lies in the complementary use of established techniques within a common simulation campaign, rather than in a new algorithm for any of the individual components.}

Overall, this study seeks to address the following Research Question:
\begin{itemize}
\item[\emph{RQ}] How can the integration of Statistical Model Checking and Process Mining enhance the reliability and interpretability of simulation-based performance evaluation of AVS/RS?
\end{itemize}


To answer this RQ, we exploit SMC to obtain statistically reliable estimates of AVS/RS performance, while PM is used to \come{reconstruct and analyse the operational} behaviour \come{recorded in} simulation \come{event} logs. This combination supports both a rigorous evaluation of AVS/RS performance under alternative stocking and picking policies and an interpretation of performance drivers \come{of the observed KPI trends, including  bottlenecks, change in the process behaviour and performance, and workload distribution.}

\section{Related Work}
This section reviews relevant contributions in three main areas: 
 the application of discrete-event simulation (DES) in AVS/RS (Section~\ref{sec:des}), the integration of Statistical Model Checking (SMC) in warehouse systems (Section~\ref{sec:smc}), and the application of Process Mining (PM) techniques to warehousing (Section~\ref{sec:pm}). 
These are summarized in Table~\ref{tab:related_work}, highlighting how the present study extends the state of the art.
The analysis highlights both the advances and the gaps that motivate the integrated framework proposed in this study.

\begin{table*}[t]
\centering
\caption{Summary of relevant studies on AVS/RS DES, SMC, and PM.}
\label{tab:related_work}
\scalebox{0.67}{
\begin{tabular}{@{}p{1cm} p{3.5cm} p{3.8cm} p{3.4cm} p{2cm} p{1.1cm} p{1.1cm} p{1.5cm}@{}}
\toprule
\textbf{Ref.} & 
\textbf{\comee{AVS/RS analysed}} & 
\textbf{Goal} & 
\textbf{Method/SW} & 
\textbf{Data used} & 
\textbf{SMC} & 
\textbf{PM} & 
\textbf{Code avail.} \\ 
\midrule
\citep{Ekren2011} & AVS/RS Single-deep & Throughput optimization & DES (ARENA) & Synthetic (simulation scenarios) & No & No & No \\
\citep{Chan2011} & Manual Multi-level & Order-picking productivity with class-based storage & DES & Real (industrial data) & No & No & No \\
\citep{Guerrazzi2019} & AVS/RS Multi-deep & Energy consumption and recovery & DES (MATLAB) & Real & No & No & No \\
\citep{ekren2020} & SBS/RS & Energy consumption and recovery & DES (ARENA) & Synthetic & No & No & No \\
\citep{Jerman2021} & AVS/RS Movable-lifts & Throughput analysis & DES (ARENA) & Synthetic & No & No & No \\
\citep{Ferrari2024} & Multi-level multi-deep AS/RS & DT model validation & DES (AnyLogic) + DoE & Real (laboratory data) & Yes & No & No \\
\citep{Baouya2024} & Multi-robot warehouse system & Functional assurance and orchestration verification & MDD (BIP) + SMC-BIP & Synthetic + real (Robotnik platform) & Yes & No & No \\
\citep{Mahendrawathi2015} & Warehouse Management (SAP WM) & Conformance and bottleneck analysis & PM (ProM – Heuristic Miner) & Real (ERP logs) & No & Yes & No \\
\citep{Knoll2019} & Internal logistics & Value stream mapping and performance analysis & PM + value stream mapping & Real (internal logistics event logs) & No & Yes & No \\
\citep{Napieraj2025} & Logistics warehouse process & End-to-end process discovery, conformance and performance analysis & PM (ProM – Inductive/Heuristic/Alpha Miner, Petri nets) & Real (WMS event logs) & No & Yes & No \\
\midrule
\textbf{This work} & Multi-deep AVS/RS & Picking and stocking cycle times, inventory consolidation & DES (MATLAB) + \mv{} & Synthetic & \textbf{Yes} & \textbf{Yes} & \textbf{Yes} \\
\bottomrule
\end{tabular}
}
\end{table*}

\subsection{Discrete-Event Simulation in AVS/RS}\label{sec:des}

DES has long been a cornerstone for the design and performance evaluation of AVS/RS as it allows for the detailed representation of their complex interactions and dynamic behaviour \citep{dehrami2020}. It enables engineers to test various control strategies and layout configurations under realistic operating conditions without disrupting real operations \citep{Agalianos2020} or even at the design stage \citep{Marchet2013}, allowing to evaluate ``what-if'' scenarios.

A recent review \citep{Kizi2024} provides a comprehensive overview of AVS/RS studies based on DES, categorizing them by design and performance objectives. This bibliometric analysis showed that DES remains the dominant method for capturing the stochastic and dynamic nature of AVS/RS operations. The review also identifies significant research gaps concerning stocking and picking policies, highlighting the need for more integrative DES frameworks that encompass physical layout and operational control. %
In the following, we revisit the studies that most closely resemble the system proposed in this work and the adopted methodological framework.

One of the earliest \comee{studies was} proposed by Ekren \citep{Ekren2011}, \comee{who} analysed several AVS/RS \comee{configurations} to \comee{assess} their impact on throughput \comee{using} the commercial software ARENA\cite{arena}. 

In the context of manual multi-level rack distribution warehouses, Chan and Chan \citep{Chan2011} use discrete-event simulation to assess how class-based storage policies affect order-picking productivity, further illustrating the suitability of DES for studying storage and retrieval strategies across different warehouse technologies. 
Other works used DES to evaluate the energy consumption and recovery within warehousing processes, demonstrating substantial energy savings compared to traditional CBAS/RS \citep{Guerrazzi2019, ekren2020}.
More recently, several AVS/RS configuration \comee{variants have} been studied, \comee{including} the movable lifts \comee{configuration} proposed \comee{in} \citep{Jerman2021}, which shows \comee{that} separating input and output stations \comee{can increase} throughput and \comee{improve} resource balance under the same number of vehicles.
In parallel, related shuttle-based storage and retrieval systems (SBS/RS) have been analysed through DES to investigate control policies, such as the balancing rules studied by \citep{Guerrazzi2024}.
Collectively, these works confirm DES as an effective approach for evaluating performance, energy, and scalability trade-offs in AVS/RS designs.


\subsection{Integrating Statistical Model Checking in warehousing}\label{sec:smc}

While DES provides valuable performance insights, \come{observations obtained from a finite set of simulation runs do not by themselves quantify the uncertainty associated with the estimated properties}. SMC addresses this limitation with formal verification, enabling quantitative estimation of temporal logic properties with user-defined statistical confidence \citep{legay2010,Agha2018}
This approach is particularly useful in complex domains, such as automated warehouses.

Yet, as highlighted by Ferrari et al. \citep{Ferrari2024}, rigorous validation of simulation models is a prerequisite for any form of trustworthy statistical or simulation-based analysis. Their work demonstrates how a Design of Experiment (DoE) methodology can be effectively employed to calibrate and validate an AS/RS discrete-event model against real laboratory data, ensuring that the digital replica accurately reflects the stochastic behaviour of the physical system. 

Beyond traditional validation, formal verification has recently emerged as a complementary strategy to ensure the reliability of warehouse automation systems.
In this direction, Baouya et al. \citep{Baouya2024} introduced the \textit{BRAIN-IoT} framework, which integrates Model-Driven Design with SMC through the so-called behaviour–Interaction–Priority formalism. The framework enables the formal specification and probabilistic verification of orchestrated robotic fleets, offering functional assurance prior to deployment.

\come{In the context of AVS/RS, however, SMC has not yet been integrated with existing DES models used directly by practitioners and researchers in the area. Black-box SMC frameworks such as \mv{} \citep{Vandin2022} can enable this, allowing for the statistically rigorous evaluation of AVS/RS performance under different layouts and control policies while maintaining compatibility with legacy simulators. } %
%
\mv{} showed its great generality and effectiveness in several domains and simulators such as crowd steering scenarios~\cite{DBLP:conf/hpcs/PianiniSV14}, public transportation systems~\cite{gilmore2014analysis,ciancia2016tool}, lending pools in decentralized finance~\cite{DBLP:conf/isola/BartolettiCJLMV22}, highly-configurable systems~\cite{DBLP:journals/tse/BeekLLV20,DBLP:conf/fm/VandinBLL18}, business process modelling~\cite{DBLP:journals/jss/CorradiniFPRTV21}, security threat modelling~\cite{ter2021quantitative,10.1007/978-3-031-25383-6_18}, adaptive robotic systems~\cite{DBLP:conf/birthday/BelznerNVW14,DBLP:journals/scp/BruniCGLV15,10.1007/978-3-031-73709-1_13}, collective adaptive systems in general \cite{DBLP:conf/wsc/GalpinGLV18,DBLP:conf/isola/CasaluceTV24}, agent-based models~\cite{Vandin2022,10.1007/978-3-031-75434-0_26,DBLP:journals/corr/abs-2604-04543,blando2026agenticagentbasedmodelsfeasibility}. %
%
\come{ Notably, \mv{} is also natively integrated in the popular frameworks for agent-based models Mesa \cite{DBLP:conf/vecos/Vandin24} and NetLogo
\cite{DBLP:journals/corr/abs-2509-10977}.}

\subsection{Process Mining in Warehousing}\label{sec:pm}
PM  techniques allow the extraction, reconstruction, and analysis of event-driven process models from system logs.
Initially introduced by \citep{Aalst2007WorkflowMining} and later systematized in \citep{Aalst2016ProcessMining}, PM has become a key tool for bridging the gap between model-based analysis and real-world system behaviour.
By uncovering how processes are actually executed, PM enables the detection of deviations, bottlenecks, and inefficiencies that would otherwise remain hidden in traditional modelling approaches.

In industrial environments, PM has been increasingly adopted to assess performance and conformance of manufacturing and logistics processes \citep{Garcia2019, Alnahas2023}.
A notable example in warehousing is provided by \citep{Mahendrawathi2015}, who applied process mining to the Warehouse Management module of SAP in a manufacturing company.
Using the \textit{Heuristic Miner} algorithm implemented in \textit{ProM} \citep{prom}, they reconstructed the material movement process and compared it with the company’s standard procedure.
The analysis revealed additional nonstandard activities, capacity-related bottlenecks, and violations of the FIFO rule, leading to recommendations for warehouse resource balancing and planning stability.
In a related context of internal logistics, Knoll et al. \citep{Knoll2019} integrate multidimensional process mining with value stream mapping to analyse in-plant material flows and identify performance bottlenecks.

The more recent contribution of Napieraj \citep{Napieraj2025} extended the application of PM to large-scale logistics operations. 
They conducted a comprehensive analysis of a real industrial warehouse process using 
\textit{ProM}. Their workflow included event-log preprocessing, process discovery, conformance checking, 
temporal and organizational mining, and error analysis. 
By modelling the process as a Petri net and experimenting with multiple discovery algorithms 
(\textit{Inductive Miner}, \textit{Heuristic Miner}, and \textit{Alpha Miner}), they demonstrated how PM can 
effectively diagnose deviations, quantify performance, and detect organizational inefficiencies. 
The study also emphasizes practical aspects such as event-log quality, noise filtering, and concept drift, 
highlighting PM as a valuable operational tool for continuous improvement in logistics environments.


\medskip
Taken together, the reviewed studies reveal a clear research gap.
While DES has been extensively used to evaluate performance and design variants of AVS/RS, few works have pursued statistically rigorous validation, and none have combined such analyses with behavioural interpretation.
SMC provides the missing formal layer to quantify uncertainty and ensure result reliability, whereas PM complements it by exposing the underlying operational dynamics hidden in simulation data, but they lack rigorous studies on AVS/RS operations.
The convergence of these three perspectives thus offers a promising path toward transparent, reproducible, and interpretable performance evaluation of AVS/RS.

\section{Methodology}
\begin{figure}[t]
    \centering
    \includegraphics[width=0.85\linewidth]{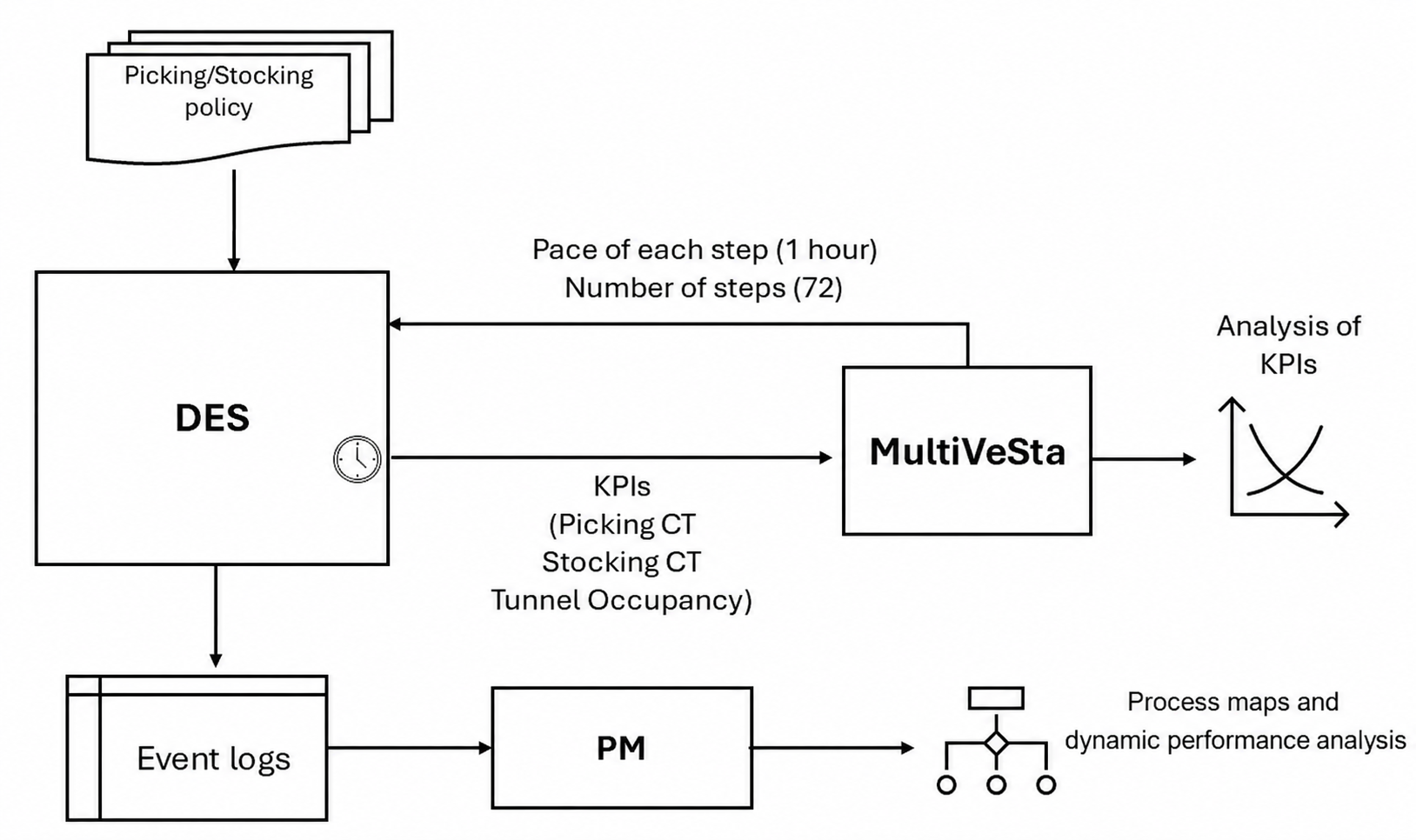}
    \caption{Proposed methodology integrating DES modelling, SMC, and PM.}
    \label{fig:methodology}
\end{figure}

The proposed methodology integrates DES, SMC, and PM into a 
\come{coordinated analysis} framework for the quantitative and behavioural analysis of AVS/RS. 
\come{Accordingly, the workflow should be understood as a tool-supported integration of established methods rather than as a new SMC, PM, or DES technique.}
The workflow, summarized in Figure~\ref{fig:methodology}, \comee{comprises} three main phases that jointly ensure reproducibility, statistical rigor and interpretability of the simulation results: modelling the system in a DES tool (Section~\ref{sec:modelling}); integrating the model with
\mv{} (Section~\ref{sec:methmv}); and performing Process Mining analysis on the
collected event logs (Section~\ref{sec:methpm}).

\subsection{System modelling}\label{sec:modelling}
\begin{figure}[t]
    \centering
    \includegraphics[width=\linewidth]{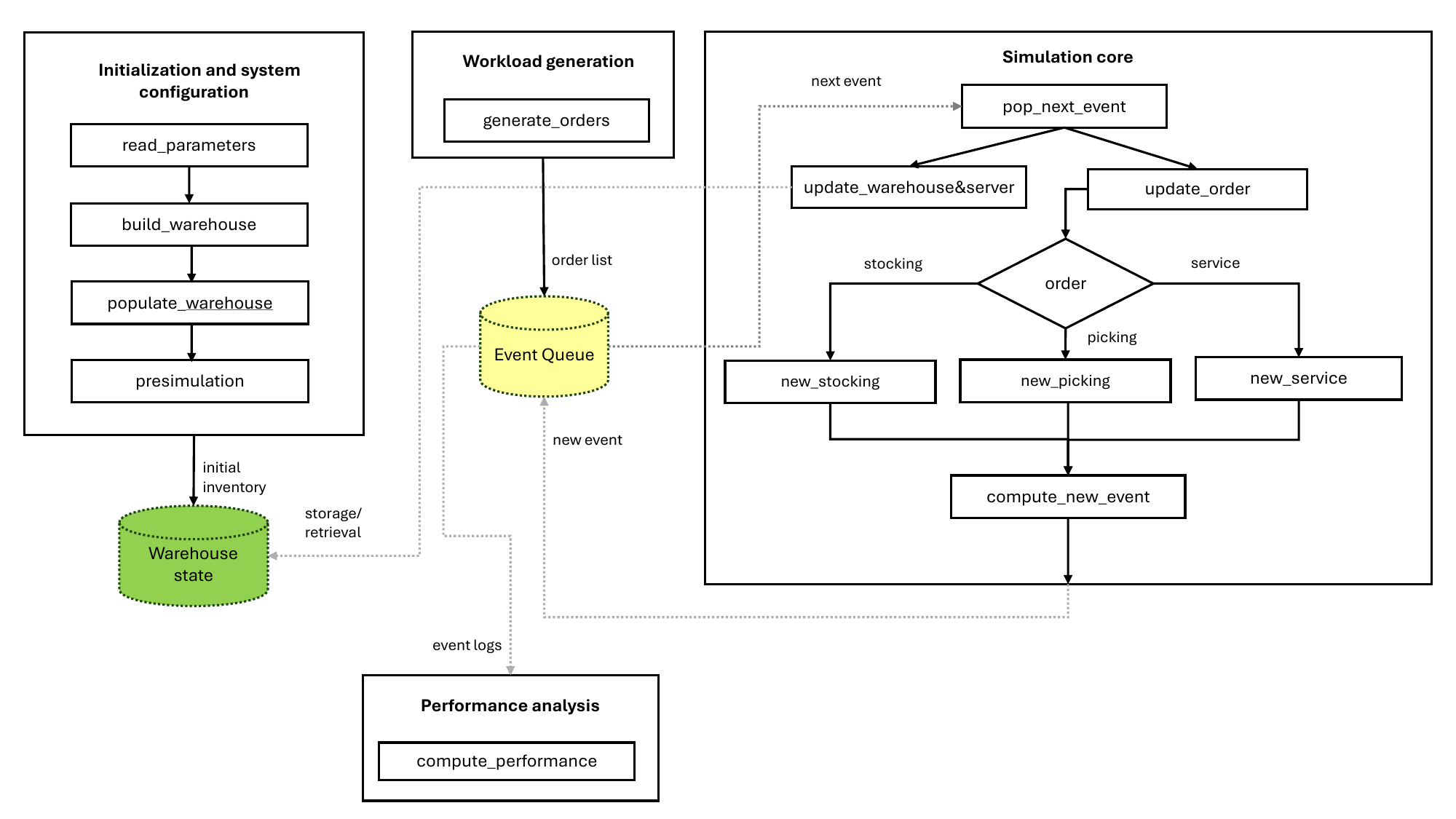}
    \caption{Software architecture of DES}
    \label{fig:system_model}
\end{figure}

The system modelling proposed in this work relies upon the development of a DES model due to its intrinsic ability to capture the dynamic and complex nature of storage and retrieval tasks carried out inside a highly parallelized system such as AVS/RS \citep{Derhami2020}.
The AVS/RS considered in this study is analogous to the one depicted in Figure \ref{fig:avsrs_layout}.
We developed a DES model using \comee{MATLAB}, following a modular workflow and according to the well-known methodology of Marchet et al. \citep{Marchet2013}.
In this section we focus only on the software architecture of the model, and on its main components, while we provide details on the considered scenario in Section ~\ref{sec:casestudy}.
Figure~\ref{fig:system_model} summarizes the software architecture of the DES.

\subsubsection*{Initialization and system configuration}
The first component of the software architecture allows users to define the AVS/RS parameters such as geometric configuration (i.e., number and geometric properties of aisles, levels, channels), kinematic data (i.e., velocity, acceleration, and deceleration profiles for lifts, shuttles and satellites), and operational logic (i.e., selection of storage and retrieval policies) through the \textit{read\_parameters} function.

Afterwards, the static environment of the warehouse is constructed by using (\textit{build\_warehouse}), initializing the coordinate system of storage cells and populating the rack with an initial inventory snapshot of different SKUs (\textit{populate\_warehouse}).
Finally, all shuttles and lifts are created and correctly positioned in their starting position of the warehouse (\textit{presimulation}).
In the output of this module, we have the current state of the warehouse, with the items placed in their initial locations.

\subsubsection*{Workload generation}
The second module of this DES (\textit{generate\_orders}) is responsible \come{for generating the workload that populates the event queue. The simulator supports two complementary workload-generation modes, depending on the objectives of the experimental campaign. 
The first corresponds to the controlled synthetic scenario used as the baseline throughout this work.
In this configuration, the workload is generated using statistical distributions designed to reproduce realistic warehouse demand patterns while maintaining full control over the experimental conditions.}
Specifically, an exponential distribution is used to generate the number of different SKUs, exploiting its intrinsic skewness, which mimic demand where a small set of items is ordered very frequently while many items are ordered rarely \citep{Merschforman2019}.
The Poisson distribution is the standard model for the count of events occurring in a fixed interval of time or space under the assumptions of independence and a constant average rate, making it a natural choice for modelling arrival counts such as customer orders per day \citep{Lorenz2025}.
Finally, to reproduce intraday congestion effects due to demand peak, the normal distribution is used to simulate a peak of activity around 12:30 PM with a spread of ±2 hours, truncated to stay within shift hours (07:00--20:00). This generates a pronounced activity peak around lunchtime and fewer orders in the early morning and late afternoon,  yielding a more realistic scenario \citep{liu2020}.

\come{The second mode supports the experiment in which the workload is calibrated using operational statistics reported in \citep{Lupi2024}. In this case, aggregate workload characteristics (e.g., number of SKUs, daily storage and retrieval missions, and vehicle kinematics) are derived from the reference study, while the detailed order stream is reconstructed through modelling assumptions whenever the original execution trace is unavailable. In particular, because the published industrial case does not provide timestamped transaction logs, initial inventory snapshots, or SKU allocation histories, order arrivals are assumed to be uniformly distributed within the daily operating window, while SKU requests follow a Zipf distribution. These modelling assumptions were necessary because the original timestamped industrial trace and inventory history are not publicly available. The specific calibration adopted in this work is described in Section~\ref{sec:casestudy}.

Regardless of the adopted workload-generation mode, customer orders are generated as collections of SKUs and subsequently decomposed into atomic SKU-level missions}. 
This allows the model to handle the retrieval of multiple items within a single order asynchronously, mirroring real-world parallel processing capabilities.
The generated order list is then used to populate the initial state of the event queue.

\subsubsection*{Simulation core}
The third component is the essence of the whole DES and relies on the well-known mechanism called Next Event Time Advances, which determines the time of occurrence of futuristic events based on a list \citep{10.5555/554952,Agalianos2020}. In this way, the simulation clock is updated only when the next event occurs. The mechanism resolves the dynamic evolution of the system through a global event list, which sorts pending events chronologically in order to accomplish the total amount of orders.

The main simulation loop functions as follows:
\begin{itemize}
    \item \textbf{Event Fetching}: the earliest scheduled event is extracted from the future event list using the \textit{pop\_next\_event} procedure.
    \item \textbf{Clock Advancement}: The simulation clock is advanced to the timestamp associated with the extracted event.
    \item \textbf{State Update}: The system state is updated to reflect the effects of the event (e.g., a shuttle reaching a storage bay or a lift completing a vertical movement). This update is handled by the \textit{update\_warehouse\_and\_server} function, which is responsible for maintaining the warehouse state as well as the positions and availability of shuttles and lifts.
    \item \textbf{Order state progression}: The corresponding order is updated through the \textit{update\_order} function, which advances the order workflow by one processing step (e.g., a stocking order transitioning from ``Shuttle arrival'' to ``Cell arrival'').
    \item \textbf{Next event computation}: Based on the order type, the subsequent event is determined. This event may correspond to a stocking operation, a picking operation, or a service operation. Service events represent internal system movements, such as a lift repositioning a shuttle across different storage levels without directly serving a picking or stocking request. This sub-component includes the availability evaluation of constrained resources (i.e., shuttles and lifts).
\end{itemize}

\come{Resource allocation is performed after the selected storage or retrieval policy has determined a feasible target location and, consequently, the target aisle and tier. The policies therefore determine where a mission is executed, whereas shuttle reservation, tier exclusivity, lift contention, pending-event handling, and event scheduling follow a common dispatching mechanism across all policy configurations.

Once the target aisle and tier have been identified, the dispatcher first searches for an available shuttle already located at the corresponding aisle-tier. If no such shuttle is available, the first available shuttle in the target aisle is selected according to the fixed internal shuttle-list order. The latter provides a deterministic tie-breaking rule among otherwise eligible shuttles; no additional nearest-shuttle or workload-balancing criterion is applied. 

A shuttle may be relocated to another tier only if the destination aisle-tier is not occupied by another shuttle, thereby enforcing a one-shuttle-per-aisle-tier constraint. Shuttle availability is reserved immediately upon assignment and is restored only when the associated operation reaches its resource-release stage. Lift availability is managed through the same exclusive-reservation principle. If a required shuttle or lift is unavailable, the corresponding event is moved to a pending set and reconsidered after subsequent system-state updates.

Picking, stocking, and shuttle-relocation service events are not assigned explicit priority classes. All events are processed through the chronologically ordered future-event list, and events with identical timestamps retain their insertion order. Service events represent auxiliary relocations required by already assigned picking or stocking missions; they are not independent customer missions and do not receive preferential treatment.}

This approach mirrors realistic warehouse operations, where batch processing of orders can improve throughput and reduce shuttle travel distances \citep{Xu2025}.

During the simulation, all movements, including horizontal shuttle travel, vertical lift operations, and loading/unloading of SKUs, are executed according to the defined kinematic constraints and operational logics.

\subsubsection*{Performance assessment}
Upon completion of the simulation run, a post-processing routine aggregates the collected data to provide a comprehensive assessment of the AVS/RS system under the given operational parameters. 
In particular, order-level cycle times are computed separately for the stocking and picking processes, based on the timestamps associated with the main operational activities. 
In addition, the tunnel occupancy is evaluated, which reflects \come{the inventory consolidation} of the warehouse and to provide insight into the saturation level of the storage system and its evolution over time under the adopted control policies.
\come{At simulation time t, tunnel occupancy is defined as the percentage of storage tunnels containing at least one stored unit load:

\[
TO(t) = 100 \cdot \frac{\text{non-empty tunnels at time } t}{N_a \, N_l \, N_c}
\]

Unlike cell occupancy, this metric does not quantify the fraction of occupied storage locations. Instead, it measures how inventory is distributed across the available tunnels. Consequently, lower values indicate a stronger consolidation of inventory into fewer tunnels, whereas higher values indicate a more dispersed allocation of stock \citep{Battarra2022}. 
}

Descriptive statistics, specifically the number of instances, and the mean and standard deviation of all the mentioned metrics are calculated to assess the system's responsiveness, efficiency and variability under a specific workload.
Finally, a granular event log is exported: this dataset records every state transition within the system, enabling further offline analyses such as throughput estimation, resource \comee{utilisation} (e.g., shuttle idle/busy time), and bottleneck identification without requiring simulation re-runs.


\come{Finally, a granular event log is exported, recording every relevant state transition of the system. In addition to supporting process discovery and bottleneck analysis, these logs are used in the present study to reconstruct shuttle busy intervals from reservation and resource-release events. This enables the computation of time-based shuttle \comee{utilisation}, concurrent fleet loading, and resource-level workload concentration without requiring additional simulation runs.}

\subsection{Integrating DES with \mv{}}\label{sec:methmv}
\begin{figure}[t]
    \centering
    \includegraphics[width=0.85\linewidth]{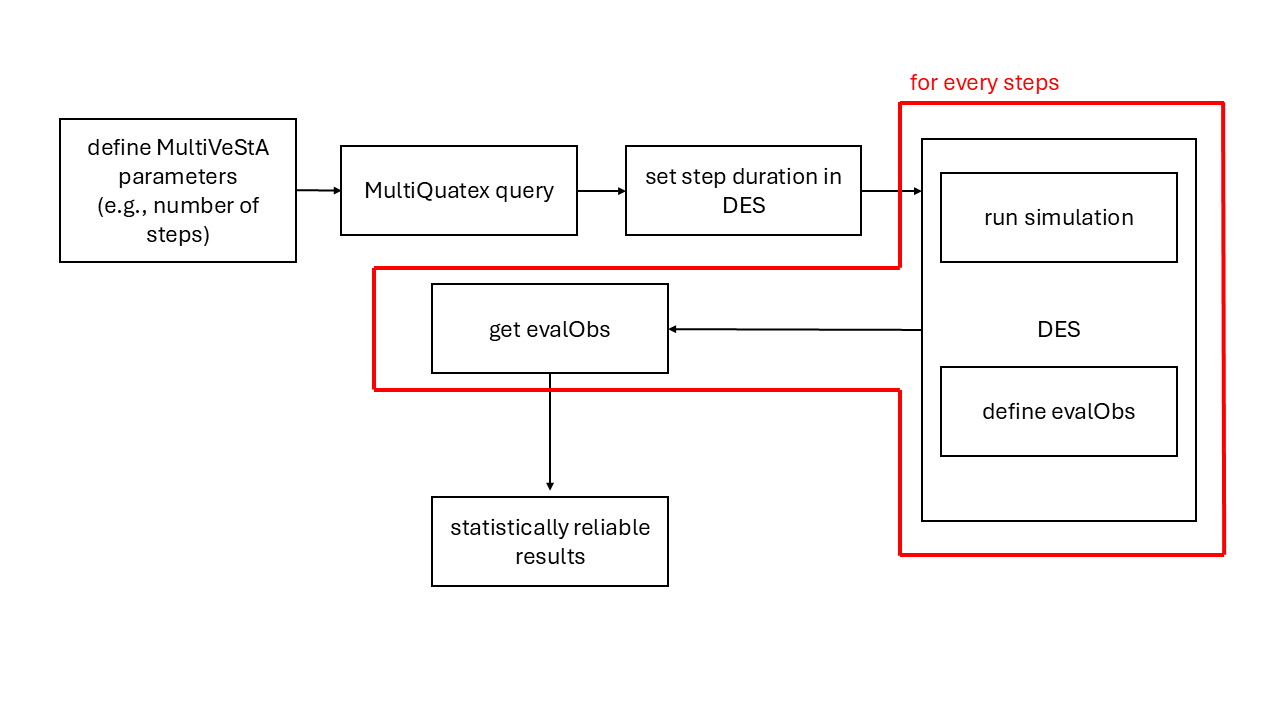}
    \caption{MultiVeStA tool-chaining with DES}
    \label{fig:multivesta_model}
\end{figure}

To enable automated SMC and ensure statistically reliable results, we integrated the DES model of AVS/RS, developed in \comee{MATLAB}, with \mv{}, which is a tool-agnostic framework capable of estimating quantitative properties of a system by driving multiple independent simulation runs.
The integration follows a black-box approach \comee{in which} \mv{} acts as an external coordinator, while the simulator functions as a stepped execution engine.
As illustrated in Figure \ref{fig:multivesta_model}, which shows the tool-chaining between \mv{} and the DES, this interaction is mediated by a generic Java-based adapter interface. 
This wrapper abstracts the internal complexity of the \comee{MATLAB} model, exposing only the three fundamental primitives required by the statistical engine:
\begin{itemize}
    \item \textit{Reset}: Initializes a new simulation run. This method handles the seeding of the random number generator to ensure reproducibility and reloads the preconfigured workspace parameters.
    \item \textit{Next Step}: Advances the simulation state by one discrete step (invoking the logic described in the previous section).
    \item \textit{Evaluate Observation}: Queries the current state of the system to extract specific metrics (e.g., current aisle occupancy, task completion time). 
\end{itemize}

In this workflow, \mv{} orchestrates the analysis by repeatedly triggering the simulation steps shown in Figure~\ref{fig:system_model}, and by performing online SMC over user-defined properties expressed in a specific language named MultiQuaTEx. The analysis is parametric with respect to the simulation step at which observations are collected, allowing the evaluation of multiple performance indicators, i.e., namely picking cycle time, stocking cycle time, and \come{inventory dispersion} at specific stages of the system evolution.  
For each parameter configuration, \mv{} estimates the expected value of the selected observables, enabling a statistically sound comparison of system behaviour across different simulation horizons.
Batch processing of replications is supported, with parameters controlling the number of independent runs, batch size, and confidence intervals for the statistical estimates. This setup enables comprehensive evaluation of system performance under varying operational conditions, while ensuring reproducibility and modularity of the simulation workflow.
\come{The number of replications is therefore determined through a prespecified confidence-and-precision criterion rather than fixed retrospectively. The adopted confidence level, precision threshold, batch size, and maximum simulation budget, together with their rationale, are reported in Section 5.1.}


\subsection{Process Mining analysis}\label{sec:methpm}
PM is employed in the final phase of the workflow to complement the statistical results produced by MultiVeStA with a behavioural interpretation of the simulated system. While SMC provides quantitative estimates of performance indicators, it does not explain \emph{why} certain trends emerge or which operational patterns drive the observed results. Recently, an integration of SMC and PM has been proposed to solve this issue~\cite{Casaluce2024}.

In this paper, we use PM to fill this gap by reconstructing the execution logic of the AVS/RS from the logs generated during simulation, enabling the discovery of process models and the identification of bottlenecks or suboptimal behaviours. The simulation trace is recorded in a tabular format where each row represents a discrete event in the AVS/RS system. The columns are defined as follows:
\begin{itemize}
    \item \textit{Order\_ID}: Unique identifier for the order being processed;
    \item \textit{SKU\_ID}: Identifier of the SKU;
    \item \textit{Operation type}: 0 for Stocking, 1 for Picking, 2 for intermediate lift movements;
    \item \textit{Location (Aisle, Level, Channel, Cell)}: Coordinates determining the storage or retrieval location within the rack system;
    \item \textit{Strategy}: The storage/retrieval strategy applied (0: Random, 1: Deepest, 2: Nearest, \come{3: AFB-ROR}). Each strategy is explained in Section \ref{sec:casestudy};
    \item \textit{Shuttle\_ID}: Unique identifier of the shuttle vehicle assigned to the task;
    \item \textit{Activity}: Description of the specific event occurring (e.g., arrival at I/O, shuttle reservation);
    \item \textit{Timestamp}: The simulation time (YYYY-MM-DD HH:MM:SS) at which the event occurred.
\end{itemize}

An extract of the event log is provided by Table \ref{tab:trace_log}.

\begin{table}[ht]
    \centering
    \caption{Fragment of the AVS/RS simulation event log for a stocking operation.}
    \label{tab:trace_log}
    \resizebox{\textwidth}{!}{%
    \begin{tabular}{| c | c | c | c | c | c | c | c | c | c | c |}
        \hline
        \textbf{Order\_ID} & \textbf{SKU\_ID} & \textbf{Type} & \textbf{Aisle} & \textbf{Level} & \textbf{Ch.} & \textbf{Cell} & \textbf{Strat.} & \textbf{S\_ID} & \textbf{Activity} & \textbf{Timestamp} \\
        YYBEJB & AXX & 0 & - & - & - & - & - & - & Stocking arrival at I/O & 2025-06-15 09:10:00 \\
        YYBEJB & AXX & 0 & 5 & 1 & 22 & 1 & 0 & 5 & Shuttle reservation & 2025-06-15 09:10:05 \\
        YYBEJB & AXX & 0 & 5 & 1 & 22 & 1 & 0 & 5 & Lift reservation & 2025-06-15 09:10:07 \\
        YYBEJB & AXX & 0 & 5 & 1 & 22 & 1 & 0 & 5 & Lift arrival & 2025-06-15 09:10:35 \\
        YYBEJB & AXX & 0 & 5 & 1 & 22 & 1 & 0 & 5 & Shuttle arrival & 2025-06-15 09:11:15 \\
        YYBEJB & AXX & 0 & 5 & 1 & 22 & 1 & 0 & 5 & Cell arrival & 2025-06-15 09:12:00 \\
        \hline
    \end{tabular}%
    }
\end{table}

\subsubsection{Control-flow and performance drift}\label{sec:pmmetrics}

\come{For each scenario-policy combination, all 20 independent simulation
replications are analysed separately. The hourly event-log files produced by
the simulator are cumulative snapshots; therefore, for each replication, only
the final cumulative snapshot generated at step~72 is retained. Concatenating
the hourly snapshots would otherwise duplicate events already recorded in
earlier files. Configuration-level results are obtained only after computing
the Process Mining measures independently for each replication.

Before the analysis, the selected logs are checked for timestamp validity,
chronological consistency, duplicate rows, trace completeness, operation types,
and missing spatial or resource attributes. Each mission is locally identified
by \textit{Order\_ID}. When logs from different scenario-policy configurations
are combined for \comee{visualisation}, the globally unique case identifier is
constructed as}
\[
\texttt{scenario|policy|Order\_ID}.
\]

\come{Replications are never concatenated during the quantitative analysis, so
a run identifier is not required in the case identifier. Service events are
retained within the mission trace to which they belong. Activity labels undergo
only whitespace and case normalization, while the original label is preserved
to maintain traceability to the raw event log.

As an additional model Verification and Validation step, the event logs were subjected to deterministic rule-based checks against the specified storage and retrieval logic. The checks verified tunnel feasibility, SKU homogeneity, direction-dependent intra-tunnel accessibility, and the policy-specific selection and fallback conditions. In particular, no recorded mission bypassed an obstructing unit load or selected an inaccessible internal cell. The selected tunnels were also consistent with the Deepest consolidation/depletion rule, the Nearest proximity rule, and the AFB-ROR zoned or global least-cost fallback. No violation of these rules was detected in the analysed logs.

To study the temporal evolution of system behaviour, each replication is
partitioned into twelve consecutive, non-overlapping six-hour windows, starting
from the first observed event. The six-hour duration was selected a priori as a
compromise between temporal resolution and case support. Each complete trace is
assigned to the window containing its first event, ensuring that individual
missions are never split across windows. A drift score is computed only between
two directly adjacent windows that both contain at least 30 cases. Windows below
this threshold remain documented, but they are excluded from drift computation
and are not bypassed to compare non-adjacent windows.

PM4Py~\citep{pm4py_website} is used to derive frequency and performance
directly-follows graphs (DFGs) for the complete event log and for each temporal
window. In a DFG, nodes represent activities, whereas directed edges represent
directly-follows relations between consecutive activities within the same
trace. The duration of an edge is computed as the timestamp difference between
the two consecutive events.

Control-flow and performance drift are evaluated separately. Let
$\mathbf{p}_{t}$ and $\mathbf{p}_{t+1}$ denote the normalized DFG
edge-frequency vectors of two consecutive supported windows, constructed over
the union of the edges observed in the two windows. Edges absent from one
window are assigned zero frequency. Control-flow drift is measured through the
base-2 Jensen--Shannon distance:}
\[
D_{\mathrm{CF}}(t,t+1)
=
\sqrt{
\frac{1}{2}
KL_{2}\!\left(\mathbf{p}_{t}\middle\|\mathbf{m}\right)
+
\frac{1}{2}
KL_{2}\!\left(\mathbf{p}_{t+1}\middle\|\mathbf{m}\right)
},
\qquad
\mathbf{m}
=
\frac{\mathbf{p}_{t}+\mathbf{p}_{t+1}}{2}.
\]

\come{The resulting score ranges from 0 to 1. A value close to 0 indicates
similar directly-follows frequency distributions in the two windows, whereas
larger values indicate stronger changes in the relative occurrence of process
transitions.

Performance drift instead evaluates changes in the execution times of
recurring transitions. Let $E^{*}_{t,t+1}$ be the set of edges observed at
least ten times in both windows. For each eligible edge, the analysis compares
its median duration in the two windows. Performance drift is defined as}
\[
D_{\mathrm{P}}(t,t+1)
=
\frac{
\displaystyle
\sum_{e \in E^{*}_{t,t+1}}
\left(f_{e,t}+f_{e,t+1}\right)
\left|
\ln
\left(
\frac{\widetilde{d}_{e,t+1}}
     {\widetilde{d}_{e,t}}
\right)
\right|
}{
\displaystyle
\sum_{e \in E^{*}_{t,t+1}}
\left(f_{e,t}+f_{e,t+1}\right)
},
\]
\come{where $f_{e,t}$ is the frequency of edge $e$ in window $t$, and
$\widetilde{d}_{e,t}$ is its median duration. Edges with missing, zero, or
negative duration values are excluded. The frequency weights ensure that
changes affecting recurrent transitions contribute more strongly than changes
affecting rarely observed edges.

Both drift measures are descriptive and dimensionless. Control-flow drift
quantifies changes in the relative frequency of process paths, whereas
performance drift quantifies changes in the transition-duration profile. They
are not interpreted as anomaly scores, percentage variations, or statistical
significance tests.

For each temporal window, performance bottlenecks are ranked according to the
fraction of the total observed DFG time attributable to each edge:}
\[
C_{e,t}
=
\frac{
\displaystyle
\sum_{i \in e,t} d_i
}{
\displaystyle
\sum_{e' \in E_t}
\sum_{i \in e',t} d_i
},
\]
\come{where $d_i$ denotes the observed duration of an occurrence of the
corresponding directly-follows relation. The numerator therefore represents the
total time accumulated by edge $e$ in the window. This ranking jointly accounts
for transition frequency and duration, avoiding the classification of an
isolated but slow transition as the dominant process bottleneck.}
\comee{To support the cross-policy interpretation of the SMC results, full-log DFG edge frequencies and mean durations are  computed independently for each replication and then summarized across the 20 runs using arithmetic means and two-sided 90\% Student-\(t\) confidence intervals.}

\subsubsection{Shuttle \comee{utilisation} and workload distribution}
\come{The event logs are used to analyse both shuttle \comee{utilisation} and the distribution of missions across the available resources. These two dimensions are evaluated separately because shuttle \comee{utilisation} is a time-based measure, whereas workload distribution is based on the number of missions assigned to each shuttle.

A shuttle is considered busy from its reservation timestamp until the
corresponding resource-release timestamp recorded by the simulator. For stocking missions, the busy interval also includes the modelled empty satellite return following cell arrival. Let \(B_{i,r}^{\mathrm{open}}\) denote the total busy time of shuttle \(i\) in simulation run \(r\), considering only the portions of its busy intervals that overlap the daily operating period from 07:00 to 20:00. Since each run covers three simulated days, the total operating time is \(H_{\mathrm{open}}=39\) hours. The operating-hour \comee{utilisation} of shuttle \(i\) in run \(r\) is defined as}

\[
U_{i,r}^{\mathrm{open}}
=
100 \cdot
\frac{B_{i,r}^{\mathrm{open}}}
     {H_{\mathrm{open}}}.
\]

\come{Thus, \(U_{i,r}^{\mathrm{open}}\) represents the percentage of the available operating time during which shuttle \(i\) is reserved and busy. It is a time-based measure and should not be confused with the percentage of missions assigned to that shuttle.

The mean fleet \comee{utilisation} $U_{\mathrm{fleet}}$ in run \(r\) is computed as}
\[
U_{\mathrm{fleet},r}^{\mathrm{open}}
=
\frac{1}{N_s}
\sum_{i=1}^{N_s}
U_{i,r}^{\mathrm{open}},
\]
\come{where \(N_s\) is the number of available shuttles. For each
scenario-policy combination, this metric is computed independently for all
20 simulation runs and summarized through its mean and a two-sided 90\%
Student-\(t\) confidence interval.}

\come{The distribution of missions among the shuttles is evaluated separately using Jain's index. Let \(M_{i,r}\) denote the total number of missions assigned to shuttle \(i\) during simulation run \(r\). The shuttle workload-distribution index for run \(r\) is defined as} \[ J_{\mathrm{S},r} = \frac{ \left( \displaystyle\sum_{i=1}^{N_s} M_{i,r} \right)^2 }{ N_s \displaystyle\sum_{i=1}^{N_s} M_{i,r}^{2} }. \]

\come{Provided that at least one mission is performed during the run, \(J_{\mathrm{S},r}\) lies between \(1/N_s\) and \(1\). A value of \(1\) is obtained when all shuttles receive the same number of missions. Progressively lower values indicate that missions are concentrated on a smaller subset of shuttles; the theoretical minimum \(1/N_s\) is obtained when all missions are assigned to a single shuttle.} 
\come{Jain's index is employed here as a normalized measure of workload distribution, rather than as a normative fairness criterion. Moreover, because it is based on mission counts, it does not directly measure the amount of time for which the shuttles are busy. A policy may therefore produce a relatively low mean fleet \comee{utilisation} while simultaneously concentrating most missions on only a few shuttles.}
\come{For each scenario-policy combination, \(J_{\mathrm{S},r}\) is computed independently for each of the simulation runs. 
\come{The configuration-level workload-distribution index is then defined as
the arithmetic mean of the replication-level values:}
\[
\come{{J}_{S} =
\frac{1}{R}\sum_{r=1}^{R} J_{S,r},}
\]
\come{where \(R\) corresponds to the number of independent simulation replications.
Computing \(J_{S,r}\) separately for each run before averaging preserves the
replication structure and avoids pooling mission counts from independent
simulation runs. The uncertainty associated with \(\overline{J}_{S}\) is summarized
through a two-sided 90\% Student-\(t\) confidence interval across the \(R\)
replication-level values.}
The workload-distribution index is thus the arithmetic mean of the resulting replication-level values. Computing the index separately for each run preserves the replication structure and avoids pooling mission counts from independent simulation runs.}

\come{The complete 72-hour event logs are retained for process discovery, temporal Process Mining analysis, and the computation of mission counts used in Jain's index. The restriction to the 07:00--20:00 operating periods applies only to the computation of shuttle \comee{utilisation}, in order to prevent inactive night-time hours from artificially diluting the measured resource loading.} 

\come{Table~\ref{tab:pm_configuration} summarizes the main configuration parameters of the Process Mining, shuttle-\comee{utilisation}, and workload-distribution analyses.}

\begin{table}[ht] 
\centering 
\caption{Configuration of the Process Mining, shuttle-\comee{utilisation}, and workload-distribution analyses.} 
\label{tab:pm_configuration} 
\small 
\renewcommand{\arraystretch}{1.15} 
\begin{tabular}{p{0.48\linewidth} p{0.40\linewidth}} 
\hline \textbf{Parameter} & \textbf{Value} \\ 
\hline Event-log horizon & Complete 72 h run \\ 
Window size & 6 h \\ 
Minimum cases per window & 30 \\ 
Treatment of low-support windows & Retained descriptively and excluded from drift computation \\ 
Minimum edge frequency in both windows & 10 \\ 
Case assignment & Trace start time \\ 
Simulation replications & 20 per scenario-policy combination \\ 
Shuttle busy interval & Reservation to resource release \\ 
\comee{Utilisation} reference period & 07:00--20:00, 39 h over three days \\ 
\comee{Utilisation} uncertainty & Two-sided 90\% Student-\(t\) confidence interval \\ 
Workload-distribution input & Mission counts assigned to individual shuttles \\ 
Jain's index horizon & Complete 72 h run \\ 
Jain's index aggregation & Computed per run and averaged over 20 replications \\ 
Jain's index interpretation & Values closer to 1 indicate a more even mission distribution \\ 
\hline 
\end{tabular} 
\end{table}

\come{The resulting drift measures, and bottleneck rankings are presented in Section~\ref{sec:pm_results}} \comee{and are computed independently over all 20 replications of each
scenario-policy combination before being aggregated at configuration level.}. Shuttle \comee{utilisation} \comee{and workload distribution are likewise evaluated over all
independent runs. The Disco process maps described in Section~\ref{sec:pm_analysis} are a separate illustrative visualisation: for these maps only, one deterministically selected representative baseline run per policy is used to display the operational patterns}.

\subsubsection{Process-map analysis}
\label{sec:pm_analysis}

\come{Following the replicated drift and bottleneck analyses, Disco
\citep{fluxicon_disco} is used \comee{to produce} an exploratory \comee{visualisation} \comee{of selected event logs}. For each scenario-policy combination, one representative simulation run is selected deterministically according to its proximity to the median run-level characteristics. This selection \comee{concerns the illustrative figures only: all quantitative PM measures reported in the study are computed over the complete set of 20 independent replications for each scenario-policy combination}.

The complete event log of the representative run, together with selected pre-critical, critical, and post-critical portions surrounding the most relevant drift episode, can be imported into Disco to generate frequency and performance process maps. In these maps, nodes represent activities, while directed edges represent directly-follows relations between consecutive activities within the same trace~\cite{Aalst2016ProcessMining}. Frequency annotations show how often activities and transitions occur, whereas performance annotations support the inspection of the elapsed times associated with the process paths.

The purpose of these maps is to provide analysts and managerial users with an intuitive representation of the operational patterns underlying the quantitative drift and bottleneck results. In particular, they support the visual comparison of recurrent paths, changes in transition frequencies, and time-consuming process segments across policies and temporal portions of the event log.

The Disco maps are used for interpretation and illustration only. They are not used to estimate prevalence across simulation replications, establish statistical significance, infer causal relationships, or provide empirical validation of the
simulation model.}

\section{Case Study}\label{sec:casestudy}
\begin{figure}[t]
    \centering
    \includegraphics[width=\textwidth]{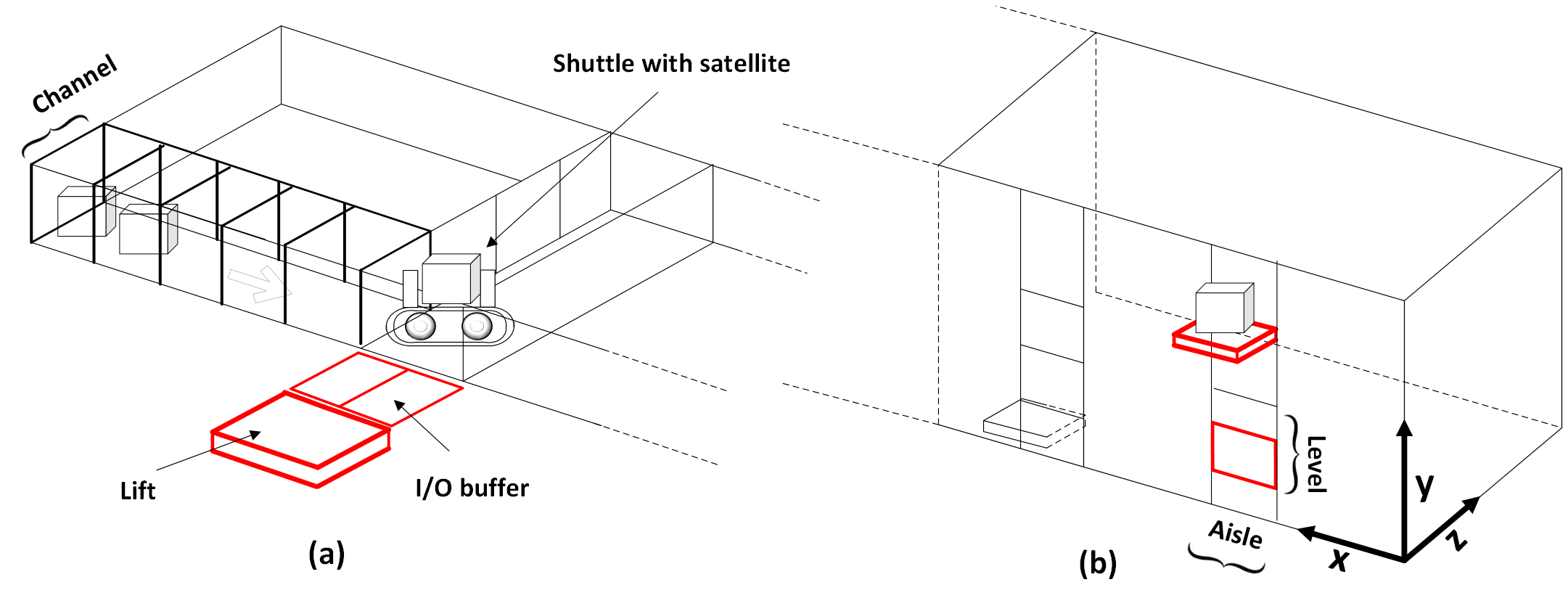}
    \caption{Overview of the studied AVS/RS. Panel (a) shows the shuttle operating within a channel, while panel (b) illustrates the lift executing vertical movements between levels. Adapted from \citep{Guerrazzi2024}}
    \label{fig:avsrs-overview}
\end{figure}

The considered AVS/RS is based on shuttle technology. 
Figure~\ref{fig:avsrs-overview} shows the two main operating components: (a) the shuttle travelling within a channel, and (b) the lift responsible for vertical movements between levels.



\come{The warehouse layout follows the architecture introduced in Figure  ~\ref{fig:avsrs_layout}}. The system can accommodate a variable number of shuttles operating along both the aisles and channels, providing scalable storage capacity.

The shuttle vehicles are capable of moving horizontally along channels and vertically between levels via lifts to transport goods. 
Their movement is characterised by controlled acceleration and deceleration, and can be modulated depending on whether the shuttle is carrying a load. Lifts enable vertical movement across tiers, with dedicated inbound and outbound zones for goods handling.

The storage/retrieval strategy determines how storage locations are selected for incoming items and how retrieval requests are \comee{prioritised} within an automated warehouse system. In this case study, four strategies are considered for stocking and picking operations:

\begin{itemize}
    \item \textbf{Random} \\
    Storage and retrieval locations are selected randomly among all feasible positions. This strategy is simple to implement and provides a baseline for comparison, but it generally leads to suboptimal travel times and inefficient system \comee{utilisation} \cite{lehmann2021travel}.
    
    \item \textbf{Deepest} \\
    \come{This strategy implements tunnel-consolidation and tunnel-depletion logic. For stocking, it selects the feasible tunnel containing the largest number of units of the requested SKU, thereby concentrating homogeneous inventory. For picking, it selects the tunnel containing the smallest number of units of that SKU, favouring the depletion of lightly occupied tunnels. Ties are resolved randomly, and stocking falls back to Random when no compatible partially filled tunnel is available.} This policy aims to increase storage density and reduce fragmentation, but may increase travel times for handling equipment \cite{Lerher2024, lerher2010travel}.
    
    \item \textbf{Nearest} \\
    \come{Nearest is a shuttle-local proximity policy rather than a global minimization of the distance from the I/O station. It first considers feasible tunnels on the aisle and level currently served by an available shuttle and ranks them by horizontal distance. If none is available, tunnels on other levels of the same aisle are evaluated using a composite measure that also accounts for the estimated lift-transfer effort. The policy falls back to Random when no feasible proximity-based choice is available. This approach generally reduces travel and cycle times \cite{mcinerney2022increasing}, but may concentrate workload on a subset of shuttles and distribute inventory across more channels.}

    \item \textbf{AFB-ROR} \\
    \come{  
     This policy, named \textit{Access-Frequency-Based Stocking and Route-Optimized Retrieval} is an operational synthesis inspired by frequency-based storage assignment and path-oriented retrieval rules for AVS/RS \citep{Kaczmarek2014}. It combines frequency-based storage assignment with cost-oriented stocking and retrieval rules. Before each simulation run, SKUs are ranked by expected access frequency and assigned to fixed A/B/C classes comprising 20\%, 30\%, and 50\% of the SKU catalogue, respectively. Storage capacity is allocated to each class proportionally to its cumulative demand, and zoning is performed at tunnel level according to static access cost. For stocking, the policy selects the lowest-cost feasible tunnel in the zone associated with the SKU class; if no such tunnel is feasible, it applies a global lowest-cost feasible fallback. 
     For retrieval, it selects the stored unit of the requested SKU with the lowest retrieval cost; ties are resolved first by FIFO order and then by location identifier. The use of expected access frequency to support storage assignment is also consistent with the broader storage-assignment literature \citep{Battini2016}. AFB-ROR does not modify the first-come-first-serve order-handling logic, shuttle or lift allocation, or resource queues.
    }
\end{itemize}

\come{
The four policies govern the selection of feasible storage and retrieval locations. Once a target aisle and tier have been determined, all policies are executed under the common shuttle- and lift-dispatching rules described in Section~3.1. Consequently, the experimental comparison does not confound policy-specific location selection with different resource-priority or reservation mechanisms. More specifically, for Random picking, the shuttle identified during the randomized candidate-selection step is retained as a preferred candidate when it remains feasible. If it is no longer feasible, the common selection rule is used.

All policies use the same feasibility and intra-channel accessibility rules. A tunnel is feasible for stocking only if it is empty or already contains the requested SKU and has at least one free cell, thereby preserving SKU homogeneity within each deep lane. Once a tunnel has been selected, the storage or retrieval position is determined through a common direction-dependent scan from the relevant access side. Consequently, no policy can bypass an obstructing unit load or arbitrarily select an internal cell. The policies therefore differ only in tunnel selection, while sharing the same intra-tunnel handling logic.
}

Fixed handling times are considered for loading and unloading operations, ensuring realistic simulation of operational performance \citep{Ferrari2024, Lupi2024}. 

Overall, the system design supports scalability and configurability, allowing adjustment of key parameters such as number of aisles, levels, channels, and shuttles to meet different operational requirements.

\subsection{Scenarios}
\come{As mentioned, one of the contributions of this paper is a new open-source AVS/RS simulation model, designed following guidelines from the literature concerning warehouse layout structure, vehicle kinematics, and operational assumptions \citep{Battarra2022, Marolt2023, Lupi2024}. The experimental evaluation follows a one-factor-at-a-time design and comprises four complementary scenarios. 
\comee{These scenarios provide a computational robustness and sensitivity assessment, even if they do not constitute empirical validation of an industrial installation.}
The first is a controlled synthetic baseline based on the configuration proposed by \citep{Marolt2023}. The second is a literature-informed workload scenario derived from the aggregate operational statistics reported by \citep{Lupi2024}. Since the published industrial case does not include the original timestamped transaction logs, SKU allocation history, or initial inventory snapshot, this scenario should be interpreted as a literature-informed robustness evaluation rather than as a full empirical validation.
The third scenario isolates the effect of vehicle kinematics by replacing only the baseline velocity, acceleration, and deceleration parameters with the values reported by \citep{Battarra2022}; all other model components are kept unchanged. Finally, a fourth scenario isolates the effect of the initial warehouse state by increasing the initial proportion of non-empty tunnels from the stochastic baseline level, while preserving the baseline demand, geometry, \comee{shuttle} fleet, \comee{number and capacity of lifts}, kinematics, and SKU logic \comee{as well as the resource-allocation, dispatching, and queueing rules. This scenario is therefore an initial-occupancy stress test, rather than a systematic analysis of shuttle or lift congestion}. This controlled design allows the sensitivity of the four storage-management policies to workload, vehicle dynamics, and initial tunnel-level inventory \comee{occupancy} to be assessed without conflating their effects.
Table~\ref{tab:parameters} summarizes the parameters adopted in all the scenarios.}

\clearpage
\begingroup

\footnotesize
\setlength{\tabcolsep}{2.5pt}
\renewcommand{\arraystretch}{1.13}

\setlength{\LTleft}{\fill}
\setlength{\LTright}{\fill}
\setlength{\LTpre}{0pt}
\setlength{\LTpost}{6pt}
\setlength{\LTcapwidth}{\textwidth}

\begin{longtable}{
@{}
L{3.00cm}
C{1.75cm}
C{1.75cm}
C{1.75cm}
C{1.75cm}
L{4.45cm}
@{}
}

\caption{Scenario design and simulation parameters. The baseline configuration
is inspired by the synthetic setting of \protect\citep{Marolt2023}; the
Lupi-inspired workload uses aggregate operational statistics reported by
\protect\citep{Lupi2024}; and the Battarra-inspired kinematic parameters follow
\protect\citep{Battarra2022}.}
\label{tab:parameters}\\

\toprule
\textbf{Category / parameter} &
\textbf{Baseline} &
\makecell{\textbf{Lupi-}\\\textbf{inspired}} &
\makecell{\textbf{Battarra-}\\\textbf{inspired}} &
\makecell{\textbf{Peak}\\\textbf{congestion}} &
\textbf{Description} \\
\midrule
\endfirsthead

\multicolumn{6}{@{}l}{
  \footnotesize\itshape Table~\thetable\ continued from the previous page
}\\[2pt]

\toprule
\textbf{Category / parameter} &
\textbf{Baseline} &
\makecell{\textbf{Lupi-}\\\textbf{inspired}} &
\makecell{\textbf{Battarra-}\\\textbf{inspired}} &
\makecell{\textbf{Peak}\\\textbf{congestion}} &
\textbf{Description} \\
\midrule
\endhead

\midrule
\multicolumn{6}{r@{}}{
  \footnotesize\itshape Continued on the next page
}\\
\endfoot

\bottomrule
\endlastfoot

\tablesection{Scenario design}

Varied factor &
Reference configuration &
Workload only &
Vehicle kinematics only &
Initial inventory state only &
The scenarios follow a one-factor-at-a-time robustness design. \\

\tablesection{Warehouse structure and resources}

$(N_a,N_l,N_c,N_d)$ &
$(6,6,30,6)$ &
\asbaseline &
\asbaseline &
\asbaseline &
Number of aisles, levels per aisle, channels per aisle--level, and storage
positions per channel. \\

Storage capacity &
6,480 &
\asbaseline &
\asbaseline &
\asbaseline &
Total number of storage locations. \\

$(W,D,H)$ [m] &
$(0.7,1.5,1.0)$ &
\asbaseline &
\asbaseline &
\asbaseline &
Channel width, storage-position pitch, and level height. \\

Shuttles / lifts &
10 / 6 &
\asbaseline &
\asbaseline &
\asbaseline &
One lift serves each aisle. \\

\tablesection{Vehicle kinematics}

\makecell[l]{
Satellite speed\\
$V_{x,1}/V_{x,0}$ [m/s]
} &
1.5 / 1.5 &
\asbaseline &
1.0 / 1.2 &
\asbaseline &
Loaded / unloaded speed along the channel. \\

\makecell[l]{
Satellite acceleration\\
and deceleration\\
$A_{x,1}/A_{x,0}={}$\\
$D_{x,1}/D_{x,0}$ [m/s$^2$]
} &
1.0 / 1.0 &
\asbaseline &
1.0 / 1.2 &
\asbaseline &
Loaded / unloaded acceleration and deceleration. \\

\makecell[l]{
Shuttle speed\\
$V_{y,1}/V_{y,0}$ [m/s]
} &
3.0 / 3.0 &
\asbaseline &
3.0 / 3.0 &
\asbaseline &
Loaded / unloaded speed along the aisle. \\

\makecell[l]{
Shuttle acceleration\\
and deceleration\\
$A_{y,1}/A_{y,0}={}$\\
$D_{y,1}/D_{y,0}$ [m/s$^2$]
} &
2.0 / 2.0 &
\asbaseline &
0.8 / 1.2 &
\asbaseline &
Loaded / unloaded acceleration and deceleration. \\

\makecell[l]{
Lift speed\\
$V_{z,1}/V_{z,0}$ [m/s]
} &
1.0 / 1.0 &
\asbaseline &
1.0 / 1.0 &
\asbaseline &
Loaded / unloaded vertical speed. \\

\makecell[l]{
Lift acceleration\\
and deceleration\\
$A_{z,1}/A_{z,0}={}$\\
$D_{z,1}/D_{z,0}$ [m/s$^2$]
} &
0.5 / 0.5 &
\asbaseline &
0.35 / 0.35 &
\asbaseline &
Loaded / unloaded vertical acceleration and deceleration. \\

$t_{\mathrm{load}}$ [s] &
1.5 &
\asbaseline &
\asbaseline &
\asbaseline &
Fixed loading or unloading time. \\

\tablesection{Order workload}

Simulation horizon &
72 h &
\asbaseline &
\asbaseline &
\asbaseline &
Three simulated working days. \\

Operating window &
07:00--20:00 &
\asbaseline &
\asbaseline &
\asbaseline &
Daily order-generation window. \\

Order-SKU catalogue &
20 &
212 &
20 &
20 &
SKU variants eligible for order generation. \\

Initial physical SKU variants &
5 &
5 &
5 &
5 &
SKU variants physically present at simulation start. \\

SKU popularity distribution &
Skewed exponential &
Zipf $(s=1)$ &
\asbaseline &
\asbaseline &
Distribution used for order-SKU selection. \\

Arrival profile &
Truncated normal &
Uniform &
\asbaseline &
\asbaseline &
For the baseline, $\mu=12.5$ h and $\sigma=2$ h. \\

Daily stocking / picking rate &
500 / 500 &
703 / 826 &
\asbaseline &
\asbaseline &
Poisson means, expressed in missions per day. \\

\tablesection{Initial inventory and control policies}

\makecell[l]{
Baseline placement \\ parameters
$(\beta,\gamma)$
} &
$(0.70,1.00)$ &
\asbaseline &
\asbaseline &
NA &
For each aisle--level pair, $\beta$ determines the number of candidate
channels, while $\gamma$ defines the maximum number of unit loads that can be
initially assigned to each candidate channel. \\

Initial tunnel occupancy &
Stochastic, $\approx 31$--$32\%$ &
\asbaseline &
\asbaseline &
Exactly 65\% &
Percentage of non-empty tunnels over 1,080 total tunnels. This is the reported
tunnel-occupancy KPI at $t=0$. \\

Expected initial cell occupancy &
$\approx 18.3\%$ &
\asbaseline &
\asbaseline &
$\approx 37.9\%$ &
Expected percentage of occupied storage cells after initialization; this
quantity is distinct from tunnel occupancy. \\

Initial inventory generation &
11 channels per aisle-level pair &
\asbaseline &
\asbaseline &
702 channels overall &
Each selected channel receives a random number between 0 and 6 unit
loads. All the selected channels from Peak scenario are non-empty. \\

Storage and retrieval policies &
\multicolumn{4}{c}{
  \makecell{
    Random / Deepest / Nearest / AFB--ROR
  }
} &
The same four material-handling policies are evaluated in every scenario. \\

\end{longtable}


\endgroup
\clearpage

\input{analysis}

\section{Conclusions \& Future Works}
This work demonstrates the value of combining DES, MultiVeStA, and Process Mining for the analysis of AVS/RS systems. 
\come{The experimental results show that storage and retrieval policies substantially affect both picking and stocking cycle times and the distribution of inventory across storage tunnels. No policy dominates all the considered performance objectives. AFB-ROR achieves the lowest numerical picking and stocking times across the four evaluated scenarios, although its stocking-time advantage over Nearest is not statistically distinguishable under all operating conditions. Deepest consistently provides the strongest inventory consolidation, as indicated
by the lowest \comee{tunnel occupancy}, but generally incurs longer cycle times. Random does not achieve the best endpoint result for any of the evaluated KPIs.}

The integration of SMC and PM \come{provides complementary forms of evidence. MultiVeStA produces confidence-controlled estimates of aggregate performance indicators across independent simulation replications, while PM analyses the resulting event logs to expose the dynamic patterns associated with those outcomes.}
This combined quantitative and behavioural perspective enhances the interpretability and transparency of simulation results, especially in systems as complex and configuration-rich as AVS/RS. \come{In future work, we will consider further PM techniques specifically tailored to quantitative behavioural studies, such as stochastic conformance checking \citep{DBLP:journals/is/IncertoVA25}. }

\come{From a decision-support perspective} the proposed methodology equips warehouse managers and system designers with actionable insights, enabling informed decisions on operational and tactical strategies. In practice, policies could be alternated dynamically according to business needs, balancing efficiency and space usage over time.
Moreover, the framework is general and can be easily transferred to other automated warehousing systems and logistics environments based on DES.

\come{The study nevertheless has limitations. The experimental campaign remains simulation-based and does not constitute a full empirical validation against a physical AVS/RS facility. To broaden the original synthetic evaluation, we introduced three controlled robustness scenarios that separately vary workload characteristics, vehicle kinematics, and the initial inventory state. One scenario is calibrated using aggregate operational evidence from an industrial case study, while another adopts vehicle parameters reported in the literature. However, the complete timestamped transaction logs, initial inventory snapshots, SKU-allocation histories, and resource-state data required to reproduce an industrial execution are not publicly available. The additional scenarios should therefore be interpreted as literature-informed robustness assessments rather than as substitutes for validation against real operational data. This limitation affects the external generalizability of the numerical findings, but not the internal consistency of the controlled comparisons performed under common experimental conditions.
\comee{Moreover, the high-initial-occupancy scenario varies only one initial tunnel-occupancy level while keeping the fleet size, lift configuration, dispatching rules, and demand profile unchanged. It should therefore not be interpreted as a systematic congestion analysis of shuttle or lift resources. Future work should examine the combined effects of these factors under a broader range of operating conditions}

A further limitation concerns the 72-hour simulation horizon. Although sufficient to compare the policies after the initial transient and across three simulated operating days, it does not capture seasonal demand patterns, long-term inventory evolution, supply-chain disruptions, or gradual performance degradation. Extending the analysis over longer periods would provide a more comprehensive assessment of policy robustness under evolving operational conditions.

Future work will therefore focus on validating the proposed workflow using complete industrial event logs or observations from a physical AVS/RS installation, supporting empirical validation of the simulator and its integration within digital-twin-based decision-support systems. Additional research will investigate hybrid and adaptive storage policies, broader sensitivity analyses across demand, fleet, layout, and saturation parameters, and richer integration between confidence-controlled KPI estimation and process-level behavioural evidence. In particular, future studies will further explore how bottlenecks, resource-concentration patterns, and process deviations influence aggregate system performance, while tighter integration with online monitoring could enable increasingly responsive and data-driven warehouse control.}

\bibliographystyle{elsarticle-num} 
\bibliography{references,bibliography_a_mano}

\end{document}

%% file: analysis.tex
\section{Results}
In this section, we present the quantitative results obtained from  the statistical analysis of the outputs produced by the combined use of \mv{} and DES (Section \ref{sec:\mv{}-results}), and application of PM to the event logs generated by DES (Section \ref{sec:pm_results}).

\subsection{\mv{} Results}\label{sec:\mv{}-results}

To evaluate the impact of different storage and retrieval strategies on system performance, we first estimated key performance indicators \come{for the four material handling policies (i.e., Random, Deepest, Nearest, and AFB-ROR) under all four scenarios}. 
Then, we \comee{performed} pairwise \comee{comparisons of} the results using statistical t-tests to check whether the different settings led to different estimations. This is known as counterfactual analysis \cite{Vandin2022}. 


\subsubsection*{Experimental setup and query definition}

The analysis was specified using the MultiQuaTEx query language. We defined a parametric query shown in Listing~\ref{lst:multiquatex}. This query estimates the expected values of three key performance indicators (picking time, stocking time, and tunnel occupancy) at hourly intervals over a simulation horizon of 72 hours. 

\begin{lstlisting}[
    language=Java, 
      float=t,
    morekeywords={if, then, else, fi, eval, parametric, s, rval, obsAtStep},
    keywordstyle=\color{blue}\bfseries,
    stringstyle=\color{red},
    commentstyle=\color{green!60!black},
    caption={MultiQuaTEx parametric query used for the analysis}, 
    label={lst:multiquatex}, 
    basicstyle=\ttfamily\small, 
    numbers=left,
    frame=single
]
obsAtStep(x,obs) =
 if ( s.rval("steps") == x )
  then s.rval(obs)
  else # obsAtStep(x,obs)
 fi ;

eval parametric(
 E[ obsAtStep(x, "mean_picking_time") ],
 E[ obsAtStep(x, "mean_stocking_time") ],
 E[ obsAtStep(x, "tunnel_occupancy") ],
 x, 1, 1, 72
);
\end{lstlisting}

In lines 7-12 we ask \mv{} to estimate the expected value of the three key performance indicators at each step \texttt{x}, varying from \texttt{1} to \texttt{72} hours, in increments of \texttt{1}.
For each simulation, we will obtain a real value per combination of step and KPI of interest. 
In particular, \texttt{obsAtStep} is a so-called recursive temporal operator defined in lines 1-5: for each step of interest \texttt{x}, and each KPI \texttt{obs}, we check whether the simulator (\texttt{s}) has currently performed exactly \texttt{x} steps (line 2). If it is the case, for this simulation, we return the current value of the KPI (line 3). Otherwise, with the \texttt{\#} keyword, we ask \mv{} to re-evaluate the current \texttt{x,obs} case in the next simulation step. 


\mv{} has two alternative stopping criteria to decide the required number of simulations \citep{Vandin2022, DBLP:conf/ifm/GilmoreRV17,sebastio2013multivesta}. The first is based on a maximum admitted confidence interval (CI) specified by the user. This is given in terms of $\alpha$ and $\delta$: roughly speaking, we perform enough simulations such that around each estimation given by \mv{} we build an interval with statistical confidence ($100- \alpha)\%$, and width at most $\delta$. In this way, for each time step $x$, \mv{} automatically orchestrated the necessary number of independent simulation replications to estimate the expected values with a ($100- \alpha)\%$ CI.

The second stopping criterion is simply a maximum number \come{of batches, each composed by $N$ simulations}. In the latter case, we nevertheless report the confidence interval associated with each estimate obtained with the given simulations ($\alpha$ shall be specified anyway). 

In this experiment, we follow the first approach, setting $\alpha=10\%$, \come{corresponding to a 90\% confidence level, and} $\delta=5$. \come{The 90\% level was selected a priori as a compromise between uncertainty quantification and computational cost, considering that the campaign estimates three KPIs at 72 hourly observations for 16 scenario-policy combinations. This confidence level is also consistent with previous \mv{}-based simulation analyses \citep{Vandin2022}.
The parameter $\delta$ denotes the maximum admitted full confidence-interval width in the native unit of each observable: 5 seconds for picking and stocking cycle times and 5 percentage points for tunnel occupancy. It should therefore be interpreted as a conservative stopping bound, rather than as an assumed measurement error or a minimum detectable policy difference.
Replications were processed in batches of $N=20$, consistently with the block size adopted in the \mv{} transient-analysis procedure \citep{Vandin2022}, with a maximum of three batches, corresponding to 60 replications per configuration. The upper limit was introduced as a computational safeguard; the effective number of replications was determined by the confidence-and-precision stopping rule and is reported below.
} 

\subsubsection*{Performance overview}
\come{
Table~\ref{tab:scenario_policy_performance} summarizes the performance obtained for all 16 scenario-policy combinations. The comparison jointly considers mean picking and stocking cycle times, together with tunnel occupancy. This broader experimental design makes it possible to assess whether the relative behaviour of the policies remains consistent when workload intensity, vehicle kinematics, and initial \comee{tunnel occupancy} conditions are varied. Lower picking and stocking times indicate faster order processing, whereas lower tunnel occupancy values capture how each policy distributes inventory across storage tunnels. Lower values correspond to stronger inventory consolidation, while higher values indicate a more dispersed allocation.
For readability and consistency with the reported confidence intervals, all values are reported to one decimal place.
Boldface identifies the lowest numerical value within each scenario and performance indicator.

\begin{table*}[t]
    \centering
    \caption{Performance overview at the end of the 72-hour simulation horizon across scenarios and stocking/picking policies. Values are reported as estimated means \(\pm\) the half-width of the corresponding two-sided 90\% confidence interval. Boldface identifies the lowest numerical value within each scenario and KPI.}
    \label{tab:scenario_policy_performance}

    \small
    \setlength{\tabcolsep}{8pt}
    \renewcommand{\arraystretch}{1.15}

    \begin{tabular}{llcc>{\centering\arraybackslash}p{2.5cm}}
        \toprule
        \textbf{Scenario} &
        \textbf{Policy} &
        \textbf{Picking (s)} &
        \textbf{Stocking (s)} &
        \textbf{Tunnel occupancy (\%)} \\
        \midrule

        Baseline & Random
        & $32.9 \pm 0.1$
        & $29.2 \pm 0.1$
        & $50.5 \pm 2.2$ \\
        
        Baseline & Deepest
        & $34.5 \pm 0.7$
        & $30.8 \pm 0.8$
        & $\mathbf{22.3 \pm 0.7}$ \\
        
        Baseline & Nearest
        & $26.3 \pm 0.4$
        & $21.0 \pm 0.5$
        & $37.8 \pm 0.8$ \\
        
        Baseline & AFB-ROR
        & $\mathbf{22.7 \pm 0.4}$
        & $\mathbf{20.6 \pm 0.3}$
        & $33.3 \pm 0.6$ \\
        
        \midrule
        
        Lupi-inspired & Random
        & $32.4 \pm 0.2$
        & $28.9 \pm 0.1$
        & $45.4 \pm 2.5$ \\
        
        Lupi-inspired & Deepest
        & $31.6 \pm 0.1$
        & $29.2 \pm 0.1$
        & $\mathbf{19.9 \pm 0.9}$ \\
        
        Lupi-inspired & Nearest
        & $27.9 \pm 0.1$
        & $21.1 \pm 0.1$
        & $37.9 \pm 1.8$ \\
        
        Lupi-inspired & AFB-ROR
        & $\mathbf{23.1 \pm 0.1}$
        & $\mathbf{20.3 \pm 0.2}$
        & $29.8 \pm 1.1$ \\
        
        \midrule
        
        Battarra-inspired & Random
        & $38.6 \pm 0.2$
        & $34.3 \pm 0.1$
        & $50.6 \pm 2.1$ \\
        
        Battarra-inspired & Deepest
        & $41.3 \pm 1.0$
        & $37.5 \pm 1.1$
        & $\mathbf{22.3 \pm 0.7}$ \\
        
        Battarra-inspired & Nearest
        & $30.7 \pm 0.4$
        & $24.9 \pm 0.5$
        & $37.4 \pm 0.8$ \\
        
        Battarra-inspired & AFB-ROR
        & $\mathbf{26.3 \pm 0.5}$
        & $\mathbf{24.1 \pm 0.4}$
        & $33.5 \pm 0.6$ \\
        
        \midrule
        
        Peak congestion & Random
        & $32.7 \pm 0.2$
        & $29.1 \pm 0.1$
        & $80.9 \pm 1.4$ \\
        
        Peak congestion & Deepest
        & $34.8 \pm 1.4$
        & $31.4 \pm 1.6$
        & $\mathbf{50.9 \pm 1.5}$ \\
        
        Peak congestion & Nearest
        & $26.7 \pm 0.4$
        & $21.5 \pm 0.4$
        & $70.8 \pm 0.9$ \\
        
        Peak congestion & AFB-ROR
        & $\mathbf{23.1 \pm 0.5}$
        & $\mathbf{21.0 \pm 0.4}$
        & $65.2 \pm 0.5$ \\

        \bottomrule
    \end{tabular}
\end{table*}
}

\begin{figure}[p]
    \centering

    \begin{subfigure}{\textwidth}
        \centering
        \includegraphics[width=0.63\linewidth]
        {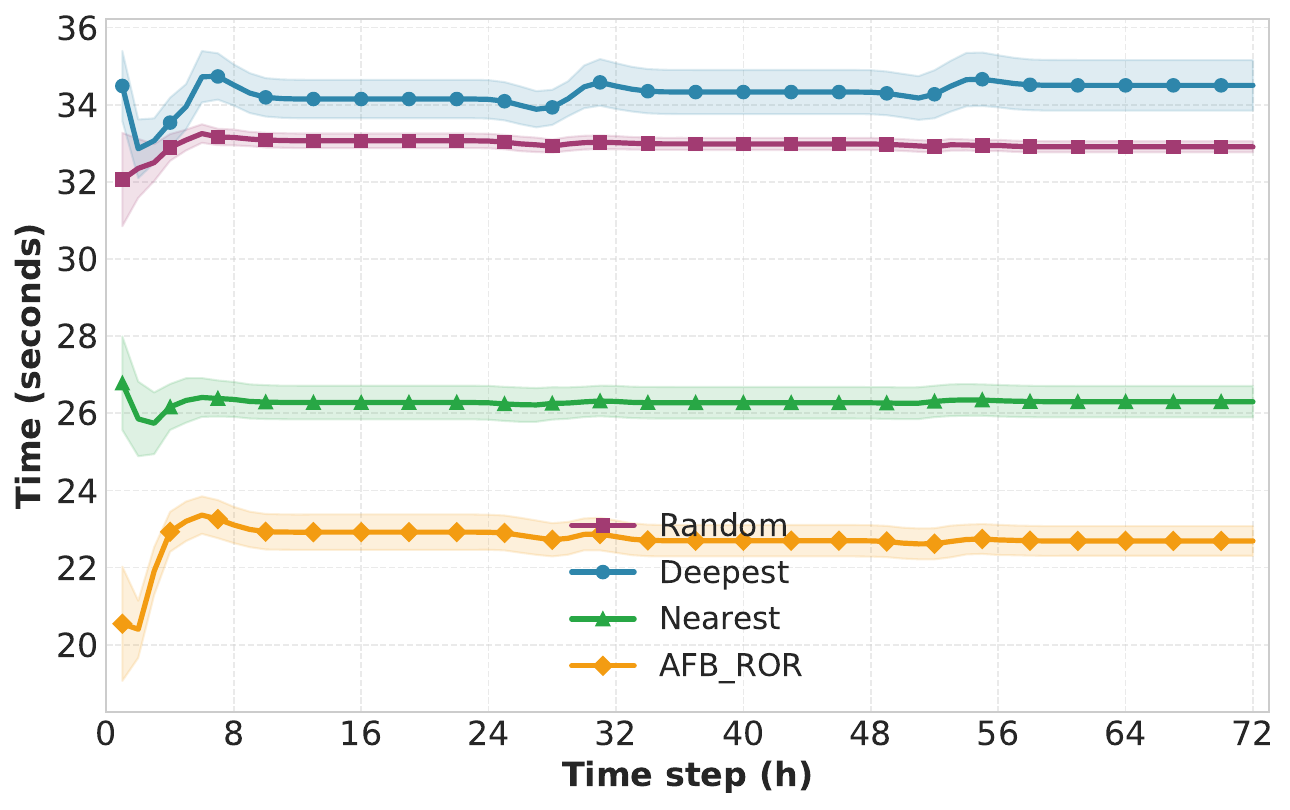}
        \caption{Mean picking time}
        \label{fig:mv-baseline-picking}
    \end{subfigure}

    \smallskip

    \begin{subfigure}{\textwidth}
        \centering
        \includegraphics[width=0.63\linewidth]
        {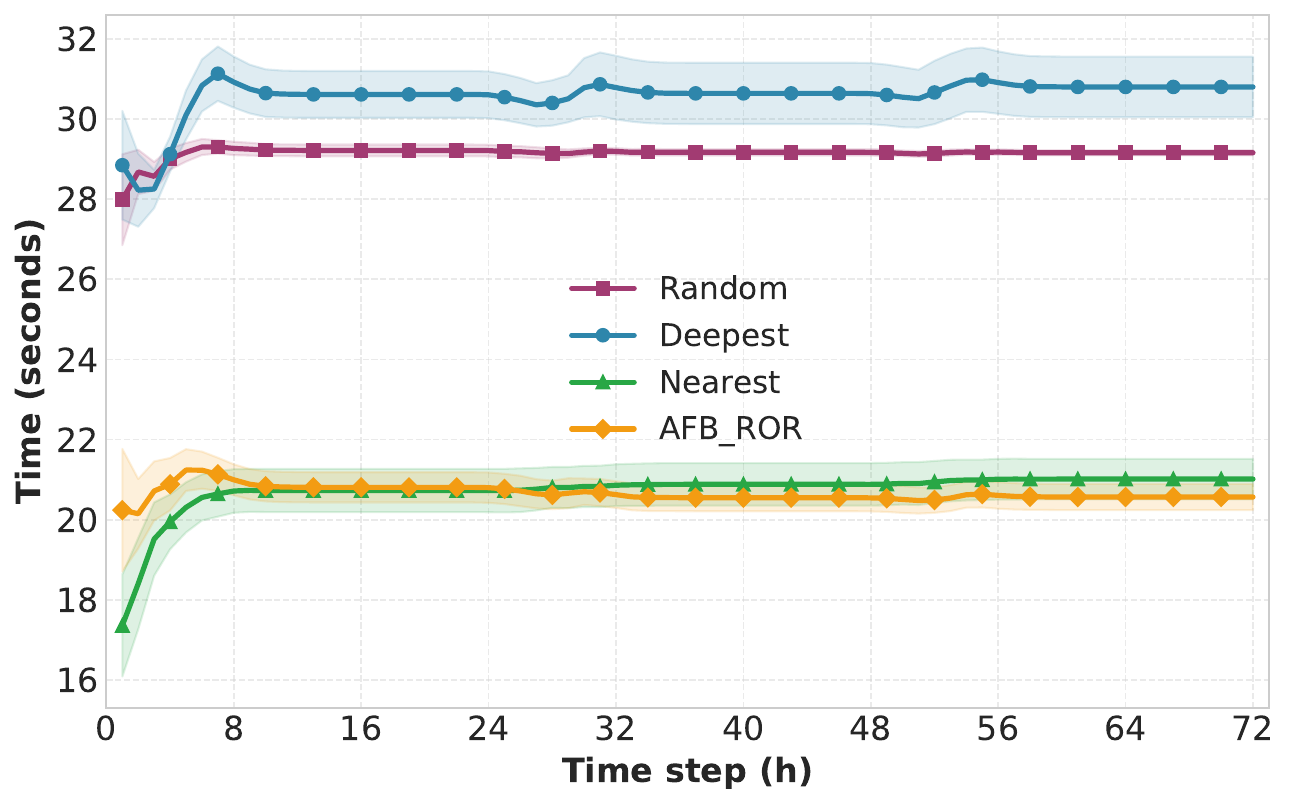}
        \caption{Mean stocking time}
        \label{fig:mv-baseline-stocking}
    \end{subfigure}

    \smallskip

    \begin{subfigure}{\textwidth}
        \centering
        \includegraphics[width=0.63\linewidth]
        {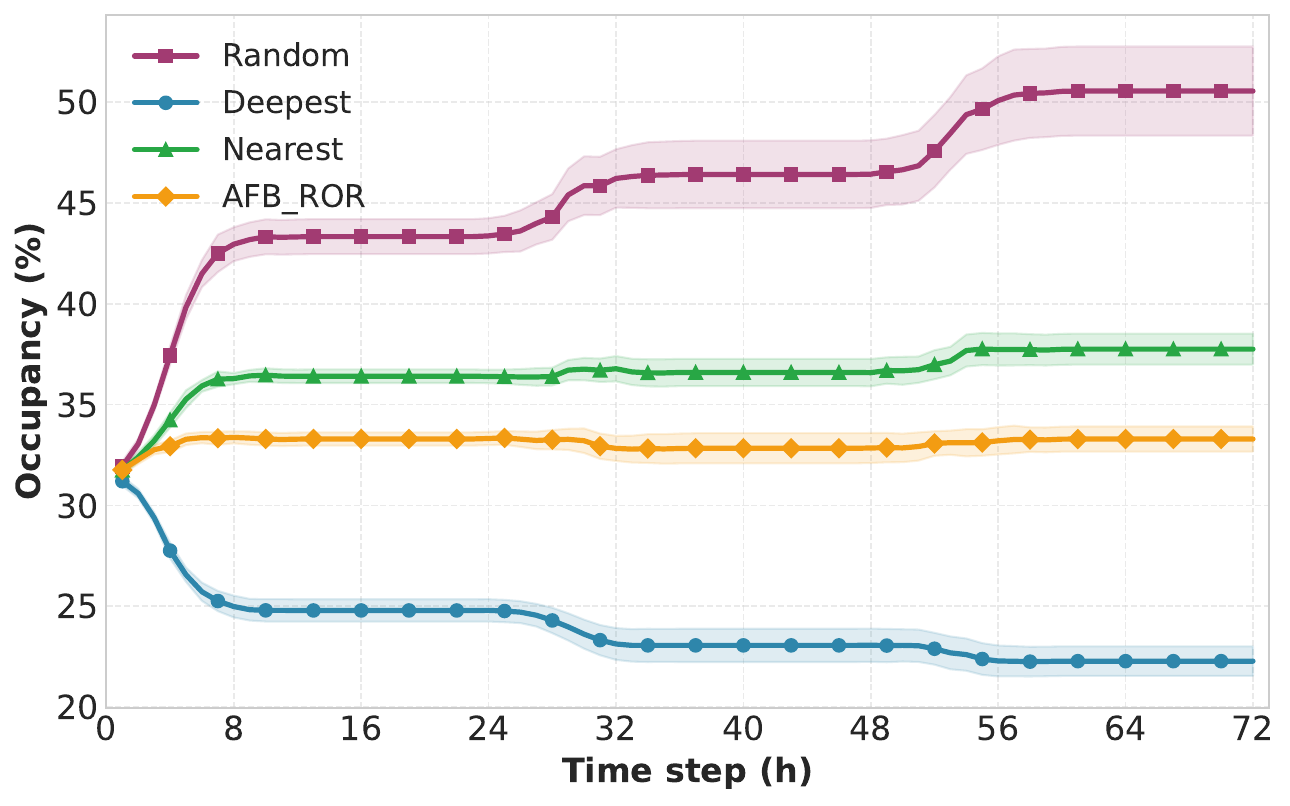}
        \caption{tunnel occupancy}
        \label{fig:mv-baseline-tunnel-occupancy}
    \end{subfigure}

    \caption{Temporal evolution of the three KPIs under the four storage
    and retrieval policies in the baseline scenario. Shaded bands represent pointwise
    two-sided 90\% confidence intervals.}
    \label{fig:mv-baseline}
\end{figure}

\begin{figure}[p]
    \centering

    \begin{subfigure}{\textwidth}
        \centering
        \includegraphics[width=0.63\linewidth]
        {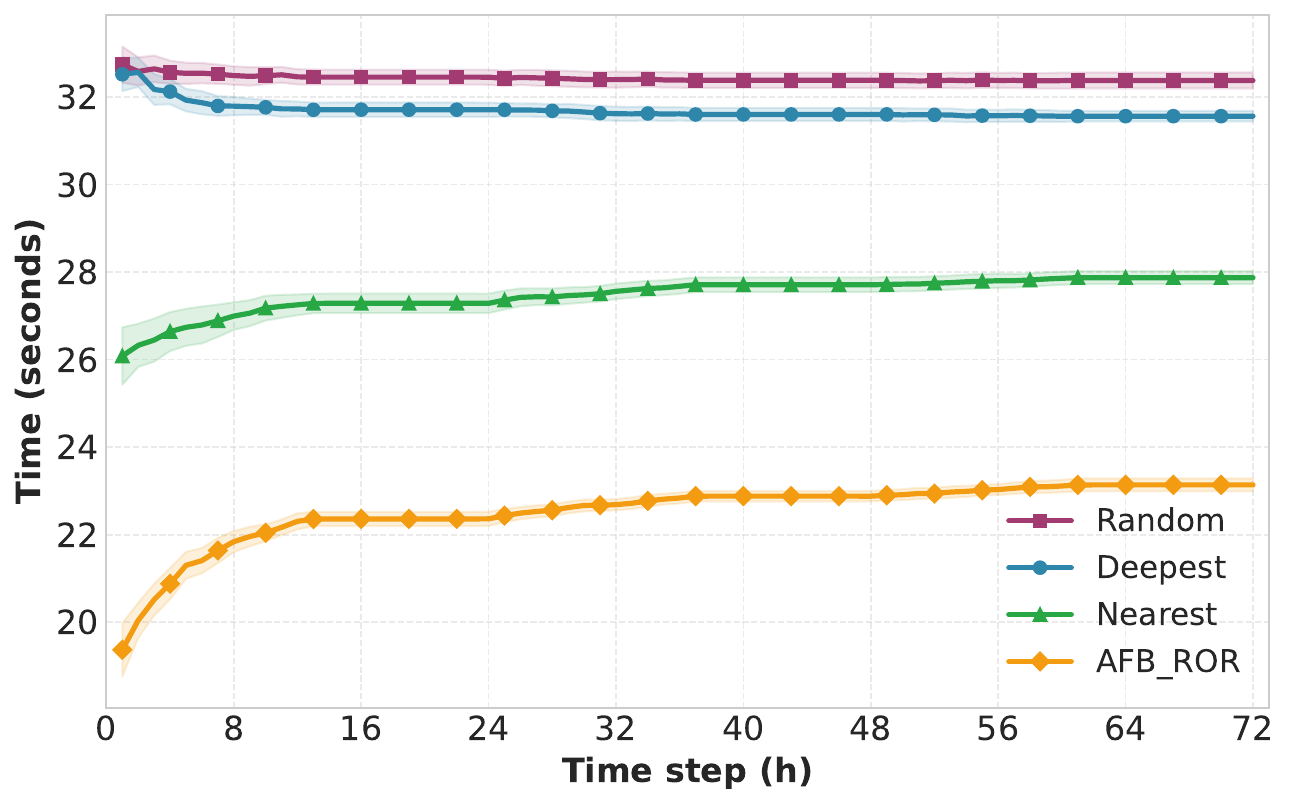}
        \caption{Mean picking time}
        \label{fig:mv-lupi-picking}
    \end{subfigure}

    \smallskip

    \begin{subfigure}{\textwidth}
        \centering
        \includegraphics[width=0.63\linewidth]
        {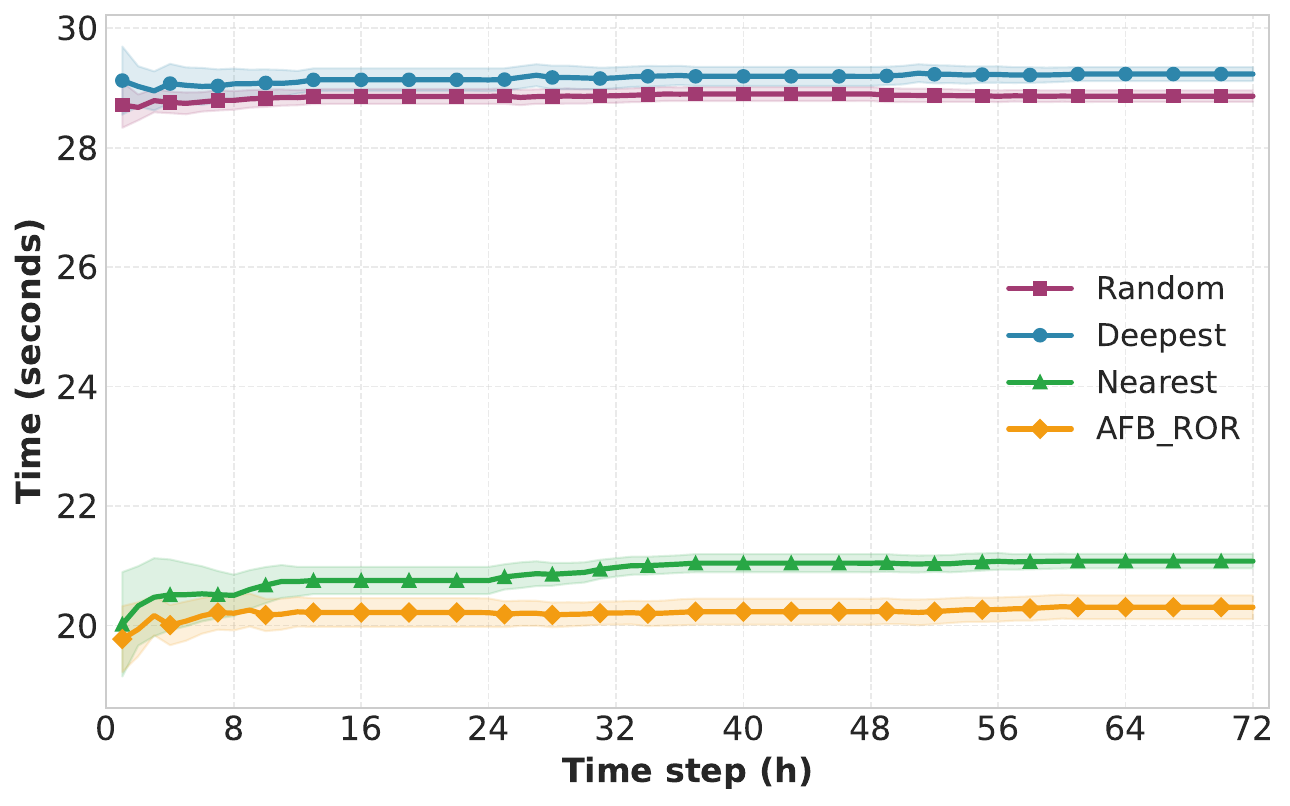}
        \caption{Mean stocking time}
        \label{fig:mv-lupi-stocking}
    \end{subfigure}

    \smallskip

    \begin{subfigure}{\textwidth}
        \centering
        \includegraphics[width=0.63\linewidth]
        {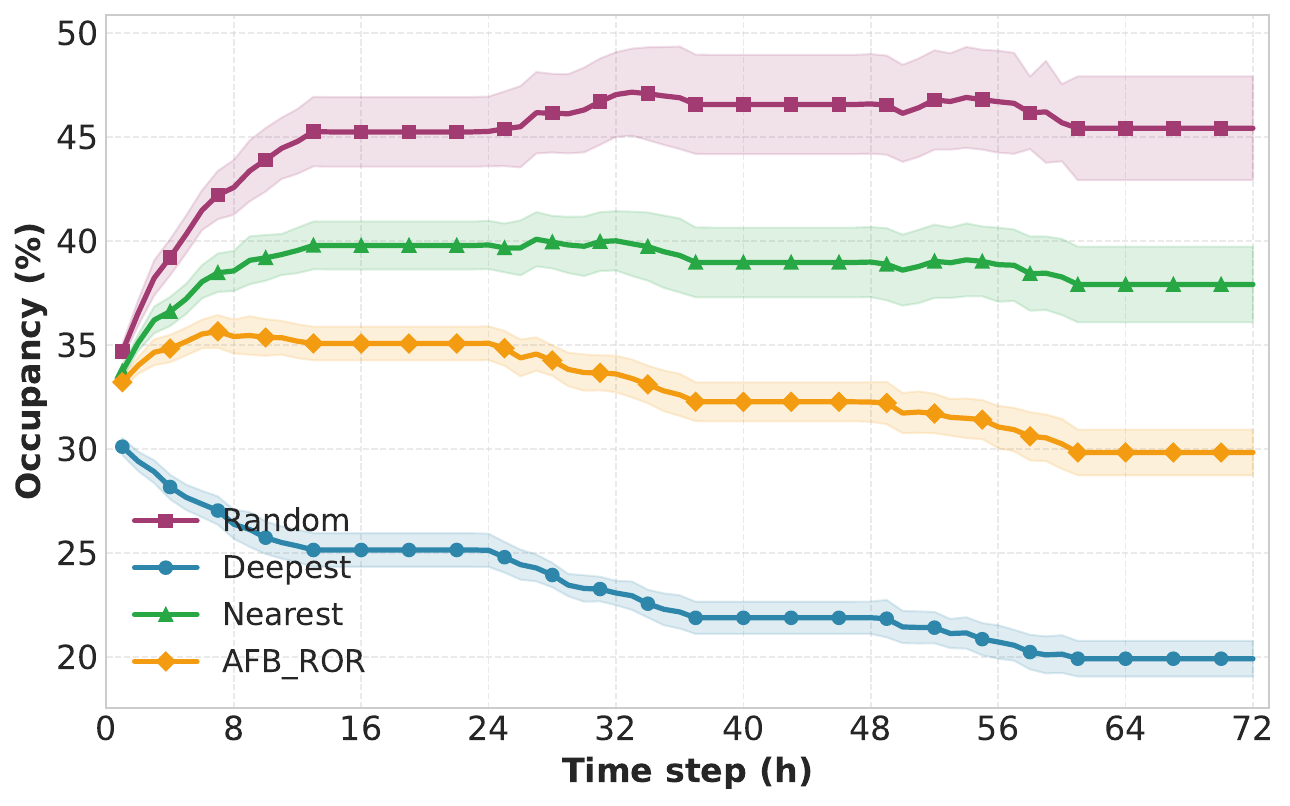}
        \caption{tunnel occupancy}
        \label{fig:mv-lupi-tunnel-occupancy}
    \end{subfigure}

    \caption{Temporal evolution of the three KPIs under the four storage
    and retrieval policies in the Lupi-inspired workload scenario. Shaded
    bands represent two-sided 90\% confidence intervals.}
    \label{fig:mv-lupi}
\end{figure}

\begin{figure}[p]
    \centering

    \begin{subfigure}{\textwidth}
        \centering
        \includegraphics[width=0.63\linewidth]
        {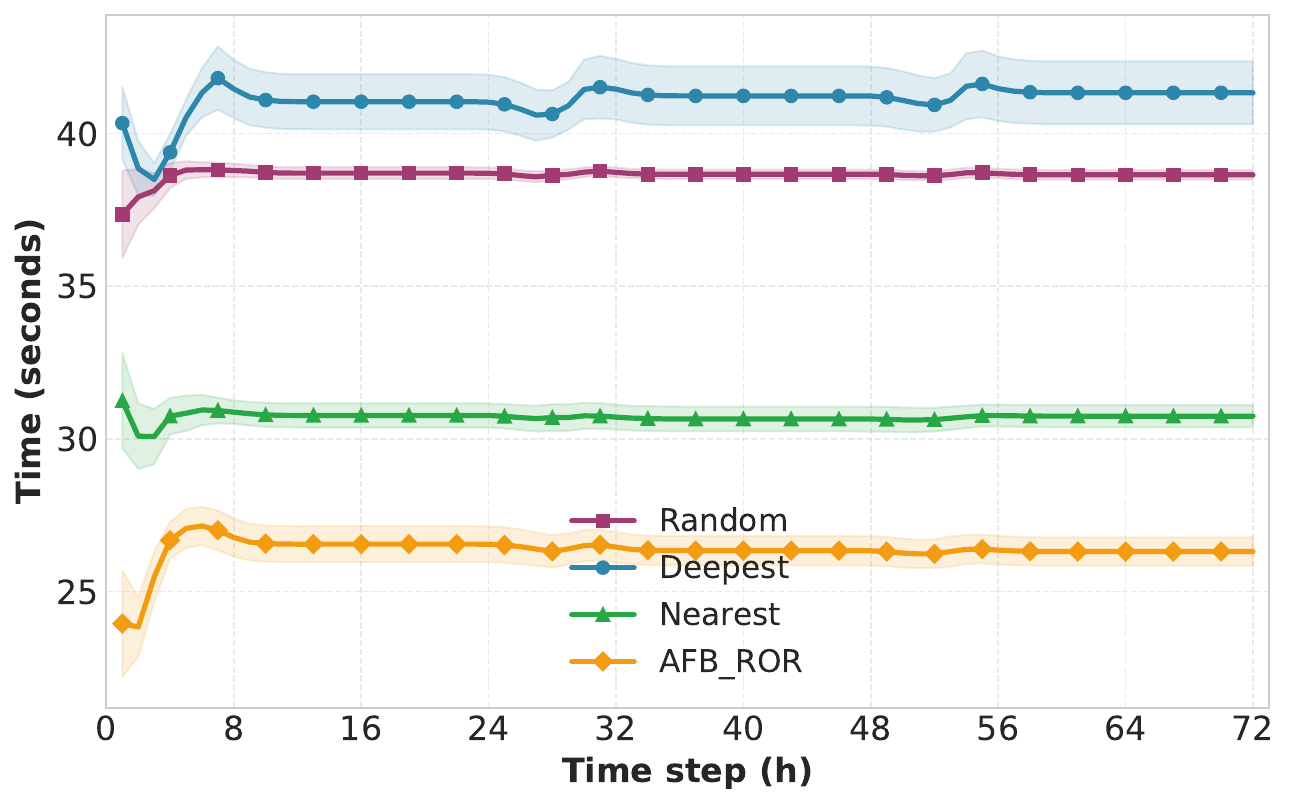}
        \caption{Mean picking time}
        \label{fig:mv-battarra-picking}
    \end{subfigure}

    \smallskip

    \begin{subfigure}{\textwidth}
        \centering
        \includegraphics[width=0.63\linewidth]
        {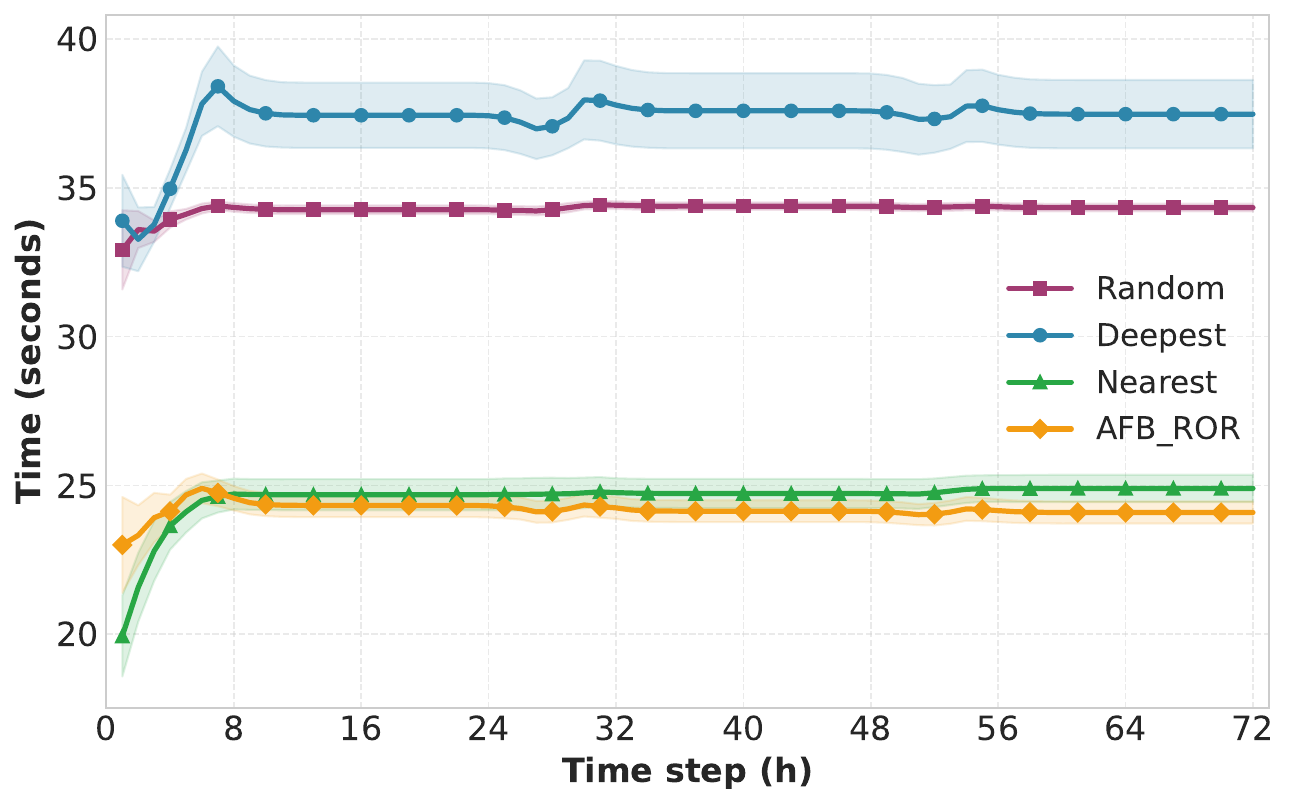}
        \caption{Mean stocking time}
        \label{fig:mv-battarra-stocking}
    \end{subfigure}

    \smallskip

    \begin{subfigure}{\textwidth}
        \centering
        \includegraphics[width=0.63\linewidth]
        {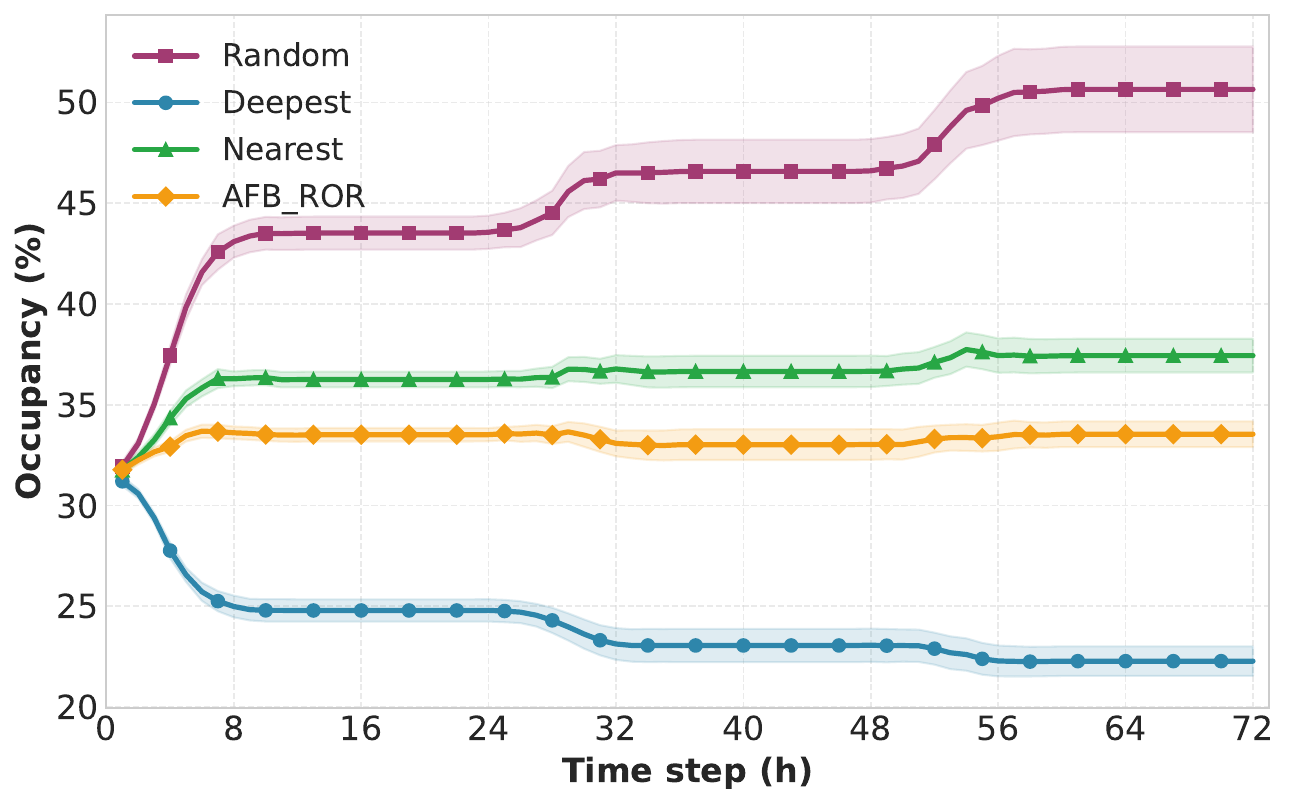}
        \caption{tunnel occupancy}
        \label{fig:mv-battarra-tunnel-occupancy}
    \end{subfigure}

    \caption{Temporal evolution of the three KPIs under the four storage
    and retrieval policies in the Battarra-inspired kinematic scenario.
    Shaded bands represent two-sided 90\% confidence intervals.}
    \label{fig:mv-battarra}
\end{figure}

\begin{figure}[p]
    \centering

    \begin{subfigure}{\textwidth}
        \centering
        \includegraphics[width=0.63\linewidth]
        {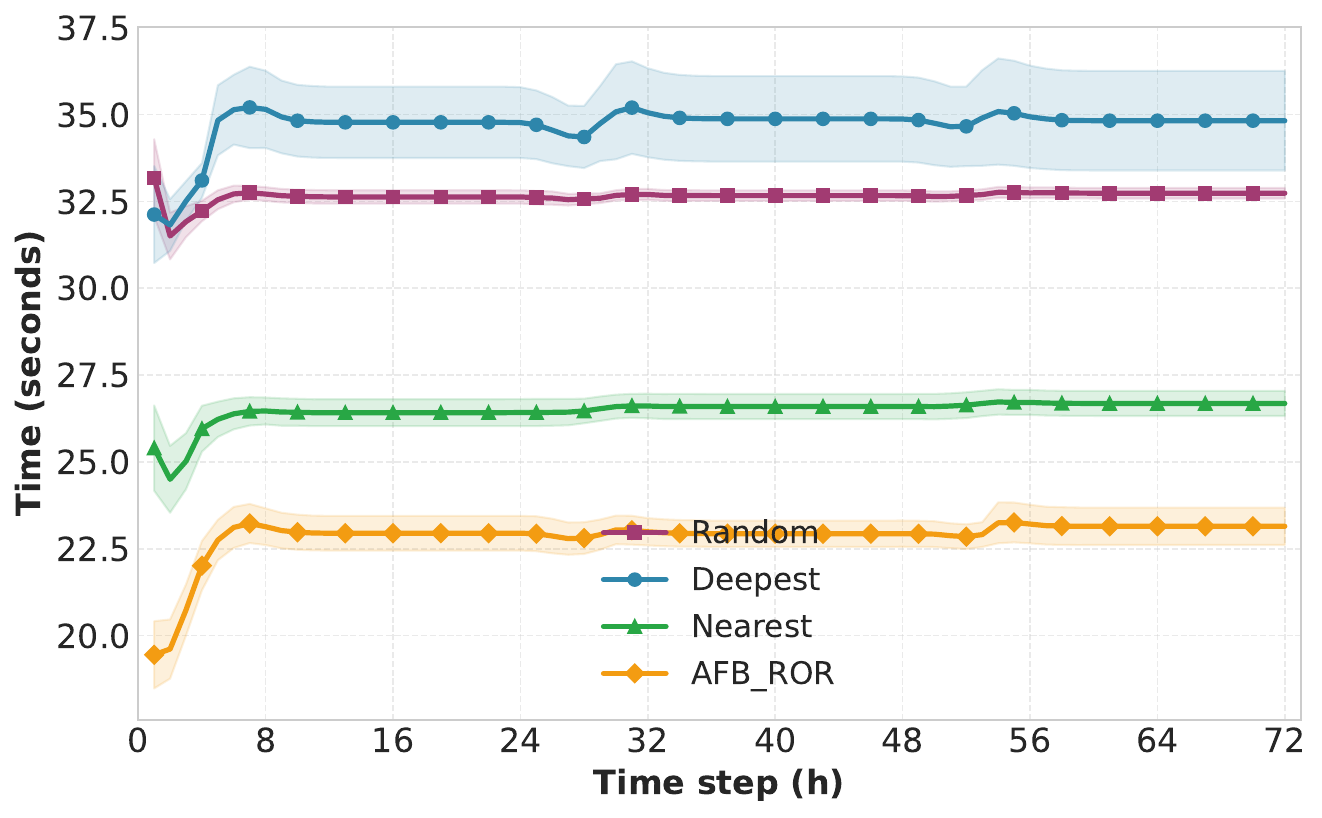}
        \caption{Mean picking time}
        \label{fig:mv-peak-picking}
    \end{subfigure}

    \smallskip

    \begin{subfigure}{\textwidth}
        \centering
        \includegraphics[width=0.63\linewidth]
        {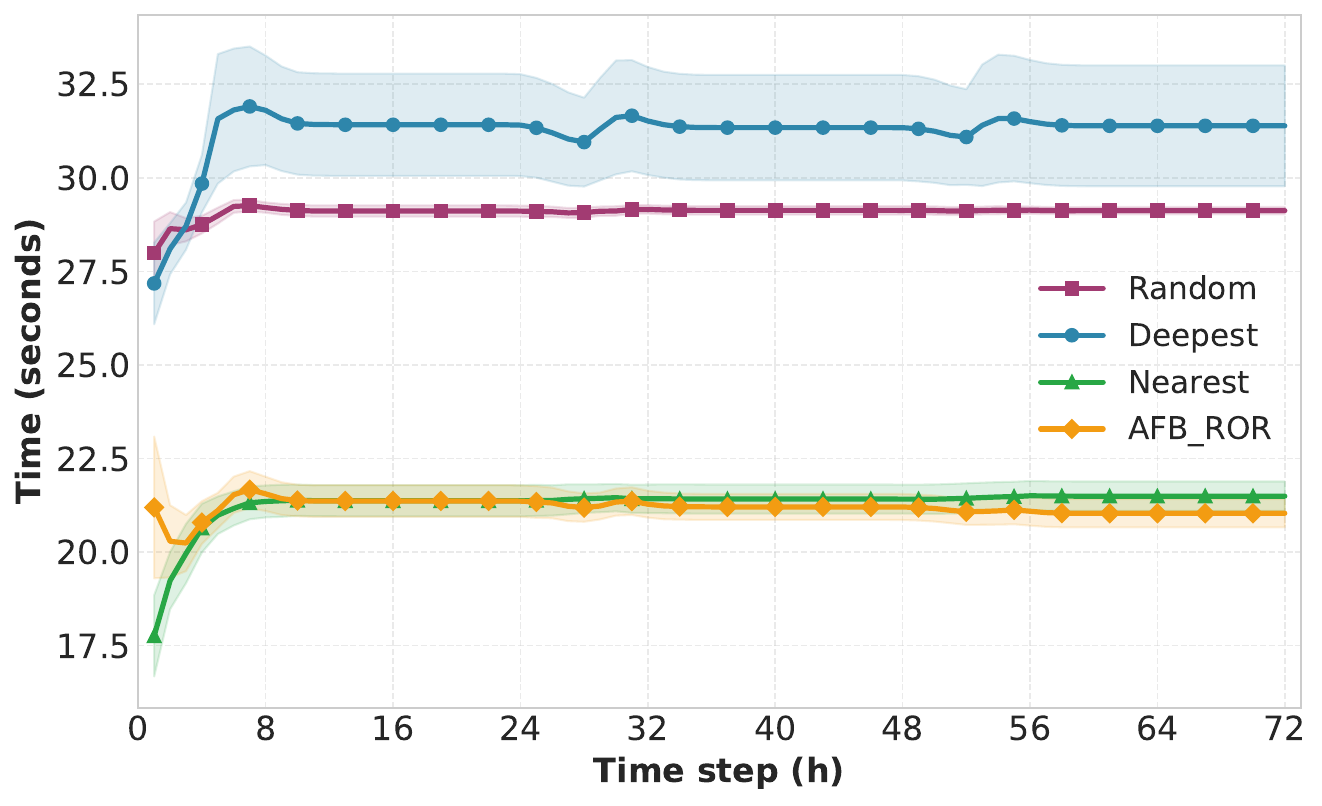}
        \caption{Mean stocking time}
        \label{fig:mv-peak-stocking}
    \end{subfigure}

    \smallskip

    \begin{subfigure}{\textwidth}
        \centering
        \includegraphics[width=0.63\linewidth]
        {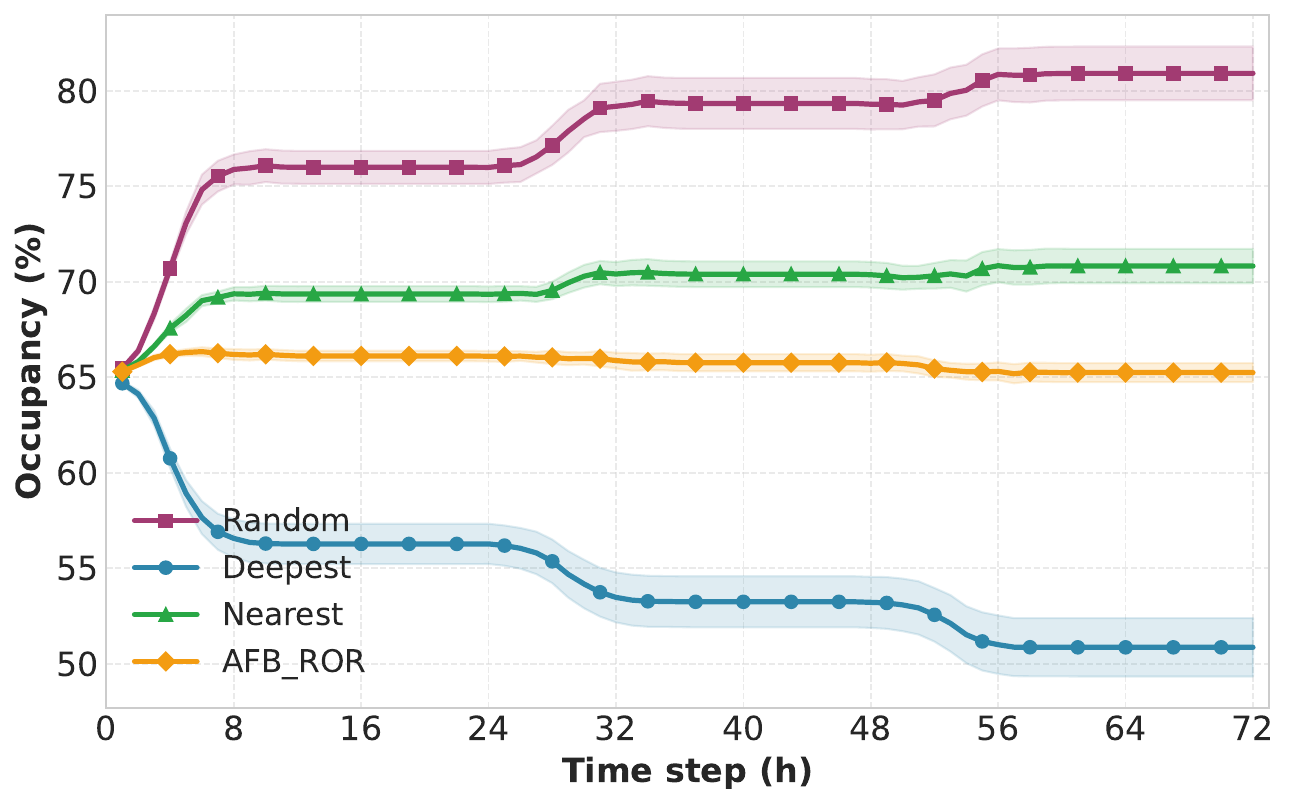}
        \caption{tunnel occupancy}
        \label{fig:mv-peak-tunnel-occupancy}
    \end{subfigure}

    \caption{Temporal evolution of the three KPIs under the four storage
    and retrieval policies in the peak-congestion scenario. Shaded bands
    represent two-sided 90\% confidence intervals.}
    \label{fig:mv-peak}
\end{figure}

\come{Figures~\ref{fig:mv-baseline}--\ref{fig:mv-peak} show the temporal evolution of the three KPIs under the four material-handling policies for each experimental scenario. In each figure, panels~(a), (b), and~(c) respectively report mean picking time, mean stocking time, and tunnel occupancy.} Solid lines represent the estimated mean values, whereas the shaded bands indicate the corresponding two-sided 90\% confidence intervals.

\come{
The prescribed confidence and precision requirements were satisfied for all
KPI--hour observations across the 16 scenario--policy combinations.
\mv{} evaluated the stopping criterion independently for each hourly
observable. Most estimates required a single batch of \(N=20\) independent
replications. Only the tunnel-occupancy estimates for the Lupi-inspired Random
configuration at hours~58 and~60 required a second batch, resulting in
\(N=40\). The maximum budget of three batches was therefore never reached.

The endpoint intervals reported in
Table~\ref{tab:scenario_policy_performance} further show that the achieved
precision was generally tighter than the prescribed upper bound. At hour~72,
the confidence-interval half-widths range from 0.1 to 1.6 seconds for the
cycle-time KPIs and from 0.5 to 2.5 percentage points for tunnel occupancy.
Equivalently, the corresponding full interval widths range from 0.2 to
3.2 seconds and from 1.0 to 5.0 percentage points, respectively. Thus, all
endpoint estimates satisfy the prescribed \(\delta=5\) requirement, while
most are substantially more precise. The limited need for additional batches
reflects the relatively low variability between replications  observed after the
initial transient phase.
}




\come{
The results reveal a clear and persistent trade-off between cycle-time performance and inventory consolidation. After the initial transient, the relative ordering of the policies remains largely stable throughout the 72-hour horizon. AFB-ROR achieves the lowest numerical endpoint mean for both picking and stocking in all four scenarios, followed by Nearest. Relative to Random, the endpoint reductions obtained by AFB-ROR range from approximately 28.5\% to 31.9\% for picking and from 27.8\% to 29.8\% for stocking cycle time. Random and Deepest instead exhibit longer cycle times, although their relative ordering depends on the scenario. In particular, Deepest is slower than Random in the baseline, Battarra-inspired, and peak-congestion scenarios, but slightly faster for picking under the Lupi-inspired workload.

A different ranking emerges for tunnel occupancy. Deepest consistently produces the lowest value across all four scenarios, reaching 22.3\% in the baseline, 19.9\% in the Lupi-inspired scenario, 22.3\% in the Battarra-inspired scenario, and 50.9\% under peak congestion at the end of the simulation horizon. 

This confirms that its tunnel-selection criterion concentrates inventory into a smaller portion of the available tunnels. AFB-ROR provides an intermediate result: its endpoint tunnel occupancy is always lower than that of Random and Nearest, but higher than that of Deepest. It therefore combines the shortest observed cycle times with a stronger degree of inventory consolidation than Nearest, without reaching the space-consolidation performance of Deepest.

The controlled scenario variations further clarify the effects of the experimental factors. The Battarra inspired scenario changes vehicle kinematics while preserving the remaining configuration. Compared with the baseline, this produces endpoint cycle-time increases of approximately 3.5-6.8 seconds, depending on the policy and operation type. Conversely, the corresponding tunnel occupancy trajectories remain almost unchanged, consistently with the fact that the tunnel-selection logic is unaltered.

Under the Lupi-inspired workload, AFB-ROR and Nearest retain their advantage in cycle-time performance, while Deepest slightly outperforms Random for picking and achieves the strongest inventory consolidation. The AFB-ROR tunnel occupancy also decreases progressively over the three simulated days, reaching 29.8\% at the final observation step. This indicates that its frequency-based zoning remains effective under the different workload and SKU-demand characteristics.

The peak-congestion scenario \comee{examines the sensitivity of policy-induced allocation patterns to a higher initial inventory occupancy level}. All policies start from the same initial tunnel occupancy of approximately 65\%, but their trajectories rapidly diverge. Random disperses the inventory further, reaching 80.9\%, while Nearest increases the value to 70.8\%. AFB-ROR remains close to the initial level, ending at 65.2\%, whereas Deepest progressively consolidates the inventory and reduces tunnel occupancy to 50.9\%. Despite the higher initial \comee{tunnel occupancy}, the endpoint cycle-time ranking remains similar to the baseline, with AFB-ROR and Nearest yielding the lowest values. \comee{These findings isolate the effect of a denser initial inventory state on policy behaviour under the baseline resource configuration.}

Overall, the results do not identify a policy that is uniformly preferable for every performance objective. AFB-ROR provides the lowest numerical cycle times across the evaluated scenarios and improves inventory consolidation relative to Random and Nearest. Deepest, conversely, provides the strongest consolidation but generally incurs longer handling times. The statistical distinguishability and practical magnitude of the observed differences are
examined separately in the following comparative analysis.
}

\subsection{Counterfactual Statistical Analysis}
\label{sec:counterfactual-analysis}

\come{The temporal trajectories presented in
Figures~\ref{fig:mv-baseline}--\ref{fig:mv-peak} describe how the KPI
estimates evolve over the complete simulation horizon. However, the initial
hours include transient effects and, in some cases, temporary reversals in the
relative ordering of the policies. For example, under peak congestion,
Nearest initially yields a lower stocking time than AFB-ROR, whereas AFB-ROR
has the lower mean at the end of the horizon. Performing separate inferential
comparisons at every hour would therefore combine transient and mature
operating conditions within the same statistical summary.

For this reason, the counterfactual statistical analysis is restricted to the
predefined endpoint at hour~72. \comee{It corresponds to the terminal observation of the three-day simulation horizon and follows the initial transient phase. Restricting the inferential analysis to this common endpoint avoids the additional multiplicity associated with testing multiple time steps and provides a consistent basis for comparing the policies under mature operating conditions. The corresponding endpoint estimates are reported in }Table~\ref{tab:scenario_policy_performance} \comee{, while} the temporal plots remain important for assessing the evolution and stabilization of the estimates,

For each scenario and KPI, all six pairwise contrasts among Random, Deepest, Nearest, and AFB-ROR were evaluated using two-sided Welch two-sample $t$-tests. The Welch formulation does not assume equal variances across policy configurations. Since the six contrasts belong to the same policy-comparison family within each scenario and KPI, their endpoint $p$-values were adjusted using Holm's step-down procedure ~\cite{Holm1979} to control the family-wise error rate at $\alpha=0.025$. 
\comee{Such parameter was fixed a priori as a deliberately conservative threshold for each scenario–KPI family of six policy contrasts. Combined with Holm’s procedure, it controls the probability of at least one false positive within that family at 2.5\%. This inferential threshold is distinct from the two-sided 90\% confidence intervals used by MultiVeStA to estimate the individual KPI means.}
The tests use the effective sample count, mean, and variance reported by \mv{} for each policy at hour~72.

For each comparison, we define
\[
\Delta_{72}
=
\overline{x}_{1,72}
-
\overline{x}_{2,72},
\]
where $\overline{x}_{1,72}$ and $\overline{x}_{2,72}$ denote the endpoint
means of the first and second policy, respectively. Since lower values are
preferable for all three KPIs, a positive $\Delta_{72}$ favours the second
policy, whereas a negative value favours the first. 

Table~\ref{tab:welch_summary} reports the endpoint differences and the corresponding Holm-adjusted significance results.

Overall, 69 of the 72 endpoint contrasts remain statistically significant after Holm adjustment. The adjustment does not alter any of the significance decisions obtained from the unadjusted tests, thereby confirming the robustness of the endpoint conclusions to multiple-comparison control. The three non-significant comparisons all concern stocking time. In the baseline scenario, AFB-ROR has a mean stocking time approximately 0.45~s lower than Nearest, but the difference is not statistically detectable ($p_{\mathrm{Holm}}=0.208$). A similar result is obtained under peak congestion, where the AFB-ROR advantage over Nearest is approximately 0.46~s ($p_{\mathrm{Holm}}=0.155$). Finally, Random and Deepest are not statistically distinguishable for peak-congestion stocking time ($p_{\mathrm{Holm}}=0.0512$), despite the 2.27~s numerical difference between their endpoint means.

For picking time, AFB-ROR is statistically distinguishable from every other
policy in all four scenarios. Compared with Nearest, its endpoint advantage
ranges from 3.54~s under peak congestion to 4.73~s in the Lupi-inspired
scenario. AFB-ROR also reduces picking time relative to Random by between
9.23 and 12.34~s. Nearest likewise provides a statistically detectable
improvement over Random and Deepest in every scenario. Deepest generally
incurs the longest picking time, except under the Lupi-inspired workload,
where it performs approximately 0.81~s better than Random.

The stocking-time comparisons provide a more nuanced distinction between
AFB-ROR and Nearest. Both policies are statistically faster than Random and
Deepest in every scenario. Their direct comparison, however, depends on the
operating conditions. AFB-ROR is statistically distinguishable from Nearest
under the Lupi-inspired workload, with an endpoint advantage of 0.77~s, and
under the Battarra-inspired kinematics, with an advantage of 0.81~s. 
The latter difference remains only slightly below the adopted family-wise
significance threshold after adjustment
($p_{\mathrm{Holm}}=0.0231$), and its practical magnitude remains limited.
Under the baseline and peak-congestion scenarios, the two policies are not statistically distinguishable for stocking time. Therefore, although AFB-ROR
has the lowest numerical stocking-time mean in all four scenarios, its
advantage over Nearest is not consistently supported across operating
conditions.

The clearest separation among policies concerns tunnel occupancy.
Deepest produces a statistically lower endpoint value than Random, Nearest,
and AFB-ROR in every scenario. Relative to Random, its reduction ranges from
25.53 to 30.06 percentage points; relative to Nearest, it ranges from 15.16
to 19.97 percentage points. AFB-ROR also significantly improves inventory
consolidation compared with Nearest, reducing tunnel occupancy by
between 3.90 and 8.08 percentage points. Nevertheless, AFB-ROR remains
statistically distinguishable from Deepest, whose endpoint
\comee{tunnel} occupancy is between 9.93 and 14.39 percentage points lower.

The endpoint analysis therefore confirms the trade-off observed in the
temporal trajectories. AFB-ROR provides the lowest picking-time means in all
scenarios and the lowest stocking-time means numerically, although its
stocking advantage over Nearest is scenario-dependent. Deepest consistently
provides the strongest inventory consolidation but generally incurs longer
cycle times. Consequently, no policy dominates all three performance
objectives, and the preferred strategy depends on whether handling speed or
inventory consolidation is \comee{prioritised}.}

\begin{table*}[t]
    \centering
    \caption{Welch comparisons between the four policies at hour~72. Each cell reports
    $\Delta_{72}=\overline{x}_{1,72}-\overline{x}_{2,72}$.
    Significance levels refer to Holm-adjusted $p$-values:
    $^{*}p_{\mathrm{Holm}}<0.025$,
    $^{**}p_{\mathrm{Holm}}<0.01$, and
    $^{***}p_{\mathrm{Holm}}<0.001$.
    Unmarked differences are not statistically significant.
    R, D, N, and A denote Random, Deepest, Nearest, and AFB-ROR,
    respectively.}
    \label{tab:welch_summary}

    \scriptsize
    \setlength{\tabcolsep}{3.8pt}
    \renewcommand{\arraystretch}{1.15}

    \begin{tabular}{llrrrrrr}
        \toprule
        \textbf{Scenario} &
        \textbf{KPI} &
        \textbf{R--D} &
        \textbf{R--N} &
        \textbf{R--A} &
        \textbf{D--N} &
        \textbf{D--A} &
        \textbf{N--A} \\
        \midrule

        Baseline &
        Picking &
        $-1.59^{***}$ &
        $+6.61^{***}$ &
        $+10.22^{***}$ &
        $+8.20^{***}$ &
        $+11.81^{***}$ &
        $+3.61^{***}$ \\

        Baseline &
        Stocking &
        $-1.64^{**}$ &
        $+8.14^{***}$ &
        $+8.59^{***}$ &
        $+9.78^{***}$ &
        $+10.23^{***}$ &
        $+0.45$ \\

        Baseline &
        Tunnel occupancy &
        $+28.27^{***}$ &
        $+12.80^{***}$ &
        $+17.25^{***}$ &
        $-15.47^{***}$ &
        $-11.01^{***}$ &
        $+4.46^{***}$ \\

        \midrule

        Lupi-inspired &
        Picking &
        $+0.81^{***}$ &
        $+4.50^{***}$ &
        $+9.23^{***}$ &
        $+3.69^{***}$ &
        $+8.42^{***}$ &
        $+4.73^{***}$ \\

        Lupi-inspired &
        Stocking &
        $-0.37^{***}$ &
        $+7.78^{***}$ &
        $+8.56^{***}$ &
        $+8.15^{***}$ &
        $+8.93^{***}$ &
        $+0.77^{***}$ \\

        Lupi-inspired &
        Tunnel occupancy &
        $+25.53^{***}$ &
        $+7.52^{***}$ &
        $+15.60^{***}$ &
        $-18.01^{***}$ &
        $-9.93^{***}$ &
        $+8.08^{***}$ \\

        \midrule

        Battarra-inspired &
        Picking &
        $-2.69^{***}$ &
        $+7.91^{***}$ &
        $+12.34^{***}$ &
        $+10.59^{***}$ &
        $+15.03^{***}$ &
        $+4.44^{***}$ \\

        Battarra-inspired &
        Stocking &
        $-3.14^{***}$ &
        $+9.43^{***}$ &
        $+10.24^{***}$ &
        $+12.57^{***}$ &
        $+13.38^{***}$ &
        $+0.81^{*}$ \\

        Battarra-inspired &
        Tunnel occupancy &
        $+28.37^{***}$ &
        $+13.21^{***}$ &
        $+17.11^{***}$ &
        $-15.16^{***}$ &
        $-11.26^{***}$ &
        $+3.90^{***}$ \\

        \midrule

        Peak congestion &
        Picking &
        $-2.09^{*}$ &
        $+6.05^{***}$ &
        $+9.59^{***}$ &
        $+8.14^{***}$ &
        $+11.68^{***}$ &
        $+3.54^{***}$ \\

        Peak congestion &
        Stocking &
        $-2.27$ &
        $+7.63^{***}$ &
        $+8.09^{***}$ &
        $+9.90^{***}$ &
        $+10.36^{***}$ &
        $+0.46$ \\

        Peak congestion &
        Tunnel occupancy &
        $+30.06^{***}$ &
        $+10.09^{***}$ &
        $+15.67^{***}$ &
        $-19.97^{***}$ &
        $-14.39^{***}$ &
        $+5.58^{***}$ \\

        \bottomrule
    \end{tabular}
\end{table*}

\subsection{Process Mining Results}\label{sec:pm_results}
\come{The \mv{} analysis quantifies how the four storage and retrieval policies affect picking time, stocking time, and tunnel occupancy across the four experimental scenarios. These aggregate indicators, however, do not reveal which process paths change during system evolution or which operational transitions accumulate the largest share of execution time.

To complement the SMC results, we therefore analyse the generated event logs at process level. Control-flow drift, performance drift, and bottleneck evolution are evaluated independently over all 20 simulation replications for each of the 16 scenario-policy combinations, corresponding to 320 complete event logs. 
\comee{Only the} Disco process maps are subsequently used as an exploratory \comee{visualisation} layer on deterministically selected representative runs, while shuttle \comee{utilisation} and workload distribution are assessed separately from the same event-log data.

The rule-based validation checks did not identify any policy-implementation anomaly in the analysed event logs: all recorded location choices complied with the corresponding feasibility, accessibility, selection, and fallback conditions.

The quantitative Process Mining measures are descriptive rather than significance tests. Their purpose is to identify temporal behavioural variation and localize the operational segments associated with the aggregate performance differences observed through \mv{}.}

\subsubsection*{Control-flow and performance drift}

\paragraph*{Temporal segmentation and support criteria}

\come{A temporal Process Mining analysis was performed independently on all
20 replications of each scenario--policy combination, for a total of 320
complete event logs. For each replication, we analysed the final cumulative
snapshot of the 72-step simulation horizon; hourly snapshots were not
concatenated, thereby avoiding duplicate events.

Each log was partitioned into twelve consecutive, non-overlapping six-hour
windows. The six-hour resolution was selected as a pragmatic compromise
between temporal detail and DFG support: it yields multiple within-day
observations over the three-day horizon while retaining enough complete cases
and recurring directly-follows relations to compute frequency and performance
measures. It is not intended to represent an operational shift boundary or an
optimised temporal threshold. Windows are anchored to the first valid event of
each replication and labelled W01--W12; they therefore denote relative
positions within a run rather than absolute clock-time intervals.

Each mission is assigned to the window containing its first timestamp, and all
of its subsequent events remain in that same window. Thus, traces are never
split across temporal observations. All windows are retained in the descriptive
outputs, but a window with fewer than 30 case starts is marked as low support
and excluded from drift estimation. This rule retains six supported windows
per run in the baseline, Battarra-inspired, and peak-congestion scenarios,
typically W01--W02, W05--W06, and W09--W10. The higher Lupi-inspired
workload also supports W03, W07, and W11. Drift is calculated only when two
numerically adjacent windows both satisfy the support rule; low-support windows
are not bypassed to compare non-adjacent periods.

\paragraph*{Drift metrics}

Control-flow drift measures the Jensen--Shannon distance (see Section \ref{sec:pmmetrics}) between the
normalised directly-follows edge-frequency vectors of two eligible adjacent
windows. Performance drift instead measures changes in the median durations
of common DFG edges: for each edge observed at least ten times in both
windows, the absolute log ratio of the two median durations is computed and
weighted by the combined edge frequency. Consequently, performance drift
jointly reflects recurrent stocking, picking, and service transitions; it is not
an aggregate picking-time or stocking-time KPI. Tunnel occupancy is not
included because it is a system-state indicator rather than an event-log
transition duration.}

\come{For each replication, the maximum supported control-flow drift and the
maximum supported performance drift were identified separately.
Table~\ref{tab:pm_drift} aggregates these run-level maxima over the 20
replications, assigning equal weight to each run despite scenario-specific
differences in the number of supported windows. It also reports the modal
maximum-drift pair and the dominant bottleneck.

For the modal drift pairs, the value in parentheses reports the number of
replications, out of 20, in which that pair contains the corresponding
run-level maximum. For example, W05--W06 (8/20) means that the maximum
drift was observed between W05 and W06 in eight independent replications.

A bottleneck is defined as the DFG edge with the largest contribution to the
total transition time observed within a supported window. Its ranking therefore
accounts jointly for occurrence frequency and transition duration. In the final
column of Table~\ref{tab:pm_drift}, the ratio reports the number of supported
windows in which the indicated edge ranks first among all DFG edges.}

\begin{table*}[t]
    \centering
    \caption{Replicated temporal Process Mining results over 20 runs per
    scenario--policy combination. For each run, the maximum control-flow and
    performance drift is identified separately over supported pairs of adjacent
    six-hour windows. The table reports the mean run-level maximum and its
    two-sided 90\% confidence interval, the modal maximum-drift pair, and the
    dominant bottleneck. Bottleneck prevalence is expressed as the number of
    supported windows in which the indicated edge has the largest contribution
    to total observed DFG transition time.}
    \label{tab:pm_drift}

    \scriptsize
    \setlength{\tabcolsep}{3.3pt}
    \renewcommand{\arraystretch}{1.16}

    \resizebox{\textwidth}{!}{%
    \begin{tabular}{llccccc}
        \toprule
        \textbf{Scenario} &
        \textbf{Policy} &
        \makecell{\textbf{Mean run-level max.}\\
        $\boldsymbol{D_{\mathrm{CF}}}$ \textbf{[CI90]}} &
        \makecell{\textbf{Modal CF pair}\\
        \textbf{(runs)}} &
        \makecell{\textbf{Mean run-level max.}\\
        $\boldsymbol{D_{\mathrm{P}}}$ \textbf{[CI90]}} &
        \makecell{\textbf{Modal performance pair}\\
        \textbf{(runs)}} &
        \makecell{\textbf{Dominant bottleneck}\\
        \textbf{(supported windows)}} \\
        \midrule

        Baseline & Random &
        0.035 [0.031, 0.039] &
        W05--W06 (8/20) &
        0.131 [0.093, 0.170] &
        W09--W10 (10/20) &
        S (89/120) \\

        Baseline & Deepest &
        0.060 [0.053, 0.067] &
        W01--W02 (10/20) &
        0.136 [0.119, 0.154] &
        W01--W02 (8/20) &
        S (119/120) \\

        Baseline & Nearest &
        0.064 [0.056, 0.073] &
        W01--W02 (8/20) &
        0.258 [0.224, 0.292] &
        W09--W10 (9/20) &
        P (116/120) \\

        Baseline & AFB-ROR &
        0.058 [0.051, 0.065] &
        W01--W02 (7/20) &
        0.149 [0.129, 0.170] &
        W09--W10 (12/20) &
        S (116/120) \\

        \midrule

        Lupi-inspired & Random &
        0.063 [0.052, 0.074] &
        W06--W07 (8/20) &
        0.148 [0.112, 0.184] &
        W02--W03 (10/20) &
        L (130/180) \\

        Lupi-inspired & Deepest &
        0.070 [0.059, 0.081] &
        W02--W03 (7/20) &
        0.226 [0.204, 0.248] &
        W06--W07 (7/20) &
        S (160/180) \\

        Lupi-inspired & Nearest &
        0.088 [0.078, 0.098] &
        W01--W02 (7/20) &
        0.225 [0.184, 0.266] &
        W02--W03 (6/20) &
        P (105/180) \\

        Lupi-inspired & AFB-ROR &
        0.088 [0.076, 0.099] &
        W06--W07 (8/20) &
        0.200 [0.182, 0.219] &
        W01--W02 (8/20) &
        S (147/180) \\

        \midrule

        Battarra-inspired & Random &
        0.034 [0.030, 0.038] &
        W05--W06 (9/20) &
        0.124 [0.092, 0.157] &
        W05--W06 (9/20) &
        S (114/120) \\

        Battarra-inspired & Deepest &
        0.059 [0.051, 0.067] &
        W01--W02 (9/20) &
        0.156 [0.134, 0.178] &
        W09--W10 (10/20) &
        S (115/120) \\

        Battarra-inspired & Nearest &
        0.054 [0.047, 0.060] &
        W05--W06 (8/20) &
        0.210 [0.177, 0.242] &
        W05--W06 (7/20) &
        P (88/120) \\

        Battarra-inspired & AFB-ROR &
        0.054 [0.045, 0.063] &
        W09--W10 (9/20) &
        0.142 [0.116, 0.169] &
        W09--W10 (10/20) &
        S (120/120) \\

        \midrule

        Peak congestion & Random &
        0.035 [0.032, 0.038] &
        W01--W02 (8/20) &
        0.192 [0.157, 0.227] &
        W01--W02 (9/20) &
        S (90/120) \\

        Peak congestion & Deepest &
        0.051 [0.045, 0.056] &
        W01--W02 (8/20) &
        0.190 [0.168, 0.212] &
        W01--W02 (9/20) &
        S (114/120) \\

        Peak congestion & Nearest &
        0.065 [0.055, 0.074] &
        W01--W02 (10/20) &
        0.208 [0.170, 0.246] &
        W09--W10 (10/20) &
        P (106/120) \\

        Peak congestion & AFB-ROR &
        0.062 [0.055, 0.070] &
        W05--W06 (9/20) &
        0.148 [0.124, 0.172] &
        W01--W02 (9/20) &
        S (116/120) \\

        \bottomrule
    \end{tabular}%
    }

    \par\medskip
    \raggedright\footnotesize
    $S$: shuttle arrival during stocking $\rightarrow$ cell arrival;
    $P$: picking shuttle reservation $\rightarrow$ picking shuttle arrival;
    $L$: lift releases shuttle $\rightarrow$ picking shuttle arrival.
    CI90 describes the within-configuration mean and is not a test between
    policies. Drift values are descriptive and the two metrics are not directly
    comparable because they have different definitions and ranges.
\end{table*}

\come{The temporal location of the largest drift varies considerably across
replications and should therefore be interpreted more cautiously than its
magnitude. In 24 of the 32 scenario--policy--metric combinations, the modal
pair contains the run-level maximum in fewer than 10 of the 20 replications,
while ties and near-ties among window pairs are also frequent. The magnitude
of temporal variation is thus replicated more consistently than its exact
position within the simulation horizon. This variability is compatible both
with the stochasticity of the order streams and with the anchoring of the
temporal windows to the first valid event of each replication. Consequently,
the modal pairs reported in Table~\ref{tab:pm_drift} identify recurrent
temporal locations of change rather than deterministic change points.

\paragraph*{Control-flow drift}

Control-flow drift remains comparatively limited across all configurations,
indicating that the relative frequencies of the directly-follows relations are
generally stable over time. The mean run-level maximum ranges from 0.034 for
Battarra-inspired Random to 0.088 for both Lupi-inspired Nearest and
Lupi-inspired AFB-ROR. Random produces the lowest mean control-flow maximum
in every scenario: 0.035 in the baseline, 0.063 under the Lupi-inspired
workload, 0.034 in the Battarra-inspired scenario, and 0.035 under peak
congestion. According to this descriptive measure, Random therefore generates
the most temporally stable distribution of process paths among the evaluated
policies. By contrast, the higher values observed for Nearest and AFB-ROR
under the Lupi-inspired workload suggest that the structural composition of
their event logs is more sensitive to the increased workload. Nevertheless,
the confidence intervals overlap in some configurations, and the reported
intervals describe the uncertainty of each within-configuration mean rather
than testing pairwise differences between policies.

Individual replications can exhibit substantially stronger structural variation
than the configuration-level means suggest. The largest observed control-flow
drift is 0.174 for Lupi-inspired Nearest between W06 and W07. Inspection of
the corresponding DFGs associates this change mainly with a reduction in the
relative frequency of stocking paths involving lift reservation, lift arrival,
shuttle reservation, and the final transfer to the storage cell. The control-flow
analysis therefore reveals a temporal redistribution of operational paths that
cannot be identified from aggregate cycle-time or tunnel occupancy indicators
alone. At the same time, the rarity of similarly large values across the
remaining replications indicates that this result should be interpreted as an
informative extreme realisation rather than as a systematic change point for
the configuration.

\paragraph*{Performance drift}

Performance drift exhibits a clearer policy-dependent pattern. The mean
run-level maximum ranges from 0.124 for Battarra-inspired Random to 0.258
for Baseline Nearest. Nearest has the highest mean performance-drift maximum
in the baseline, Battarra-inspired, and peak-congestion scenarios, with values
of 0.258, 0.210, and 0.208, respectively. Under the Lupi-inspired workload,
Deepest and Nearest produce almost identical mean maxima, equal to 0.226 and
0.225, while AFB-ROR remains lower at 0.200. Random records the lowest
value in the baseline, Lupi-inspired, and Battarra-inspired scenarios, whereas
AFB-ROR has the lowest mean maximum under peak congestion. These results
indicate that policies with similar aggregate cycle-time performance can
nevertheless differ in the temporal stability of their transition-duration
profiles. In particular, AFB-ROR remains below Nearest in every scenario,
suggesting that its recurrent transition durations are generally more stable
over time under the tested conditions.

The largest individual performance-drift value is 0.484 for Lupi-inspired
Nearest between W02 and W03. The change is primarily associated with
lift-related picking and stocking transitions, including the transitions from
lift arrival to picking arrival at the I/O station, from lift arrival during
stocking to shuttle arrival, and from lift reservation by a shuttle to the lift
collecting that shuttle. This result localises the process segments responsible
for the strongest observed temporal variation, but it does not by itself indicate
an anomaly or a statistically significant regime change. More generally,
because control-flow drift is a bounded Jensen--Shannon distance whereas
performance drift is based on weighted absolute log ratios and is not bounded
to the same range, the numerical magnitudes of the two metrics are not directly
comparable. Considered jointly, however, they show that temporal variation
more often concerns the duration of recurrent transitions than a major
alteration of the available process paths.

\paragraph*{Bottleneck analysis}

The bottleneck analysis further distinguishes the operational patterns
associated with the policies. For Deepest and AFB-ROR, the dominant edge is
generally the final stocking transition from shuttle arrival during stocking to
cell arrival, denoted by (S). For Baseline Deepest, this edge ranks first in 119
of the 120 supported windows. It also ranks first in all 120 supported windows
for Battarra-inspired AFB-ROR, 116 of 120 windows for Baseline AFB-ROR,
160 of 180 windows for Lupi-inspired Deepest, 147 of 180 windows for
Lupi-inspired AFB-ROR, 115 of 120 windows for Battarra-inspired Deepest,
114 of 120 windows for Peak-congestion Deepest, and 116 of 120 windows for
Peak-congestion AFB-ROR. This repeated concentration of transition time in
the final in-channel movement is consistent with policies that apply more
deliberate tunnel selection or inventory consolidation, although the Process
Mining results identify the location of accumulated time rather than
establishing its operational cause.

Nearest produces a distinct bottleneck profile. Its dominant edge is typically
the interval between picking-shuttle reservation and picking-shuttle arrival,
denoted by (P). This transition ranks first in 116 of 120 supported baseline
windows, 105 of 180 Lupi-inspired windows, 88 of 120 Battarra-inspired
windows, and 106 of 120 peak-congestion windows. Thus, although Nearest
achieves strong aggregate cycle-time performance in the \mv{} analysis, its
event logs reveal a persistent concentration of process time in the reservation,
availability, and arrival of the selected picking shuttle.

Random exhibits a less uniform bottleneck structure. The final stocking edge
(S) dominates in the baseline, Battarra-inspired, and peak-congestion scenarios,
appearing as the leading bottleneck in 89 of 120, 114 of 120, and 90 of 120
supported windows, respectively. Under the more intensive Lupi-inspired
workload, however, the dominant edge shifts to the transition from lift release
to picking-shuttle arrival, denoted by (L), which ranks first in 130 of the 180
supported windows. This shift suggests that the increased workload changes the
principal location of accumulated transition time: rather than remaining
concentrated in the final stocking movement, it moves toward the interaction
between lift release and subsequent shuttle availability. Random therefore
maintains relatively stable control flow while exhibiting a workload-dependent
redistribution of its principal temporal bottleneck.

Overall, the replicated temporal Process Mining results complement the
\mv{} analysis in two aspects. First, the drift measures determine whether
the relative process-path frequencies and recurrent transition-duration profiles
remain stable while aggregate KPIs evolve. Second, the bottleneck ranking
identifies the specific process edge on which observed DFG time is concentrated,
accounting jointly for transition frequency and duration. The \mv{} analysis
establishes which policies differ in overall system performance, whereas the
Process Mining analysis shows where those differences materialise within the
process and whether the underlying execution patterns remain stable over time.}

\subsubsection*{Shuttle \comee{utilisation} and workload distribution}
\come{An additional analysis was conducted to quantify how shuttle \comee{utilisation} and workload distribution vary under the different stocking and picking policies. Leveraging on the event logs, it was possible to distinguish overall fleet loading from the distribution of missions among individual shuttles. These two dimensions must be considered separately: a low fleet-average \comee{utilisation} does not necessarily imply that the operational workload is evenly distributed across the available resources.

Table~\ref{tab:shuttle_utilization_distribution} jointly reports mean fleet \comee{utilisation} during the daily operating periods and Jain's Index $J_{\mathrm{S}}$ for the distribution of mission counts among the ten shuttles.
Values approaching one indicate an even distribution, whereas lower values indicate that missions are concentrated on a smaller subset of shuttles.}
        
\begin{table*}[t]
\centering
\caption{Shuttle \comee{utilisation} and workload distribution across the experimental
configurations. Values are means over 20 independent simulation replications
and are reported as mean $\pm$ half-width of the two-sided 90\% Student-$t$
confidence interval.
$U_{\mathrm{F}}$ denotes mean fleet \comee{utilisation} during warehouse operating
hours. $J_{\mathrm{S}}$ denotes Jain's Index computed from the mission counts
assigned to the ten shuttles.}
\label{tab:shuttle_utilization_distribution}

\small
\setlength{\tabcolsep}{3.5pt}
\renewcommand{\arraystretch}{1.15}

\begin{tabular}{@{}llcccc@{}}
\toprule
\textbf{Scenario} &
\textbf{Metric} &
\textbf{Random} &
\textbf{Deepest} &
\textbf{Nearest} &
\textbf{AFB-ROR} \\
\midrule

Baseline
& $U_{\mathrm{F}}$ (\%)
& $6.63 \pm 0.05$
& $6.16 \pm 0.06$
& $4.90 \pm 0.07$
& $4.36 \pm 0.08$ \\

& $J_{\mathrm{S}}$
& $0.815 \pm 0.008$
& $0.819 \pm 0.026$
& $0.255 \pm 0.004$
& $0.727 \pm 0.036$ \\

\addlinespace[3pt]

Lupi-inspired
& $U_{\mathrm{F}}$ (\%)
& $9.84 \pm 0.08$
& $9.58 \pm 0.07$
& $7.75 \pm 0.06$
& $6.97 \pm 0.06$ \\

& $J_{\mathrm{S}}$
& $0.795 \pm 0.008$
& $0.794 \pm 0.012$
& $0.300 \pm 0.009$
& $0.698 \pm 0.026$ \\

\addlinespace[3pt]

Battarra-inspired
& $U_{\mathrm{F}}$ (\%)
& $7.77 \pm 0.07$
& $7.22 \pm 0.08$
& $5.74 \pm 0.07$
& $4.99 \pm 0.08$ \\

& $J_{\mathrm{S}}$
& $0.828 \pm 0.008$
& $0.834 \pm 0.018$
& $0.279 \pm 0.005$
& $0.748 \pm 0.035$ \\

\addlinespace[3pt]

Peak congestion
& $U_{\mathrm{F}}$ (\%)
& $6.54 \pm 0.06$
& $6.11 \pm 0.05$
& $4.97 \pm 0.06$
& $4.37 \pm 0.08$ \\

& $J_{\mathrm{S}}$
& $0.805 \pm 0.008$
& $0.799 \pm 0.020$
& $0.265 \pm 0.006$
& $0.714 \pm 0.029$ \\

\bottomrule
\end{tabular}
\end{table*}

\come{Mean fleet \comee{utilisation} remains below 10\% in every scenario-policy combination. The highest \comee{utilisation} is observed under the Lupi-inspired workload, which increases shuttle loading across all four policies. Random reaches 9.84\% and Deepest 9.58\%, compared with 7.75\% for Nearest and 6.97\% for AFB-ROR. Conversely, the similarity between the baseline and peak-congestion results indicates that increasing the initial inventory footprint, without simultaneously increasing the mission workload, does not produce fleet-wide shuttle saturation.

AFB-ROR has the lowest fleet-average \comee{utilisation} in all four scenarios, followed by Nearest. This result is consistent with their shorter mission cycle times, because individual shuttles remain reserved and busy for shorter periods. Nevertheless, fleet-average \comee{utilisation} alone conceals substantial differences in how the missions are allocated among resources.}

\come{Random and Deepest produce the most even shuttle workload distributions, with $J_{\mathrm{S}}$ between approximately 0.79 and 0.83 across the four scenarios. Nearest exhibits a markedly different profile: its index ranges only from 0.255 to 0.300, showing that its relatively low aggregate \comee{utilisation} is associated with a strong concentration of missions on a limited subset of shuttles. This result is consistent with the spatial preference embedded in the policy.

AFB-ROR occupies an intermediate position. Its shuttle-distribution index ranges from 0.698 to 0.748, substantially above Nearest but below Random and Deepest. Therefore, AFB-ROR combines the lowest fleet-average \comee{utilisation} with a considerably less concentrated resource allocation than Nearest.} 


\subsubsection*{Process-map analysis}

\come{To complement the replicated temporal analysis, Disco process maps are used to provide an intuitive visual representation of the operational patterns identified in the event logs.}

\begin{sidewaysfigure}
    \centering

    \begin{subfigure}{0.235\textheight}
        \centering
        \includegraphics[width=\linewidth]
        {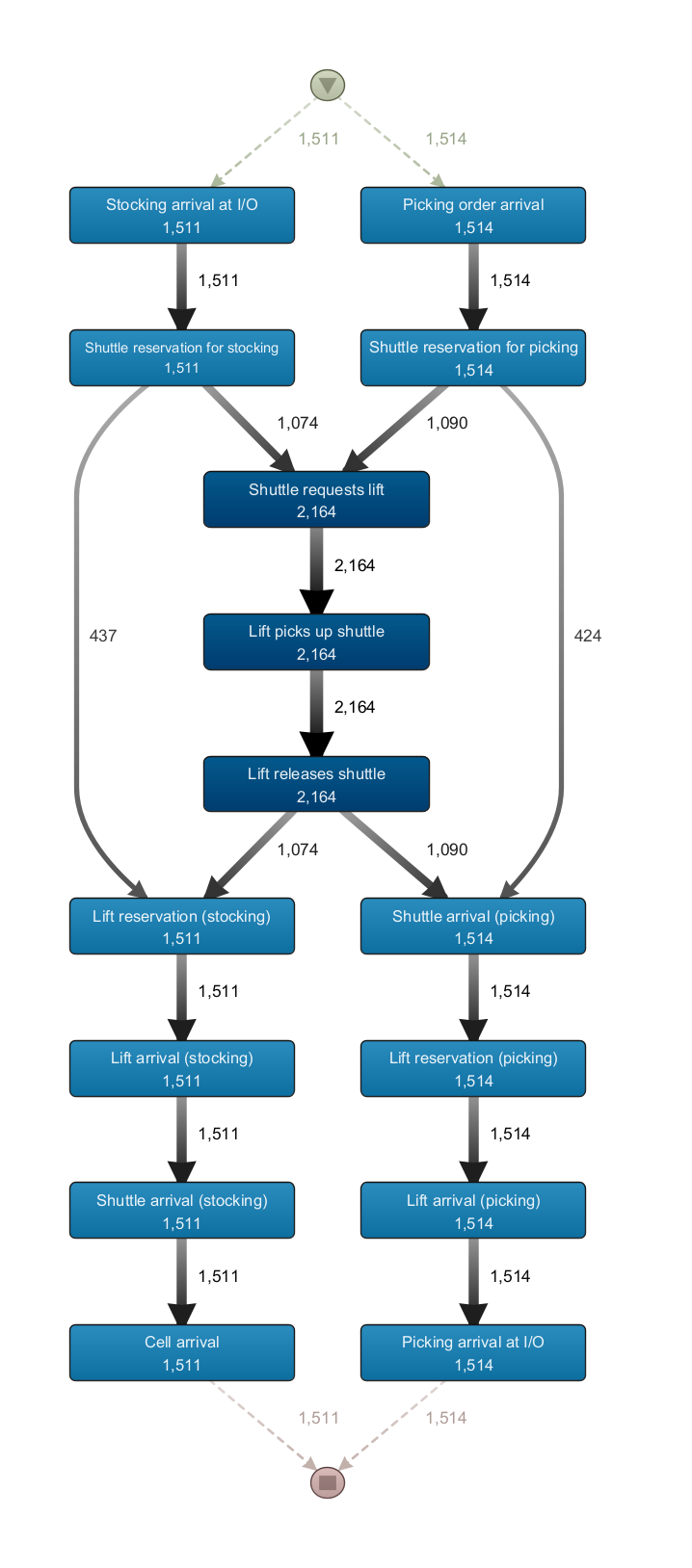}
        \caption{Random}
        \label{fig:disco_frequency_random}
    \end{subfigure}
    \hfill
    \begin{subfigure}{0.235\textheight}
        \centering
        \includegraphics[width=\linewidth]
        {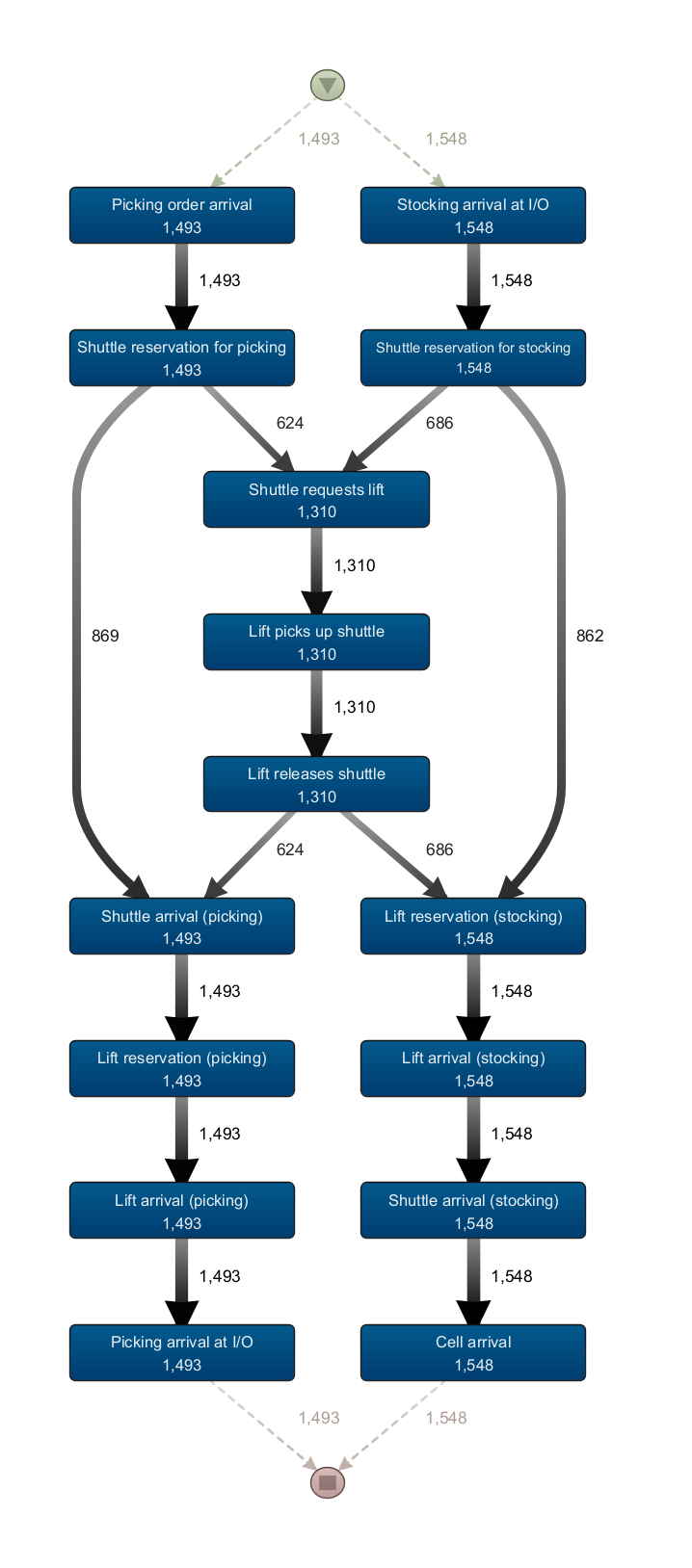}
        \caption{Deepest}
        \label{fig:disco_frequency_deepest}
    \end{subfigure}
    \hfill
    \begin{subfigure}{0.235\textheight}
        \centering
        \includegraphics[width=\linewidth]
        {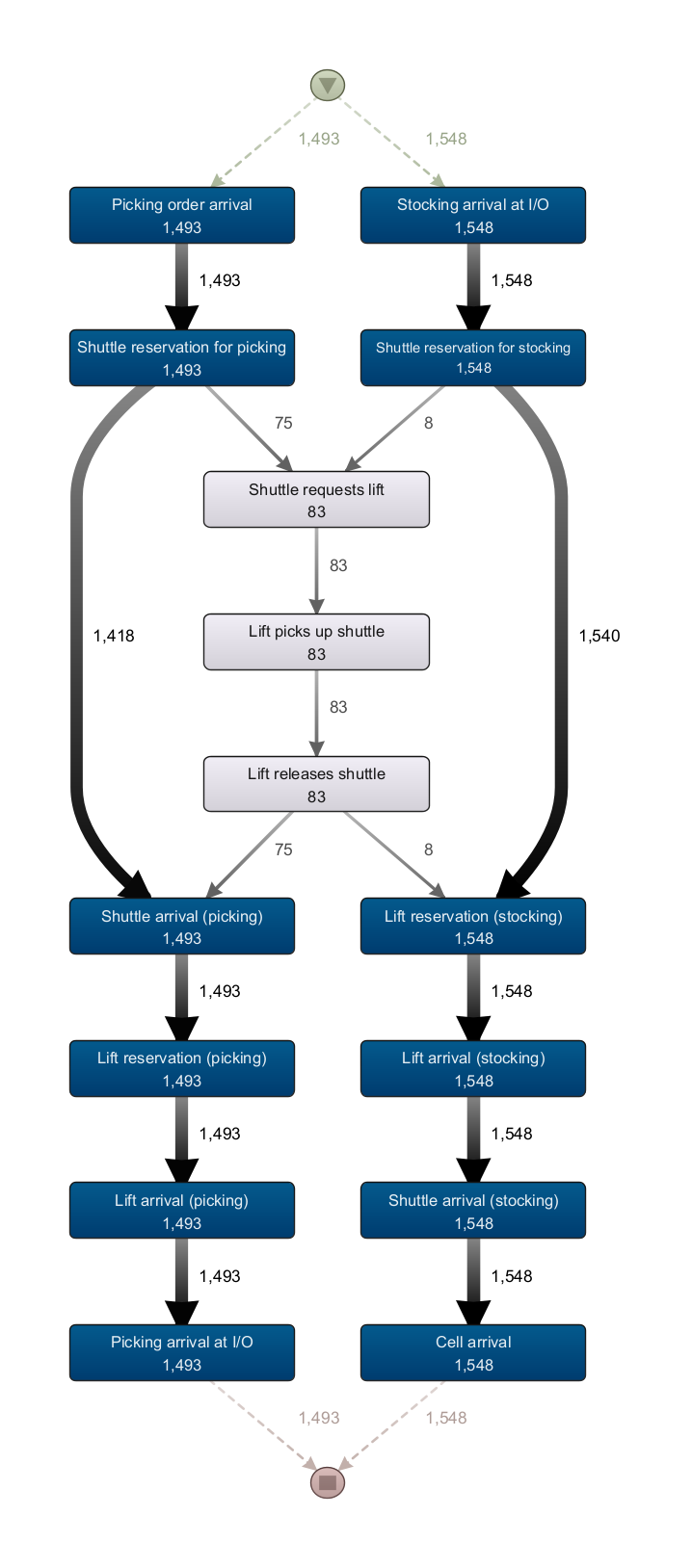}
        \caption{Nearest}
        \label{fig:disco_frequency_nearest}
    \end{subfigure}
    \hfill
    \begin{subfigure}{0.235\textheight}
        \centering
        \includegraphics[width=\linewidth]
        {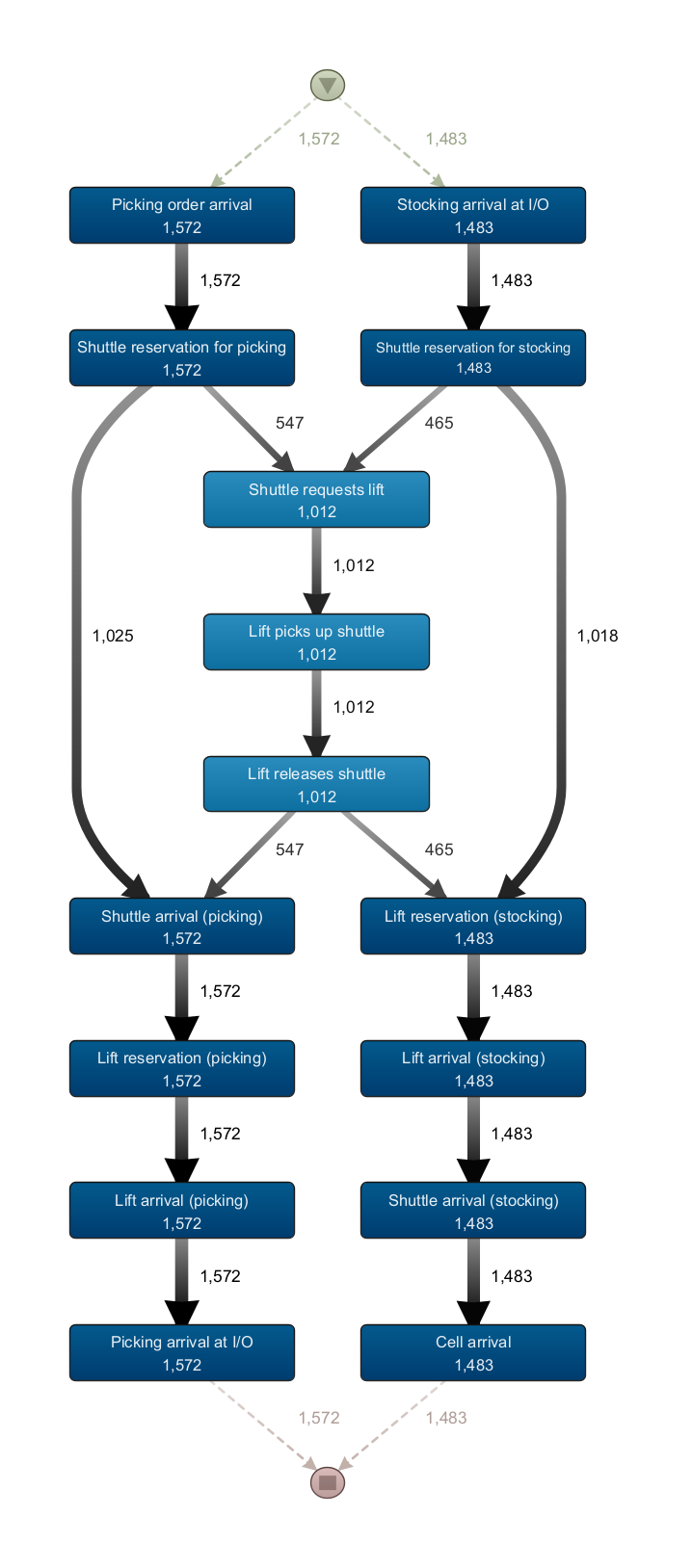}
        \caption{AFB-ROR}
        \label{fig:disco_frequency_afbror}
    \end{subfigure}

    \caption{Frequency process maps obtained from one representative baseline run for each policy. Edge annotations and thickness represent transition frequency.}
    \label{fig:process_maps}
\end{sidewaysfigure}

\begin{sidewaysfigure}
    \centering

    \begin{subfigure}{0.235\textheight}
        \centering
        \includegraphics[width=\linewidth]
        {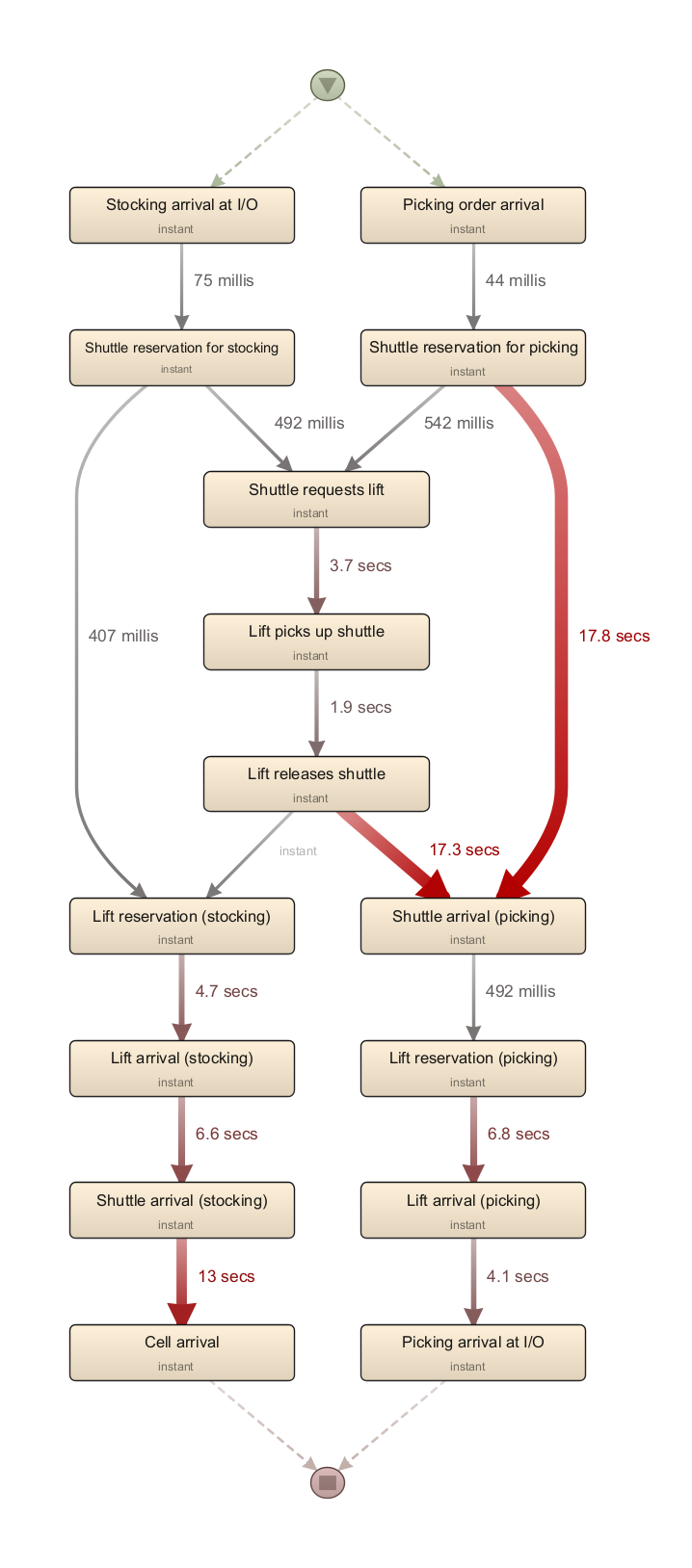}
        \caption{Random}
        \label{fig:performance_random}
    \end{subfigure}
    \hfill
    \begin{subfigure}{0.235\textheight}
        \centering
        \includegraphics[width=\linewidth]
        {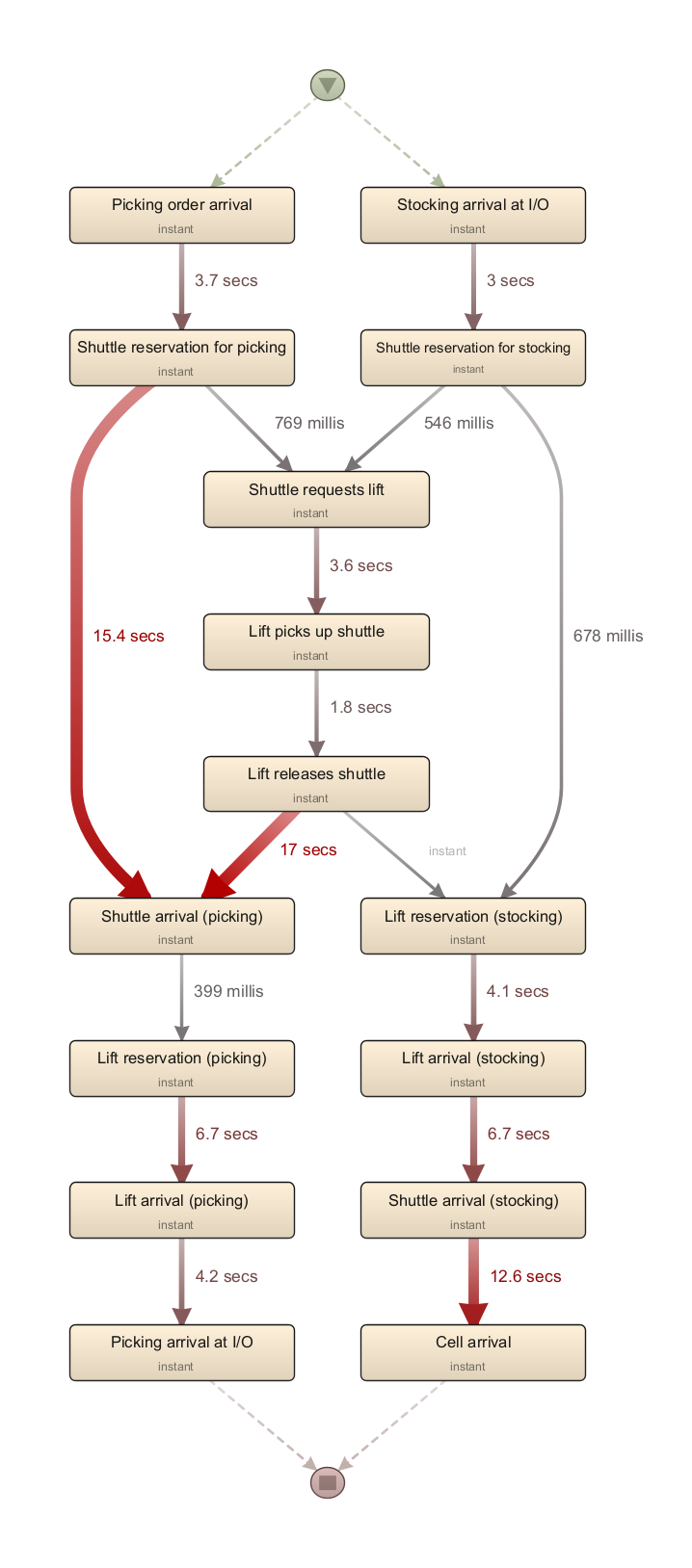}
        \caption{Deepest}
        \label{fig:performance_deepest}
    \end{subfigure}
    \hfill
    \begin{subfigure}{0.235\textheight}
        \centering
        \includegraphics[width=\linewidth]
        {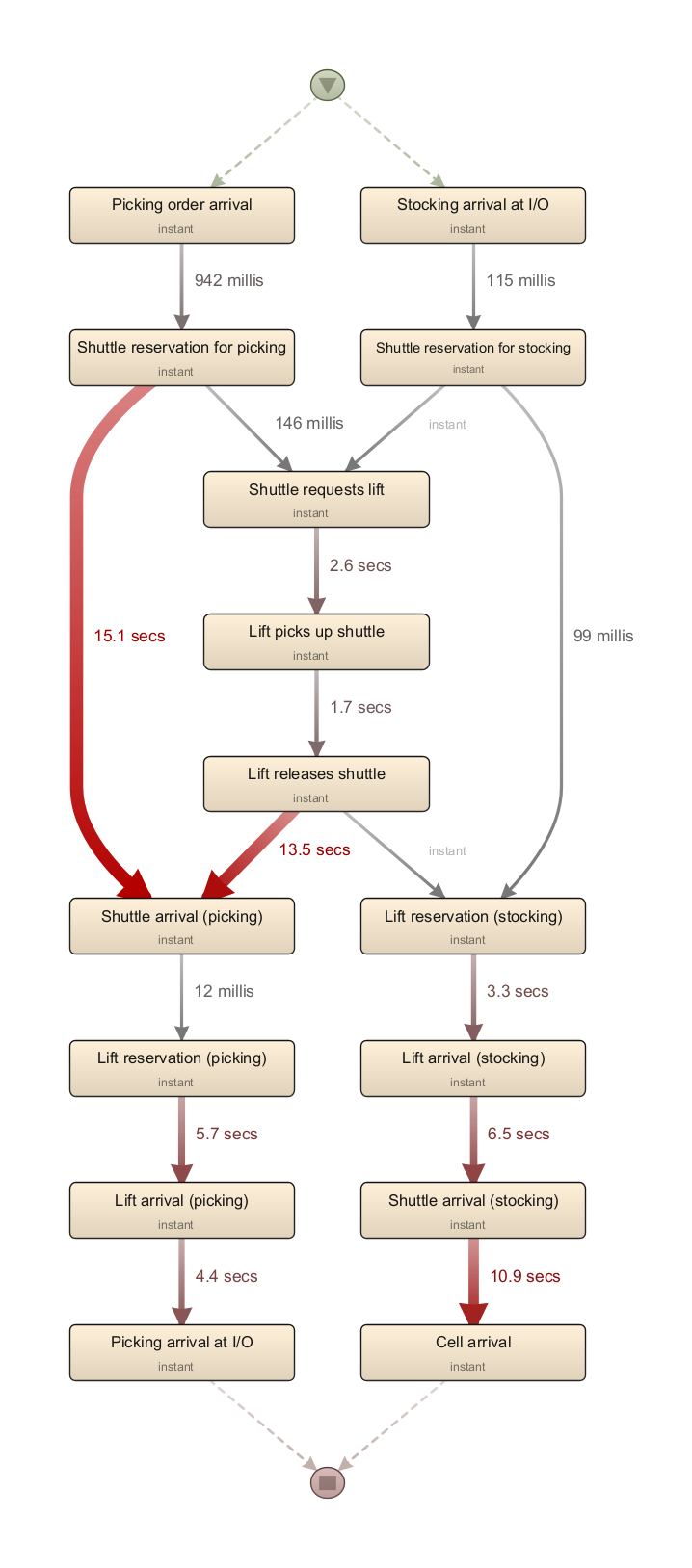}
        \caption{Nearest}
        \label{fig:performance_nearest}
    \end{subfigure}
    \hfill
    \begin{subfigure}{0.235\textheight}
        \centering
        \includegraphics[width=\linewidth]
        {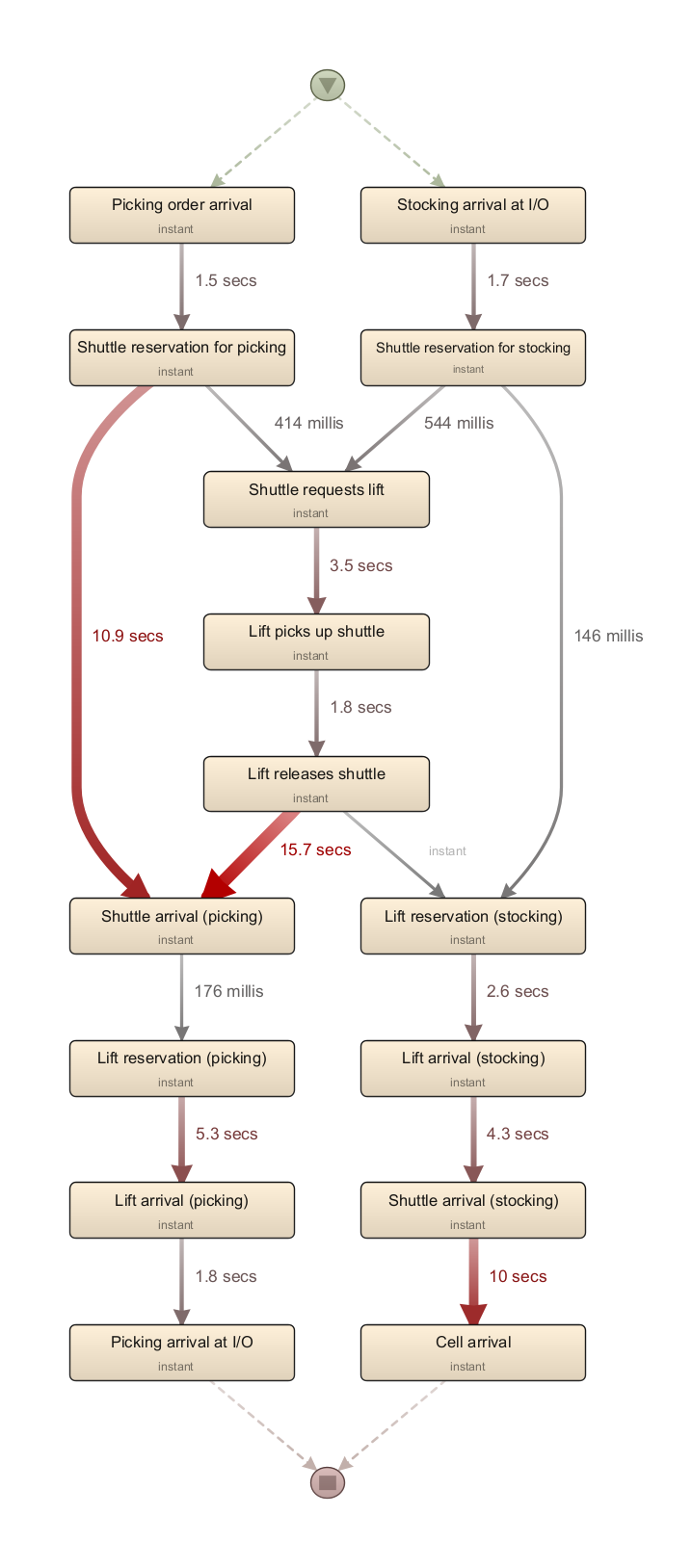}
        \caption{AFB-ROR}
        \label{fig:performance_afbror}
    \end{subfigure}

    \caption{Mean-duration process maps obtained from one representative baseline run for each policy. Edge annotations report the mean elapsed time between consecutive activities.}
    \label{fig:performance_pm}
\end{sidewaysfigure}

\come{Figures~\ref{fig:process_maps} and~\ref{fig:performance_pm} provide an
exploratory \comee{visualisation} of the process behaviour observed in one representative
baseline run for each policy. The baseline is used as a common illustrative setting so that the four policies can be compared under the same workload, geometry, and kinematic conditions.

The frequency maps in Figure~\ref{fig:process_maps} show that all policies share the same core stocking and picking structure, as expected because they are executed by the same simulation model. In the maps, node and edge annotations report occurrence counts, while edge thickness represents transition frequency. The central three-activity branch corresponds to the lift-assisted relocation of a shuttle. The main visual difference concerns the central shuttle-relocation path, consisting of the sequence \emph{Shuttle requests lift}, \emph{Lift picks up shuttle}, and \emph{Lift releases shuttle}. This path represents an additional relocation of the shuttle through the lift and should not be interpreted as the total use of lift-related activities within a mission.

In the representative baseline logs, the lift-assisted shuttle-relocation path appears in approximately 72\% of the missions under Random, 43\% under Deepest, 33\% under AFB-ROR, and only 3\% under Nearest. Thus, Nearest proceeds almost always through the direct path from shuttle reservation to shuttle arrival, whereas Random requires the intermediate relocation sequence in roughly seven missions out of ten. Deepest and AFB-ROR exhibit intermediate behaviour.}

\come{The mean-duration maps in Figure~\ref{fig:performance_pm} complement the
frequency view by reporting the mean elapsed time between consecutive activities.
The maps show that some of the longest transitions occur on the picking side,
between shuttle reservation or lift release and the subsequent shuttle arrival.
For the direct transition from picking-shuttle reservation to shuttle arrival, the
representative mean duration is 17.8~s under Random, 15.4~s under Deepest,
15.1~s under Nearest, and 10.9~s under AFB-ROR. When shuttle relocation is
required, the transition from lift release to picking-shuttle arrival takes,
respectively, 17.3, 17.0, 13.5, and 15.7~s.

The maps also highlight the final stocking transition from shuttle arrival to cell
arrival, whose representative mean duration is 13.0~s for Random, 12.6~s for
Deepest, 10.9~s for Nearest, and 10.0~s for AFB-ROR. A long mean duration alone, however, does not define the dominant bottleneck. The ranking reported in Table~\ref{tab:pm_drift} also accounts for how frequently each edge occurs.
Consequently, the final stocking transition remains dominant for Random, Deepest, and AFB-ROR in the baseline, whereas the frequently traversed picking-shuttle reservation path dominates under Nearest.

The maps therefore provide a visual interpretation of the replicated bottleneck
results. They show not only which transitions are relatively slow, but also how
their operational relevance depends on the frequency with which the policies
traverse them. Since each map represents a single selected run, it is used for
interpretation and illustration rather than for statistical inference, prevalence
estimation, or causal attribution.}

\subsection{\comee{Integrated interpretation of SMC and PM results}} 
\label{sec:integrated}
\comee{A direct baseline comparison illustrates the complementary contribution of the two analyses. SMC shows that AFB-ROR reduces endpoint picking and stocking times by 10.22 s and 8.59 s, respectively, relative to Random, with both differences remaining significant after Holm adjustment (\(p_{\mathrm{Holm}}<0.001\)). The replicated PM analysis localizes these aggregate differences within the operational process. The lift-assisted shuttle-relocation path occurs, on average across the 20 replications, in 70.1\% of Random picking missions, compared with 34.5\% under AFB-ROR.

Moreover, the mean duration of the direct transition from picking-shuttle reservation to shuttle arrival is 16.65 s under Random and 10.83 s under AFB-ROR; when shuttle relocation is required, the transition from lift release to picking-shuttle arrival lasts 17.02 s and 13.88 s, respectively. On the stocking side, the final transition from shuttle arrival to cell arrival lasts 12.86 s under Random and 9.58 s under AFB-ROR. 

Thus, whereas SMC establishes the magnitude and statistical reliability of the aggregate performance advantage, PM associates it with less frequent shuttle relocation and shorter shuttle-arrival and in-channel movements. These PM results provide diagnostic localization of the observed difference rather than causal attribution.

A second example concerns the 3.61 s statistically significant picking-time advantage of AFB-ROR over Nearest. PM identifies the picking-shuttle reservation-to-arrival transition as the dominant Nearest bottleneck in 116 of 120 supported windows; its replicated mean duration is 14.43 s under Nearest and 10.83 s under AFB-ROR. Nearest also shows a substantially more concentrated shuttle allocation (\(J_S=0.255\), compared with \(0.727\) for AFB-ROR).
Combined with the low baseline fleet utilisation of both policies, this concentration points to a localised shuttle-assignment and arrival issue rather than to a fleet-wide capacity shortage.

These examples use the baseline as a controlled setting in which workload, warehouse geometry, and vehicle kinematics are held fixed. The same integrated interpretation can be applied to the other evaluated scenarios: SMC first identifies reliable KPI differences under the scenario-specific conditions, while PM localises the process paths, time-intensive transitions, and resource-use patterns associated with those differences.
}

\section{Discussion}

\subsection{\comee{Integrative and Application-Oriented Contributions}}
\comee{This study makes an integrative and application-oriented contribution by addressing the research question through a coordinated combination of DES, Statistical Model Checking, and Process Mining}.
\come{The contribution of this study is integrative and application-oriented rather than algorithmic. We do not introduce a new DES formalism, SMC procedure, or PM algorithm. Instead, we organize these established techniques into a coordinated analysis workflow tailored to the evaluation of AVS/RS operations. In this workflow, DES generates controlled system executions, SMC provides statistically reliable estimates of aggregate performance indicators across independent simulation replications, and PM analyses the resulting event logs to characterise the operational patterns associated with the observed performance outcomes.

The three components therefore play complementary but distinct roles. SMC quantifies how the considered stocking and picking policies affect KPIs such as cycle times and tunnel occupancy, while explicitly accounting for simulation uncertainty through confidence intervals. PM moves the analysis from aggregate indicators to trace-level system behaviour by reconstructing control-flow relations, evaluating their temporal and performance evolution, identifying recurrent bottlenecks, and characterising resource-use patterns. In this way, PM helps explain how and where the differences observed through SMC materialize within the operational workflow.

The integration should therefore be understood as an analytical protocol rather than as a new tightly coupled SMC-PM technique. SMC provides replication-level evidence about the reliability of the observed KPI differences, whereas PM provides diagnostic evidence about the behavioural mechanisms associated with those differences. \comee{Section \ref{sec:integrated} makes this integrative contribution concrete. The workflow does not merely place SMC and PM results side by side, but links statistically supported differences in system-level KPIs to replicated process-level evidence showing where those differences materialise. This cross-level interpretation can be repeated across operating scenarios without changing either the simulation model or the analysis protocol.
The quantitative PM analyses, including control-flow and performance-drift measures, bottleneck rankings, and resource-use statistics, are computed across all independent simulation replications. 
The process-map figures shown in the manuscript are instead generated from single illustrative complete simulation runs, selected deterministically to be representative of the corresponding baseline-policy configuration. They provide a visual view of execution paths, transition frequencies, and time-intensive process segments that supports the interpretation of the quantitative results. Rather, the replicated PM measures, complemented by the illustrative maps, complement the aggregate SMC results by exposing operational patterns that cannot be directly observed from KPI trajectories alone}.

With respect to the RQ, reliability is enhanced through replicated simulation and confidence-controlled estimation, whereas interpretability is enhanced by relating aggregate performance outcomes to control-flow, performance, bottleneck, and resource-use patterns extracted from the event logs. To the best of our knowledge, the domain contribution consists of the first application of a combined SMC--PM workflow to the analysis of tier-to-tier, multi-deep AVS/RS operations across multiple policies and operational scenarios.

Finally, the study provides a technical and reproducibility contribution through the release of an open-source, \comee{parameterisable} AVS/RS simulator, its black-box integration with \mv{}, and the generation of PM-ready event logs and supporting analysis artifacts. The simulator is designed according to contemporary industrial guidelines and operational parameters documented in the literature \citep{Lupi2024}. To the best of our knowledge, it is the first publicly available discrete-event simulator specifically developed for multi-deep AVS/RS configurations with configurable layout characteristics, control policies, and operational parameters.

}

\subsection{Managerial Implications}
From a managerial perspective, the proposed integrated framework provides decision-makers with actionable insights that go beyond traditional performance rankings. 

\come{The key managerial implication is that storage and retrieval policy selection should be treated as a contingent, multi-objective decision rather than as a search for a universally dominant rule. Across the evaluated operating scenarios, AFB-ROR provides the strongest cycle-time performance, whereas Deepest is preferable when inventory consolidation is the primary objective. Nearest remains competitive in terms of execution time but may concentrate missions on a limited subset of shuttles, while Random primarily serves as a benchmark and does not achieve the best endpoint result for any of the considered KPIs. SMC establishes whether the observed performance differences are statistically credible under a given configuration, while PM exposes the control-flow, temporal, bottleneck, and resource-use patterns associated with those outcomes. The combined analysis therefore clarifies not only which policy performs better under the considered conditions, but also the operational mechanisms and trade-offs underlying that result.}

\comee{In practical terms, managers prioritizing short order-fulfilment cycles may select AFB-ROR as the preferred candidate, whereas those prioritizing inventory consolidation may deliberately select Deepest and accept its associated cycle-time penalty. Nearest may remain an appropriate proximity-based alternative when rapid execution is the main objective, but its use should be accompanied by monitoring or rebalancing of shuttle assignments if concentrated resource use is undesirable.
The shuttle-\comee{utilisation} analysis further shows that, under the fixed resource configuration considered in this study, fleet-wide shuttle \comee{utilisation} remains low.}
\come{However, some policies, particularly Nearest, concentrate missions on a limited subset of resources. The combination of low fleet-average \comee{utilisation} and concentrated mission assignment indicates a \comee{localised} allocation imbalance rather than a system-wide capacity shortage. This distinction is managerially relevant because it redirects intervention from potentially unnecessary fleet expansion toward targeted dispatching, zoning, or workload-rebalancing measures.}

\comee{Moreover, PM shows that the policies are associated with different recurrent bottlenecks. For Deepest and AFB-ROR, the dominant bottleneck is generally the final transition from shuttle arrival during stocking to cell arrival, whereas for Nearest it is typically the transition from picking-shuttle reservation to picking-shuttle arrival. These patterns help managers identify the process segment to investigate when a policy does not meet the desired operational objective.
At an operational level, the process-map layer can also support case-specific diagnostic drill-down. After SMC or routine monitoring identifies an unexpected performance deterioration, managers could generate process maps for the affected day or supported temporal window to inspect the paths and transitions involved. If operational logs also record failures, maintenance interventions, or other disruption states, these views could help distinguish recurring policy-related inefficiencies from exceptional events. Such maps would support local investigation and communication, rather than replace the replicated quantitative analysis used for general conclusions. Since failures and maintenance events are not represented in the present experiments, this application remains a prospective extension of the framework.
}

\come{More broadly, the framework can be employed as a repeatable, condition-dependent decision-support instrument rather than as a one-off evaluation exercise. Managers and system designers can update the simulation configuration to reflect the current workload profile, vehicle kinematics, initial inventory distribution, resource availability, and candidate control policies. SMC can then quantify the expected performance consequences and their associated uncertainty, while PM can identify the process segments, temporal changes, and resource-use patterns responsible for those consequences. Repeating this analysis as operating conditions or managerial priorities evolve makes it possible to reassess which policy is preferable, identify when a previously effective policy becomes suboptimal, and determine whether the appropriate response is a policy change, a resource-rebalancing intervention, or an increase in system capacity. The framework thus supports recurring and explainable what-if analysis, while the process-level evidence facilitates communication between technical analysts and managerial stakeholders.}

\come{Finally, the integration of SMC and PM supports risk-aware and explainable decision-making by distinguishing robust performance differences from condition-specific effects and by linking aggregate KPI outcomes to trace-level operational evidence. Managers are therefore not presented with a single opaque performance ranking, but with an auditable basis for selecting and revising control policies according to the warehouse's current objectives, constraints, and operating conditions.}

%% file: backup_bibliography_a_mano.bib
@InProceedings{10.1007/978-3-031-73709-1_13,
	author="Casaluce, Roberto
	and Tschaikowski, Max
	and Vandin, Andrea",
	editor="Margaria, Tiziana
	and Steffen, Bernhard",
	title="White-Box Validation of Collective Adaptive Systems by Statistical Model Checking and Process Mining",
	booktitle="Leveraging Applications of Formal Methods, Verification and Validation. REoCAS Colloquium in Honor of Rocco De Nicola",
	year="2025",
	publisher="Springer",
	address="Cham",
	pages="204--222",
	isbn="978-3-031-73709-1",
    doi = {10.1007/978-3-031-73709-1\_13}
}

@article{DBLP:journals/is/IncertoVA25,
  author       = {Emilio Incerto and
                  Andrea Vandin and
                  Sima Sarv Ahrabi},
  title        = {Stochastic conformance checking based on variable-length Markov chains},
  journal      = {Inf. Syst.},
  volume       = {133},
  pages        = {102561},
  year         = {2025},
  url          = {https://doi.org/10.1016/j.is.2025.102561},
  doi          = {10.1016/J.IS.2025.102561},
  bibsource    = {dblp computer science bibliography, https://dblp.org}
}

@inproceedings{DBLP:journals/corr/abs-2604-04543,
  author       = {Stefano Blando and
                  Giorgio Fagiolo and
                  Daniele Giachini and
                  Andrea Vandin and
                  Ernest Ivanaj},
  editor       = {Maurice H. ter Beek and
                  Gregor G{\"{o}}ssler},
  title        = {Statistical Model Checking of the Island Model: An Established Economic
                  Agent-Based Model of Endogenous Growth},
  booktitle    = {Proceedings of the 7th Workshop on Models for Formal Analysis of Real
                  Systems, {MARS} 2026, Turin, Italy, 12th April 2026},
  series       = {{EPTCS}},
  volume       = {443},
  pages        = {3--22},
  year         = {2026},
  month        = apr,
  url          = {https://doi.org/10.4204/EPTCS.443.2},
  doi          = {10.4204/EPTCS.443.2},
  bibsource    = {dblp computer science bibliography, https://dblp.org}
}

@misc{blando2026agenticagentbasedmodelsfeasibility,
      title={Towards Agentic Agent-based Models: Feasibility, Performance, and Statistical Model Checking}, 
      author={Stefano Blando and Emanuele Guerrazzi and Riccardo Porcedda and Giuseppe Squillace and Max Tschaikowski and Andrea Vandin},
      year={2026},
      eprint={2607.17948},
      archivePrefix={arXiv},
      primaryClass={cs.AI},
      url={https://arxiv.org/abs/2607.17948}, 
}

@article{DBLP:journals/corr/abs-2509-10977,
	author       = {Marco Pangallo and
	Daniele Giachini and
	Andrea Vandin},
	title        = {Statistical Model Checking of NetLogo Models},
	journal      = {CoRR},
	volume       = {abs/2509.10977},
	year         = {2025},
	doi          = {10.48550/ARXIV.2509.10977},
	bibsource    = {dblp computer science bibliography, https://dblp.org}
}

@inproceedings{DBLP:conf/wsc/GalpinGLV18,
	author    = {Vashti Galpin and
	Anastasis Georgoulas and
	Michele Loreti and
	Andrea Vandin},
	editor    = {Bj{\"{o}}rn Johansson and
	Sanjay Jain},
	title     = {Statistical Analysis of {CARMA} Models: an Advanced Tutorial},
	booktitle = {Proc. of {WSC}},
	pages     = {395--409},
	publisher = {{IEEE}},
	year      = {2018},
	doi       = {10.1109/WSC.2018.8632456},
	bibsource = {dblp computer science bibliography, https://dblp.org}
}

@article{DBLP:journals/scp/BruniCGLV15,
	author       = {Roberto Bruni and
	Andrea Corradini and
	Fabio Gadducci and
	Alberto Lluch{-}Lafuente and
	Andrea Vandin},
	title        = {Modelling and analyzing adaptive self-assembly strategies with Maude},
	journal      = {Sci. Comput. Program.},
	volume       = {99},
	pages        = {75--94},
	year         = {2015}
}

@inproceedings{DBLP:conf/birthday/BelznerNVW14,
	author       = {Lenz Belzner and
	Rocco De Nicola and
	Andrea Vandin and
	Martin Wirsing},
	editor       = {Shusaku Iida and
	Jos{\'{e}} Meseguer and
	Kazuhiro Ogata},
	title        = {Reasoning (on) Service Component Ensembles in Rewriting Logic},
	booktitle    = {Specification, Algebra, and Software - Essays Dedicated to Kokichi
	Futatsugi},
	series       = {Lecture Notes in Computer Science},
	volume       = {8373},
	pages        = {188--211},
	publisher    = {Springer},
	year         = {2014}
}

@InProceedings{10.1007/978-3-031-25383-6_18,
	author="Casaluce, Roberto
	and Burattin, Andrea
	and Chiaromonte, Francesca
	and Vandin, Andrea",
	editor="Cabanillas, Cristina
	and Garmann-Johnsen, Niels Frederik
	and Koschmider, Agnes",
	title="Process Mining Meets Statistical Model Checking: Towards a Novel Approach to Model Validation and Enhancement",
	booktitle="BPM Workshops",
	year="2023",
	publisher="Springer",
	pages="243--256",
	isbn="978-3-031-25383-6"
}

@article{DBLP:journals/jss/CorradiniFPRTV21,
	author    = {Flavio Corradini and
	Fabrizio Fornari and
	Andrea Polini and
	Barbara Re and
	Francesco Tiezzi and
	Andrea Vandin},
	title     = {A formal approach for the analysis of {BPMN} collaboration models},
	journal   = {JSS},
	volume    = {180},
	pages     = {111007},
	year      = {2021},
	bibsource = {dblp computer science bibliography, https://dblp.org}
}

@inproceedings{DBLP:conf/fm/VandinBLL18,
	author    = {Andrea Vandin and
	Maurice H. ter Beek and
	Axel Legay and
	Alberto Lluch{-}Lafuente},
	title     = {{QFLan: A Tool for the Quantitative Analysis of Highly Reconfigurable
	Systems}},
	booktitle = {FM},
	year      = {2018}
}

@article{DBLP:journals/tse/BeekLLV20,
	author    = {Maurice H. ter Beek and
	Axel Legay and
	Alberto Lluch{-}Lafuente and
	Andrea Vandin},
	title     = {{A Framework for Quantitative Modeling and Analysis of Highly (Re)configurable
	Systems}},
	journal   = {{IEEE} Trans. Software Eng.},
	volume    = {46},
	number    = {3},
	pages     = {321--345},
	year      = {2020}
}

@inproceedings{gilmore2014analysis,
	title={An analysis pathway for the quantitative evaluation of public transport systems},
	author={Gilmore, Stephen and Tribastone, Mirco and Vandin, Andrea},
	booktitle={IFM},
	year      = {2014}
}

@inproceedings{ciancia2016tool,
	title={A tool-chain for statistical spatio-temporal model checking of bike sharing systems},
	author={Ciancia, Vincenzo and Latella, Diego and Massink, Mieke and Pa{\v{s}}kauskas, Rytis and Vandin, Andrea},
	year={2017},
	booktitle={{ISOLA'17}}
}

@inproceedings{DBLP:conf/hpcs/PianiniSV14,
	author    = {Danilo Pianini and
	Stefano Sebastio and
	Andrea Vandin},
	title     = {Distributed statistical analysis of complex systems modeled through
	a chemical metaphor},
	booktitle = {{HPCS}},
	pages     = {416--423},
	year      = {2014}
}

@article{ter2021quantitative,
	title={{Quantitative security risk modeling and analysis with RisQFLan}},
	author={ter Beek, Maurice H and Legay, Axel and Lafuente, Alberto Lluch and Vandin, Andrea},
	journal={Computers \& Security},
	volume={109},
	pages={102381},
	year={2021},
	publisher={Elsevier}
}

@inproceedings{DBLP:conf/isola/BartolettiCJLMV22,
	author    = {Massimo Bartoletti and
	James Hsin{-}yu Chiang and
	Tommi Junttila and
	Alberto Lluch{-}Lafuente and
	Massimiliano Mirelli and
	Andrea Vandin},
	title     = {Formal Analysis of Lending Pools in Decentralized Finance},
	booktitle = {Proc. of ISoLA 2022},
	series    = {LNCS},
	volume    = {13703},
	pages     = {335--355},
	publisher = {Springer},
	year      = {2022}
}

@inproceedings{10.1007/978-3-031-75434-0_26,
	author = {Vandin, Andrea},
	title = {Statistical Model Checking of Python Agent-Based Models: An Integration of MultiVeStA and Mesa},
	year = {2025},
	isbn = {978-3-031-75433-3},
	publisher = {Springer-Verlag},
	address = {Berlin, Heidelberg},
	url = {https://doi.org/10.1007/978-3-031-75434-0_26},
	doi = {10.1007/978-3-031-75434-0_26},
	booktitle = {Bridging the Gap Between AI and Reality: Second International Conference, AISoLA 2024, Crete, Greece, October 30 – November 3, 2024, Proceedings},
	pages = {398–419},
	numpages = {22},
	location = {Crete, Greece}
}

@inproceedings{DBLP:conf/ifm/GilmoreRV17,
	author    = {Stephen Gilmore and
	Dani{\"{e}}l Reijsbergen and
	Andrea Vandin},
	title     = {Transient and Steady-State Statistical Analysis for Discrete Event
	Simulators},
	booktitle = {{IFM}},
	pages     = {145--160},
	publisher = {Springer},
	year      = {2017},
	bibsource = {dblp computer science bibliography, https://dblp.org}
}

@book{VanderAalst2016,
	title = {{Process Mining}},
	year = {2016},
	author = {van der Aalst, Wil M.P.},
	edition = {2nd},
	publisher = {Springer},
	isbn = {9783662498507},
}

@misc{deloitte,
	title        = {Deloitte Global Process Mining Survey},
	author       = {Deloitte},
	year         = {2025},
	url          = {https://bit.ly/deloittePMIndustry},
	note         = {Accessed on 13/01/2026}
}

@inproceedings{DBLP:conf/bpm/CasaluceBCV22,
	author       = {Roberto Casaluce and
	Andrea Burattin and
	Francesca Chiaromonte and
	Andrea Vandin},
	editor       = {Cristina Cabanillas and
	Niels Frederik Garmann{-}Johnsen and
	Agnes Koschmider},
	title        = {Process Mining Meets Statistical Model Checking: Towards a Novel Approach
	to Model Validation and Enhancement},
	booktitle    = {Business Process Management Workshops - {BPM} 2022 International Workshops,
	M{\"{u}}nster, Germany, September 11-16, 2022, Revised Selected
	Papers},
	series       = {Lecture Notes in Business Information Processing},
	volume       = {460},
	pages        = {243--256},
	publisher    = {Springer},
	year         = {2022},
	url          = {https://doi.org/10.1007/978-3-031-25383-6\_18},
	doi          = {10.1007/978-3-031-25383-6\_18},
	bibsource    = {dblp computer science bibliography, https://dblp.org}
}

@inproceedings{DBLP:conf/isola/CasaluceTV24,
	author       = {Roberto Casaluce and
	Max Tschaikowski and
	Andrea Vandin},
	editor       = {Tiziana Margaria and
	Bernhard Steffen},
	title        = {White-Box Validation of Collective Adaptive Systems by Statistical
	Model Checking and Process Mining},
	booktitle    = {Leveraging Applications of Formal Methods, Verification and Validation.
	REoCAS Colloquium in Honor of Rocco De Nicola - 12th International
	Symposium, ISoLA 2024, Crete, Greece, October 27-31, 2024, Proceedings,
	Part {I}},
	series       = {Lecture Notes in Computer Science},
	volume       = {15219},
	pages        = {204--222},
	publisher    = {Springer},
	year         = {2024},
	url          = {https://doi.org/10.1007/978-3-031-73709-1\_13},
	doi          = {10.1007/978-3-031-73709-1\_13},
	bibsource    = {dblp computer science bibliography, https://dblp.org}
}

@inproceedings{DBLP:conf/damocles/CasaluceBCLV24,
	author       = {Roberto Casaluce and
	Andrea Burratin and
	Francesca Chiaromonte and
	Alberto Lluch{-}Lafuente and
	Andrea Vandin},
	editor       = {Bernardo Breve and
	Giuseppe Desolda and
	Vincenzo Deufemia and
	Lucio Davide Spano},
	title        = {Enhancing Threat Model Validation: {A} White-Box Approach based on
	Statistical Model Checking and Process Mining},
	booktitle    = {Proceedings of the First International Workshop on Detection And Mitigation
	Of Cyber attacks that exploit human vuLnerabilitiES {(DAMOCLES} 2024)
	co-located with 17th International Conference on Advanced Visual Interfaces
	{(AVI} 2024), Arenzano (Genoa), Italy, Arenzano, Italy, June 4th,
	2024},
	series       = {{CEUR} Workshop Proceedings},
	volume       = {3713},
	pages        = {9--20},
	publisher    = {CEUR-WS.org},
	year         = {2024},
	url          = {https://ceur-ws.org/Vol-3713/paper\_2.pdf},
	bibsource    = {dblp computer science bibliography, https://dblp.org}
}

@inproceedings{DBLP:conf/vecos/Vandin24,
  author       = {Andrea Vandin},
  editor       = {Bernhard Steffen},
  title        = {Statistical Model Checking of Python Agent-Based Models: An Integration
                  of MultiVeStA and Mesa},
  booktitle    = {Bridging the Gap Between {AI} and Reality - Second International Conference,
                  AISoLA 2024, Crete, Greece, October 30 - November 3, 2024, Proceedings},
  series       = {Lecture Notes in Computer Science},
  volume       = {15217},
  pages        = {398--419},
  publisher    = {Springer},
  year         = {2024},
  url          = {https://doi.org/10.1007/978-3-031-75434-0\_26},
  doi          = {10.1007/978-3-031-75434-0\_26},
  bibsource    = {dblp computer science bibliography, https://dblp.org}
}

@article{dehrami2020,
title = {A simulation-based optimization approach to design optimal layouts for block stacking warehouses},
journal = {International Journal of Production Economics},
volume = {223},
pages = {107525},
year = {2020},
issn = {0925-5273},
doi = {https://doi.org/10.1016/j.ijpe.2019.107525},
url = {https://www.sciencedirect.com/science/article/pii/S0925527319303524},
author = {Shahab Derhami and Jeffrey S. Smith and Kevin R. Gue}
}

@article{ekren2020,
title = {A simulation-based experimental design for SBS/RS warehouse design by considering energy related performance metrics},
journal = {Simulation Modelling Practice and Theory},
volume = {98},
pages = {101991},
year = {2020},
issn = {1569-190X},
doi = {https://doi.org/10.1016/j.simpat.2019.101991},
url = {https://www.sciencedirect.com/science/article/pii/S1569190X19301248},
author = {Banu Y. Ekren}
}

@book{Aalst2016ProcessMining,
  author    = {Wil M. P. van der Aalst},
  title     = {Process Mining: Data Science in Action},
  edition   = {2nd},
  year      = {2016},
  publisher = {Springer},
  address   = {Berlin, Germany},
  doi       = {10.1007/978-3-662-49851-4}
}

@article{Aalst2007WorkflowMining,
  author    = {Wil M. P. van der Aalst and Ton Weijters and Laura Maruster},
  title     = {Workflow Mining: Discovering Process Models from Event Logs},
  journal   = {IEEE Transactions on Knowledge and Data Engineering},
  volume    = {16},
  number    = {9},
  pages     = {1128--1142},
  year      = {2004},
  doi       = {10.1109/TKDE.2004.47}
}

@article{Alnahas2023,
  author       = {Alnahas, Jasim},
  title        = {Application of Process Mining in Logistic Processes of Manufacturing Organizations: A Systematic Review},
  journal      = {Sustainability},
  year         = {2023},
  volume       = {15},
  number       = {15},
  pages        = {11783},
  doi          = {10.3390/su151511783},
  url          = {https://doi.org/10.3390/su151511783}
}

@incollection{Kucukyasar2021,
title = {13 - Energy efficient automated warehouse design},
editor = {John R. Vacca},
booktitle = {Solving Urban Infrastructure Problems Using Smart City Technologies},
publisher = {Elsevier},
pages = {269-292},
year = {2021},
isbn = {978-0-12-816816-5},
doi = {https://doi.org/10.1016/B978-0-12-816816-5.00013-9},
url = {https://www.sciencedirect.com/science/article/pii/B9780128168165000139},
author = {Melis K{\"u}{\c{c}}{\"u}kya{\c{s}}ar and Banu Yetkin Ekren and Tone Lerher}
}

@article{Agha2018,
author = {Agha, Gul and Palmskog, Karl},
year = {2018},
month = {01},
pages = {1-39},
title = {A Survey of Statistical Model Checking},
volume = {28},
journal = {ACM Transactions on Modeling and Computer Simulation},
doi = {10.1145/3158668}
}

@misc{fluxicon_disco,
  title        = {Disco: Process Mining Software},
  author       = {Fluxicon},
  year         = {2025},
  url          = {https://fluxicon.com/disco/},
  note         = {Software project}
}

@misc{pm4py_website,
  title        = {PM4Py: Process Mining for Python},
  author       = {Berti, Alessandro and van Zelst, Sebastiaan J. and PM4Py Community},
  year         = {2025},
  url          = {https://pm4py.fit.fraunhofer.de/},
  note         = {Software project}
}

@ARTICLE{Guerrazzi2024,
	author = {Guerrazzi, Emanuele and Mininno, Valeria and Aloini, Davide},
	title = {Balancing picking and outbound loading efficiency in an SBS/RS through a digital twin},
	year = {2024},
	journal = {Flexible Services and Manufacturing Journal},
	doi = {10.1007/s10696-024-09554-w},
	url = {https://www.scopus.com/inward/record.uri?eid=2-s2.0-85195980418&doi=10.1007%2fs10696-024-09554-w&partnerID=40&md5=00bee5b2f2e95337d42b0975a1e72197},
	type = {Article},
	publication_stage = {Article in press},
	source = {Scopus},
	note = {Cited by: 1; All Open Access, Hybrid Gold Open Access}
}

@inproceedings{Liiv2014,
  author = {Liiv, Innar and Lepik, Ott},
  title = {Business Process Mining in Warehouses: a Case Study},
  booktitle = {Proceedings of the 11th International Baltic Conference on Databases and Information Systems},
  year = {2014},
  pages = {387--394},
  publisher = {TUT Press}
}

@inproceedings{Kedem2018,
  author = {Kedem-Yemini, Sagit and Mamon, Naor S. and Mashiah, Gal},
  title = {An Analysis of Cargo Release Services with Process Mining: A Case Study in a Logistics Company},
  booktitle = {Proceedings of the International Conference on Industrial Engineering and Operations Management},
  year = {2018},
  pages = {725--736},
  address = {Washington DC, USA}
}

@inproceedings{Montianrat2020,
  author = {Montianrat, S.},
  title = {Analysis of Warehouse Operations using Process Mining},
  year = {2020}
}

@inproceedings{Sen2005,
  author = {Sen, Koushik and Viswanathan, Mahesh and Agha, Gul},
  title = {VESTA: A Statistical Model-checker and Analyzer for Probabilistic Systems},
  booktitle = {Proceedings of the 17th Conference on Computer Aided Verification (CAV'05)},
  year = {2005},
  publisher = {Springer}
}

@misc{arena,
  title        = {Arena Simulation Software},
  author       = {{Rockwell Automation}},
  year         = {2025},
  note         = {Commercial simulation software},
  url          = {https://www.rockwellautomation.com/en-us/products/software/arena-simulation.html}
}

@misc{prom,
  title        = {ProM: Process Mining framework},
  author       = {{ProM Tools}},
  year         = {2025},
  note         = {Open-source framework for process mining},
  url          = {https://promtools.org/}
}

@article{Xu2025,
  author = {Xu, Huiyue and Shao, Juping and Sun, Yanan},
  title = {Research on order batching optimization based on improved {NSGA-II} algorithm},
  journal = {PLoS One},
  year = {2025},
  volume = {20},
  number = {2},
  pages = {e0319182},
  doi = {10.1371/journal.pone.0319182}
}

@article{Derhami2020,
title = {A simulation-based optimization approach to design optimal layouts for block stacking warehouses},
journal = {International Journal of Production Economics},
volume = {223},
pages = {107525},
year = {2020},
issn = {0925-5273},
doi = {https://doi.org/10.1016/j.ijpe.2019.107525},
url = {https://www.sciencedirect.com/science/article/pii/S0925527319303524},
author = {Shahab Derhami and Jeffrey S. Smith and Kevin R. Gue}
}

@inproceedings{liu2020,
  title     = {Research on Dynamic Prediction Model of Orders in E-Commerce Distribution Center Based on GA-BP Neural Network},
  author    = {Liu, Y. and Zhang, H. and Wang, J.},
  booktitle = {Proceedings of the 2020 International Conference on Logistics Engineering and Intelligent Transportation Systems},
  pages     = {1--6},
  year      = {2020},
  publisher = {Atlantis Press},
  doi       = {10.2991/aebmr.k.201203.001}
}

@article{Lorenz2025,
author = {Lorenz, Catherine and Otto, Alena and Gendreau, Michel},
title = {Picking Operations in Warehouses With Dynamically Arriving Orders: How Good is Reoptimization?},
journal = {Networks},
volume = {86},
number = {2},
pages = {157-173},
doi = {https://doi.org/10.1002/net.22281},
url = {https://onlinelibrary.wiley.com/doi/abs/10.1002/net.22281},
eprint = {https://onlinelibrary.wiley.com/doi/pdf/10.1002/net.22281},
year = {2025}
}

@article{Merschforman2019,
title = {Decision rules for robotic mobile fulfillment systems},
journal = {Operations Research Perspectives},
volume = {6},
pages = {100128},
year = {2019},
issn = {2214-7160},
doi = {https://doi.org/10.1016/j.orp.2019.100128},
url = {https://www.sciencedirect.com/science/article/pii/S2214716019300946},
author = {M. Merschformann and T. Lamballais and M.B.M. {de Koster} and L. Suhl}
}

@article{lehmann2021travel,
  title={Travel time model for multi-deep automated storage and retrieval system with a homogeneous allocation structure},
  author={Lehmann, T. and Hu{\ss}mann, J.},
  journal={Logistics Research},
  volume={14},
  number={5},
  year={2021},
  doi={10.23773/2021_5},
  note={Conferma l'uso della Random policy nei modelli di tempo di viaggio per sistemi multi-deep.}
}

@Inbook{Lerher2024,
author="Lerher, Tone
and Marolt, Jakob
and Sgarbossa, Fabio
and Ekren, Banu Y.
and Dukic, Goran",
editor="Manzini, Riccardo
and Accorsi, Riccardo",
title="Design and Operation of Single- and Multi-deep Shuttle-Based Storage and Retrieval Systems (SBS/RS)",
bookTitle="Warehousing and Material Handling Systems for the Digital Industry: The New Challenges for the Digital Circular Economy",
year="2024",
publisher="Springer International Publishing",
address="Cham",
pages="333--375",
isbn="978-3-031-50273-6",
doi="10.1007/978-3-031-50273-6_13",
url="https://doi.org/10.1007/978-3-031-50273-6_13"
}

@article{mcinerney2022increasing,
  title={Increasing warehouse throughput through the implementation of class-based storage},
  author={McInerney, S. E. and others},
  journal={South African Journal of Industrial Engineering},
  volume={33},
  number={2},
  year={2022},
  note={Descrive la politica "Closest-Open" (Nearest) come quella che assegna il bin vuoto più vicino.}
}

@article{lerher2010travel,
  title={Travel time model for double-deep automated storage and retrieval systems},
  author={Lerher, Tone and Sraml, Matjaz and Potrc, Iztok and Borovinsek, Matej},
  journal={International Journal of Production Research},
  volume={48},
  number={11},
  pages={3151--3172},
  year={2010},
  publisher={Taylor \& Francis},
  note={Tratta le implicazioni delle strategie multi-deep (deepest) sui tempi di viaggio.}
}

@article{Garcia2019,
  title   = {Process mining techniques and applications -- A systematic mapping study},
  author  = {Garcia, Cleiton dos Santos and Meincheim, Alex and Faria Junior, Elio Ribeiro and Dallagassa, Marcelo Rosano and Sato, Denise Maria Vecino and Carvalho, Deborah Ribeiro and Santos, Eduardo Alves Portela and Scalabrin, Edson Emilio},
  journal = {Expert Systems with Applications},
  year    = {2019},
  volume  = {133},
  pages   = {260--295},
  issn    = {0957-4174},
  doi     = {10.1016/j.eswa.2019.05.003}
}

@article{Knoll2019,
  title   = {Enabling value stream mapping for internal logistics using multidimensional process mining},
  author  = {Knoll, Dino and Reinhart, Gunther and Pr{\"u}glmeier, Marco},
  journal = {Expert Systems with Applications},
  year    = {2019},
  volume  = {124},
  pages   = {130--142},
  issn    = {0957-4174},
  doi     = {10.1016/j.eswa.2019.01.026}
}

@article{Chan2011,
  title   = {Improving the productivity of order picking of a manual-pick and multi-level rack distribution warehouse through the implementation of class-based storage},
  author  = {Chan, Felix T. S. and Chan, H. K.},
  journal = {Expert Systems with Applications},
  year    = {2011},
  volume  = {38},
  number  = {3},
  pages   = {2686--2700},
  issn    = {0957-4174},
  doi     = {10.1016/j.eswa.2010.08.058}
}

@misc{zenodo_simulator,
  author       = {Guerrazzi, Emanuele and Vandin, Andrea},
  title        = {{AVS/RS Simulator - MultiVeSta Integration}},
  year         = {2025},
  publisher    = {Zenodo},
  doi          = {10.5281/zenodo.21822783},
  url          = {https://doi.org/10.5281/zenodo.21822783}
}

@conference{Kaczmarek2014,
	author = {Kaczmarek, Sascha and Goldenstein, Jonas and Ten Hompel, Michael},
	title = {Performance analysis of autonomous vehicle storage and retrieval systems depending on storage management policies},
	year = {2014},
	journal = {IEEE International Conference on Industrial Engineering and Engineering Management},
	volume = {2015-January},
	pages = {1424 – 1428},
	doi = {10.1109/IEEM.2014.7058873},
	url = {https://www.scopus.com/inward/record.uri?eid=2-s2.0-84988273258&doi=10.1109%2fIEEM.2014.7058873&partnerID=40&md5=2996c08fb9f44ad93ded63b98e2639b3},
	type = {Conference paper},
	publication_stage = {Final},
	source = {Scopus},
}

@article{Battini2016,
title = {Human energy expenditure in order picking storage assignment: A bi-objective method},
journal = {Computers \& Industrial Engineering},
volume = {94},
pages = {147-157},
year = {2016},
issn = {0360-8352},
doi = {https://doi.org/10.1016/j.cie.2016.01.020},
url = {https://www.sciencedirect.com/science/article/pii/S0360835216300134},
author = {Daria Battini and Christoph H. Glock and Eric H. Grosse and Alessandro Persona and Fabio Sgarbossa},
}

@article{holm1979,
  title   = {A Simple Sequentially Rejective Multiple Test Procedure},
  author  = {Holm, Sture},
  journal = {Scandinavian Journal of Statistics},
  volume  = {6},
  number  = {2},
  pages   = {65--70},
  year    = {1979},
  issn    = {0303-6898},
  url     = {https://www.jstor.org/stable/4615733}
}

@misc{legay2010,      title={Statistical Model Checking : An Overview}, 
      author={Axel Legay and Benoit Delahaye},
      year={2010},
      eprint={1005.1327},
      archivePrefix={arXiv},
      primaryClass={cs.LO},
      url={https://arxiv.org/abs/1005.1327}, 
}

@inproceedings{Cosma2023DCRPetriNets,
  author    = {Cosma, Vlad Paul and Hildebrandt, Thomas T. and Slaats, Tijs},
  title     = {Transforming Dynamic Condition Response Graphs to Safe Petri Nets},
  booktitle = {Application and Theory of Petri Nets and Concurrency},
  editor    = {Gomes, Luis and Lorenz, Robert},
  series    = {Lecture Notes in Computer Science},
  volume    = {13929},
  pages     = {417--439},
  publisher = {Springer},
  year      = {2023},
  doi       = {10.1007/978-3-031-33620-1_22}
}


%% file: bibliography_a_mano.bib
@book{10.5555/554952,
author = {Law, Averill M. and Kelton, David M.},
title = {Simulation Modeling and Analysis},
year = {2015},
isbn = {978-0-07-340132-4},
publisher = {McGraw-Hill Higher Education, http://www.averill-law.com/simulation-book/},
edition = {5th}
}


%% file: references.bib
@article{Lupi2024,
   author = {Giacomo Lupi and Riccardo Accorsi and Ilaria Battarra and Riccardo Manzini and Gabriele Sirri},
   doi = {10.1007/s00170-024-13160-3},
   issn = {14333015},
   issue = {3-4},
   journal = {International Journal of Advanced Manufacturing Technology},
   month = {3},
   pages = {1449-1466},
   publisher = {Springer Science and Business Media Deutschland GmbH},
   title = {Space efficiency and throughput performance in AVS/RS under variant lane depths},
   volume = {131},
   year = {2024}
}

@article{Casaluce2024,
   author = {Roberto Casaluce and Andrea Burattin and Francesca Chiaromonte and Alberto Lluch Lafuente and Andrea Vandin},
   doi = {10.1016/j.jss.2024.111983},
   issn = {01641212},
   journal = {Journal of Systems and Software},
   month = {4},
   publisher = {Elsevier Inc.},
   title = {White-box validation of quantitative product lines by statistical model checking and process mining},
   volume = {210},
   year = {2024}
}

@article{Marchet2013,
   author = {Gino Marchet and Marco Melacini and Sara Perotti and Elena Tappia},
   doi = {10.1080/00207543.2013.778430},
   issn = {00207543},
   issue = {14},
   journal = {International Journal of Production Research},
   month = {7},
   pages = {4365-4387},
   title = {Development of a framework for the design of autonomous vehicle storage and retrieval systems},
   volume = {51},
   year = {2013}
}

@article{Marolt2023,
   author = {Jakob Marolt and Simona Šinko and Tone Lerher},
   doi = {10.1080/00207543.2022.2087568},
   issn = {1366588X},
   issue = {15},
   journal = {International Journal of Production Research},
   pages = {4991-5008},
   publisher = {Taylor and Francis Ltd.},
   title = {Model of a multiple-deep automated vehicles storage and retrieval system following the combination of Depth-First storage and Depth-First relocation strategies},
   volume = {61},
   year = {2023}
}

@article{Yang2023,
   author = {Y Yu and J Yang and X Guo},
   doi = {10.1080/00207543.2022.2158244},
   issue = {24},
   journal = {International Journal of Production Research},
   pages = {8634-8653},
   title = {Retrieval sequencing in autonomous vehicle storage and retrieval systems},
   volume = {61},
   year = {2023}
}

@article{Lerher2021,
   author = {Tone Lerher and Mirko Ficko and Iztok Palčič},
   doi = {10.1016/j.apm.2020.10.032},
   issn = {0307904X},
   journal = {Applied Mathematical Modelling},
   month = {3},
   pages = {1004-1022},
   publisher = {Elsevier Inc.},
   title = {Throughput performance analysis of Automated Vehicle Storage and Retrieval Systems with multiple-tier shuttle vehicles},
   volume = {91},
   year = {2021}
}

@article{Vandin2022,
   author = {Andrea Vandin and Daniele Giachini and Francesco Lamperti and Francesca Chiaromonte},
   doi = {10.1016/j.jedc.2022.104458},
   issn = {01651889},
   journal = {Journal of Economic Dynamics and Control},
   month = {10},
   publisher = {Elsevier B.V.},
   title = {Automated and distributed statistical analysis of economic agent-based models},
   volume = {143},
   year = {2022}
}

@article{YetkinEkren2021,
   author = {B Yetkin Ekren},
   doi = {10.1080/00207543.2020.1720927},
   issue = {4},
   journal = {International Journal of Production Research},
   pages = {1107-1126},
   title = {A multi-objective optimisation study for the design of an AVS/RS warehouse},
   volume = {59},
   year = {2021}
}

@article{Guerrazzi2019,
   author = {Emanuele Guerrazzi and Valeria Mininno and Davide Aloini and Riccardo Dulmin and Claudio Scarpelli and Marco Sabatini},
   doi = {10.3390/su11143817},
   issn = {20711050},
   issue = {14},
   journal = {Sustainability (Switzerland)},
   publisher = {MDPI},
   title = {Energy evaluation of deep-lane autonomous vehicle storage and retrieval system},
   volume = {11},
   year = {2019}
}

@misc{Boysen2019,
   author = {Nils Boysen and René de Koster and Felix Weidinger},
   doi = {10.1016/j.ejor.2018.08.023},
   issn = {03772217},
   issue = {2},
   journal = {European Journal of Operational Research},
   month = {9},
   pages = {396-411},
   publisher = {Elsevier B.V.},
   title = {Warehousing in the e-commerce era: A survey},
   volume = {277},
   year = {2019}
}

@article{Ekren2024,
   author = {Banu Yetkin Ekren and Tone Lerher and Melis Küçükyaşar and Boris Jerman},
   doi = {10.1080/00207543.2023.2199101},
   issn = {1366588X},
   issue = {5},
   journal = {International Journal of Production Research},
   pages = {1648-1662},
   publisher = {Taylor and Francis Ltd.},
   title = {Cost and performance comparison of tier-captive SBS/RS with a novel AVS/RS/ML},
   volume = {62},
   year = {2024}
}

@article{Ferrari2024,
   author = {Andrea Ferrari and Antonio Carlin and Carlo Rafele and Giovanni Zenezini},
   doi = {10.1007/s00170-023-12660-y},
   issn = {14333015},
   issue = {11},
   journal = {International Journal of Advanced Manufacturing Technology},
   month = {4},
   pages = {5369-5382},
   publisher = {Springer Science and Business Media Deutschland GmbH},
   title = {A method for developing and validating simulation models for automated storage and retrieval system digital twins},
   volume = {131},
   year = {2024}
}

@article{Ekren2011,
   author = {Banu Y. Ekren and Sunderesh S. Heragu},
   doi = {10.1016/j.simpat.2011.02.008},
   issn = {1569190X},
   issue = {7},
   journal = {Simulation Modelling Practice and Theory},
   pages = {1640-1650},
   publisher = {Elsevier B.V.},
   title = {Simulation based performance analysis of an autonomous vehicle storage and retrieval system},
   volume = {19},
   year = {2011}
}

@misc{Kizi2024,
   author = {Elif Burcu Kızılırmak and Sinan Öztaş and Nadide Çağlayan and Mahmut Tutam},
   doi = {10.1007/s11831-024-10084-9},
   issn = {18861784},
   issue = {6},
   journal = {Archives of Computational Methods in Engineering},
   month = {8},
   pages = {3455-3472},
   publisher = {Springer Science and Business Media B.V.},
   title = {Design and Performance Measures of AVS/R Systems: A Bibliometric Literature Review},
   volume = {31},
   year = {2024}
}

@article{Jerman2021,
   author = {Boris Jerman and Banu Yetkin Ekren and Melis Küçükyaşar and Tone Lerher},
   doi = {10.3390/app11052283},
   issn = {20763417},
   issue = {5},
   journal = {Applied Sciences (Switzerland)},
   month = {3},
   pages = {1-16},
   publisher = {MDPI AG},
   title = {Simulation‐based performance analysis for a novel avs/rs technology with movable lifts},
   volume = {11},
   year = {2021}
}

@article{Battarra2022,
   author = {Ilaria Battarra and Riccardo Accorsi and Riccardo Manzini and Sara Rubini},
   doi = {10.1007/s00170-022-09375-x},
   issn = {14333015},
   issue = {1-2},
   journal = {International Journal of Advanced Manufacturing Technology},
   month = {7},
   pages = {1191-1217},
   publisher = {Springer Science and Business Media Deutschland GmbH},
   title = {Hybrid model for the design of a deep-lane multisatellite AVS/RS},
   volume = {121},
   year = {2022}
}

@inproceedings{Agalianos2020,
   author = {K. Agalianos and S. T. Ponis and E. Aretoulaki and G. Plakas and O. Efthymiou},
   doi = {10.1016/j.promfg.2020.10.228},
   issn = {23519789},
   booktitle = {Procedia Manufacturing},
   pages = {1636-1641},
   publisher = {Elsevier B.V.},
   title = {Discrete event simulation and digital twins: Review and challenges for logistics},
   volume = {51},
   year = {2020}
}

@inproceedings{Mahendrawathi2015,
   author = {E. R. Mahendrawathi and Hanim Maria Astuti and Ika Rakhma Kusuma Wardhani},
   doi = {10.1007/978-3-319-19509-4_9},
   isbn = {9783319195087},
   issn = {18651348},
   booktitle = {Lecture Notes in Business Information Processing},
   pages = {115-127},
   publisher = {Springer Verlag},
   title = {Material movement analysis for warehouse business process improvement with process mining: A case study},
   volume = {219},
   year = {2015}
}

@article{Napieraj2025,
   author = {Aneta Napieraj and Natalia Pyzik},
   doi = {10.3390/app15168968},
   issn = {20763417},
   issue = {16},
   journal = {Applied Sciences (Switzerland)},
   month = {8},
   publisher = {Multidisciplinary Digital Publishing Institute (MDPI)},
   title = {Analysis of a Logistics Process Based on the Event Log},
   volume = {15},
   year = {2025}
}

@article{Baouya2024,
   author = {Abdelhakim Baouya and Salim Chehida and Saddek Bensalem and Levent Gürgen and Richard Nicholson and Miquel Cantero and Mario Diaznava and Enrico Ferrera},
   doi = {10.1007/s11227-023-05483-x},
   issn = {15730484},
   issue = {1},
   journal = {Journal of Supercomputing},
   month = {1},
   pages = {1206-1237},
   publisher = {Springer},
   title = {Deploying warehouse robots with confidence: the BRAIN-IoT framework’s functional assurance},
   volume = {80},
   year = {2024}
}

@inproceedings{sebastio2013multivesta,
	author    = {Stefano Sebastio and
	Andrea Vandin},
	title     = {{MultiVeStA}: statistical model checking for discrete event simulators},
	booktitle = {Proc. of ValueTools'13},
	pages     = {310--315},
	publisher = {{ICST/ACM}},
	year      = {2013},
	bibsource = {dblp computer science bibliography, https://dblp.org}
}
